\documentclass[twocolumn,showpacs,preprintnumbers,amsmath,amssymb]{revtex4}

\usepackage{graphicx}
\usepackage{dcolumn}
\usepackage{bm}

\begin{document}

\title{Universal fractal structures in the weak interaction of solitary waves in generalized nonlinear Schr\"{o}dinger equations}

\author{Yi  Zhu}
\email{zhuyi03@mails.tsinghua.edu.cn}
\affiliation{%
Zhou Pei-Yuan Center for Applied Mathematics, Tsinghua University,
Beijing 100084, P.R.China\\
}%

\author{Jianke Yang}
\email{jyang@cems.uvm.edu} \affiliation{
Department of Mathematics and Statistics, University of Vermont, 16 Colchester Avenue, Burlington, VT 05401, USA\\
}%


\begin{abstract}
Weak interactions of solitary waves in the generalized nonlinear
Schr\"{o}dinger equations are studied. It is first shown that these
interactions exhibit similar fractal dependence on initial
conditions for different nonlinearities. Then by using the
Karpman-Solov'ev method, a universal system of dynamical equations
is derived for the velocities, amplitudes, positions and phases of
interacting solitary waves. These dynamical equations contain a
single parameter, which accounts for the different forms of
nonlinearity. When this parameter is zero, these dynamical equations
are integrable, and the exact analytical solutions are derived. When
this parameter is non-zero, the dynamical equations exhibit fractal
structures which match those in the original wave equations both
qualitatively and quantitatively. Thus the universal nature of
fractal structures in the weak interaction of solitary waves is
analytically established. The origin of these fractal structures is
also explored. It is shown that these structures bifurcate from the
initial conditions where the solutions of the integrable dynamical
equations develop finite-time singularities. Based on this
observation, an analytical criterion for the existence and locations
of fractal structures is obtained. Lastly, these analytical results
are applied to the generalized nonlinear Schr\"{o}dinger equations
with various nonlinearities such as the saturable nonlinearity, and
predictions on their weak interactions of solitary waves are made.
\end{abstract}

\maketitle

\section{\label{sec:level1}Introduction}
\quad

Solitary wave interactions are a fascinating and important
phenomenon for both physical and mathematical reasons. Physically,
such interactions have arisen in a wide array of disciplines such as
water waves \cite{Ablowitz_Segur}, optics
\cite{Hasegawa_Kodama,Krolikowski,Anastassiou,Chen_Kivshar,
Kivshar_Agrawal, Aceves_stud}, and Josephson junctions \cite{Yukon}.
For instance, in soliton-based fiber communication systems, optical
pulses traveling in different frequency channels pass through each
other, giving rise to collisions (strong interactions) of solitary
waves. In the same frequency channel, neighboring optical pulses
interfere with each other through overlapping tails, giving rise to
weak interactions of solitary waves. Motivated by these physical
applications, solitary wave interactions has been studied
extensively in both the mathematical and physical communities. If
the system is integrable, collisions of solitons are elastic
\cite{Ablowitz_Segur}, and weak interactions of solitons exhibit
interesting yet simple behaviors
\cite{Hasegawa_Kodama,Karpman_Solovev,Gerdjikov_PRL,Gerdjikov_Yang,Yang_Manakov}.
However, in non-integrable systems, solitary wave interactions can
be far more complex. The first sign of this complexity was reported
by Ablowitz, et al. \cite{Ablowitz_Kruskal} for kink and anti-kink
collisions in the $\phi^4$ model where, inside the trapping
interval, a reflection window was found. Later extensive numerical
studies on this model by Campbell, et al. \cite{Campbell1,
Campbell2, Campbell3, Campbell4} revealed that in fact, sequences of
two- and more-bounce reflection windows exist, and the physical
mechanism for these refection windows is a resonant energy transfer
between the translational motion and internal modes of
kinks/antikinks. Anninos, et al. \cite{anninos} pointed out further
that there is a fractal structure in kink-antikink collisions. Using
a collective-coordinate (i.e., variational) approach, they derived a
set of fourth-order ordinary differential equations (ODEs) for these
collisions, and these ODEs exhibit qualitatively similar fractal
structures as in the $\phi^4$ model (a comprehensive review on
kink-antikink collisions in $\phi^4$-type equations can be found in
\cite{Kudryavtsev}). These complex dynamics turn out to be not
restricted to kink-antikink collisions. Indeed, similar phenomena
have been reported on kink-defect collisions in the sine-Gordon and
$\phi^4$ models \cite{Kivshar1, Kivshar2, Kivshar3}, as well as
vector-soliton collisions in the coupled nonlinear Schr\"{o}dinger
(NLS) equations \cite{YangTan,YangTan2,TanYang}. Furthermore,
fractal scattering has also been reported on weak interactions of
breathers in a weakly discrete sine-Gordon equation
\cite{Dmitriev_Kivshar} and weak interactions of solitary waves in a
weakly discrete NLS equation \cite{Dmitriev}. Recently, Goodman and
Haberman \cite{Goodman_Haberman1, Goodman_Haberman2,
 Goodman_Haberman3} provided a deep analysis on the
collective-coordinate models (ODEs) for kink-antikink collisions in
the $\phi^4$ model \cite{anninos}, kink-defect collisions in the
sine-Gordon model \cite{Kivshar1}, and vector-soliton collisions in
the coupled NLS equations \cite{Ueda_Kath, TanYang}, using
sophisticated dynamical system techniques. They derived analytical
formulas for the locations of reflection-window sequences, which
agree qualitatively with numerical results on the original partial
differential equations (PDEs). Their results shed much light on the
origins of these window sequences and fractal structures, especially
from a mathematical point of view.

Despite the above progress on solitary wave interactions, our
understanding on these phenomena is far from satisfactory. On the
collision of solitary waves, the analysis done so far were all based
on approximate collective-coordinate approaches, hence the reduced
ODE models can only provide qualitative results at best. Many
features reported in the ODE models can not be seen in the PDE
simulations, thus it is not possible to make a reliable prediction
on the collision dynamics based on those ODE models and their
analysis. In addition, the ODE models obtained from the
collective-coordinate approaches not only are complicated, but also
differ significantly from one PDE system to another. This forced
previous researchers to analyze each PDE and its reduced ODE systems
on an individual basis, which prevents an overall understanding on
collision processes of solitary waves. On the weak interaction of
solitary waves, the situation is even less satisfactory. The fractal
nature of this weak interaction was reported only for systems which
are weakly perturbed integrable systems (sine-Gordon and NLS
equations, to be more specific) \cite{Dmitriev_Kivshar, Dmitriev}.
It is not known yet whether similar phenomena arise in strongly
non-integrable equations. More seriously, the previous work on this
subject is largely numerical. No analysis has been attempted yet
(not even the approximate collective-coordinate studies). Thus an
analytical understanding on weak interactions of solitary waves is a
completely open question.

In this paper, we study weak interactions of solitary waves in a
whole class of generalized NLS equations (with arbitrary
nonlinearities) both analytically and numerically. These generalized
NLS equations are not weak perturbations of the NLS equation in
general. First we show by direct PDE simulations that these weak
interactions for different nonlinearities exhibit similar fractal
structures on initial parameters of solitary waves. This establishes
that fractal scattering is a common feature of weak interactions in
this class of generalized NLS equations. Next, we rigorously derive
a universal system of dynamical equations (ODEs) for the velocities,
amplitudes, positions and phases of interacting solitary waves in
this class of PDEs by the Karpman-Solov'ev method. This universal
ODE system is remarkably simple, and it contains only a single
parameter which depends on the individual PDEs (after variable
rescalings). When this parameter is zero, these dynamical equations
are integrable, and their exact analytical solutions are derived.
When this parameter is non-zero, the dynamical equations are found
to exhibit fractal structures for a wide range of initial
conditions. These fractal structures match those in the original
PDEs both qualitatively and quantitatively, thus the universal
nature of fractal scattering in the weak interaction of solitary
waves is analytically established. We further explore the origin of
these fractal structures. Our numerical studies on the ODE system
show that these fractal structures bifurcate from the initial
conditions where the solutions of the integrable dynamical equations
develop finite-time singularities. Based on this observation, we
present an analytical criterion for the existence and locations of
fractal structures. One corollary from this criterion is that when
the initial separation velocity is above a certain threshold value,
fractal structures should disappear --- a prediction which agrees
with our PDE numerics as well as previous numerics on the weakly
discrete NLS equation (see Fig. 6 in Ref. \cite{Dmitriev}). Lastly,
we apply these analytical results to the generalized NLS equations
with various nonlinearities such as the cubic-quintic, exponential
and saturable nonlinearities, and make detailed predictions on the
dynamics of their weak interactions.

This paper is structured as follows. In Sec. 2, we describe
individual solitary waves in the generalized NLS equations. In Sec.
3, we present direct PDE simulation results on weak interactions of
solitary waves in the generalized NLS equations with two different
nonlinearities, and reveal the common (universal) fractal structures
in this class of PDEs. In Sec. 4, we analytically derive a universal
system of dynamical equations (ODEs) for parameters of interacting
solitary waves using asymptotic methods, and show that these ODEs
accurately describe the weak interactions in the PDEs. In Sec. 5, we
solve this ODE system analytically when the single parameter in this
system is equal to zero (which is the integrable case). In addition,
we derive explicit conditions for the solutions of the integrable
ODE system to develop finite-time singularities. In Sec. 6, we show
that fractal structures appear in this ODE system when its parameter
is non-zero, and explore the origin of these fractal structures. In
Sec. 7, we apply the analytical results to the generalized NLS
equations with various nonlinearities. In Sec. 8, we summarize the
results of the paper and make some further remarks.

\section{Preliminaries}
The generalized NLS equation is
\begin{equation}
iU_t+U_{xx}+F(|U|^2)U=0, \label{eqn:1}
\end{equation}
where $F(\cdot)$ is a real-valued algebraic function with $F(0)=0$.
Equation (\ref{eqn:1}) supports solitary waves of the form
\begin{equation}
U=\Phi(x-Vt-x_0;\beta)e^{\frac{1}{2}iV(x-x_0)-\frac{1}{4}iV^2t+i\beta
t-i\sigma_0}, \label{eqn:solution}
\end{equation}
where $\Phi(\theta)$ is a positive function which satisfies the
following equation
\begin{eqnarray}
\Phi_{\theta\theta}+F(|\Phi|^2)\Phi-\beta\Phi=0,  \nonumber\\
\Phi\rightarrow 0, |\theta|\rightarrow \infty, \label{eqn:2}
\end{eqnarray}
and $\beta \: (>0), V, x_0, \sigma_0$ are real constants. For
convenience, we introduce the notations
\begin{eqnarray}
\xi=Vt+x_0, \quad \theta=x-\xi,\nonumber\\
\sigma=(\beta+\frac{1}{4}V^2)t-\sigma_0, \quad
\phi=\frac{1}{2}V\theta+\sigma.     \label{def_phi}
\end{eqnarray}
Physically, $\beta$ is the propagation constant which is related to
the solitary-wave amplitude (henceforth, we call $\beta$ an
amplitude parameter), $\phi$ is the phase of the solitary wave,
$\sigma_0$ is its initial phase, $\xi$ is its center position,  $V$
is its velocity, and $x_0$ is its initial position. The solitary
wave is characterized uniquely by its four parameters:
$V,\beta,\sigma_0$ and $x_0$. The asymptotic behavior of this
solution at infinity is
\begin{equation}
\Phi(\theta)\rightarrow c\: e^{-\sqrt{\beta}|\theta|}, \quad
|\theta|\rightarrow\infty, \label{eqn:asy}
\end{equation}
where $c$ is the tail coefficient which is determined by the
nonlinear function $F$ and propagation constant $\beta$. We define
the power of the solitary wave as
\begin{equation}
P(\beta)=\int^{\infty}_{-\infty}\Phi^2(\theta;\beta)d\theta,
\end{equation}
which plays an important role in the linear stability of the
solitary wave. For general functions $F$, the analytical formulas
for $\Phi, \; P$ and $c$ are not available. But for some special
nonlinearities, one can find the analytical solutions. For instance,
for the cubic-quintic nonlinearity
\begin{equation}
F(|U|^2)=\alpha|U|^2+\gamma|U|^4,  \label{cqNL}
\end{equation}
the analytical formulas for $\Phi, \; P$ and $c$ are
\cite{Kivshar_Agrawal,Kovalev}
\begin{eqnarray}
&& \Phi(\theta;\beta)=\sqrt{\frac{4B\beta/\alpha}{B+\cosh{2\sqrt{\beta}\theta}}},\label{cqNLs}\\
&&P=\frac{4B\sqrt{\beta}(\pi/2-\arctan{\frac{B}{\sqrt{1-B^2}}})}{\alpha\sqrt{1-B^2}},
\label{Pformula}
\\&&c=\sqrt{8B\beta/\alpha},  \label{cformula}
\end{eqnarray}
where
\begin{eqnarray}
B=\textrm{sgn}(\alpha)(1+\frac{16\beta\gamma}{3\alpha^2})^{-1/2}.
\end{eqnarray}
For special values of $\alpha=1,~\gamma=0$, Eq.(\ref{eqn:1}) becomes
the integrable NLS equation, and then
\begin{eqnarray}
&&B=1,\quad \Phi(\theta;\beta)=\sqrt{2\beta}\textrm{sech}(\sqrt{\beta}\theta),\nonumber\\
&&P=4\sqrt{\beta}, \quad c=\sqrt{8\beta}.
\end{eqnarray}

\section{\label{sec:PDE}Universal fractal structures in weak
interactions of solitary-waves \setcounter{equation}{0}}

When two solitary waves are placed adjacent to each other, they
would interfere through tail overlapping. In this case, the initial
condition is
\begin{eqnarray}
U(x,0)=U_{1}(x,0)+U_{2}(x,0), \nonumber\\
U_{k}(x, 0)=\Phi(x-x_{0,k};\beta_{0,k})e^{i\phi_{0,k}},
\nonumber \\
\phi_{0,k}=\frac{1}{2}V_{0,k}(x-x_{0,k})-\sigma_{0,k},
\label{pdeinitial}
\end{eqnarray}
where $\Phi$ satisfies Eq.(\ref{eqn:2}). Here "0" in the subscript
represents the initial value of the underlying parameter. For
convenience, we assign the left solitary wave with index $k=1$, and
the right solitary wave with index $k=2$. To study the weak
interaction between these two solitary waves, we require that the
two solitary waves are both stable, well separated, and having
almost the same velocities and amplitudes. Introducing notations
\begin{eqnarray}
\beta=\frac{1}{2}(\beta_1+\beta_2),\;V=\frac{1}{2}(V_1+V_2),\;\xi=\frac{1}{2}(\xi_1+\xi_2),
\end{eqnarray}
and
\begin{eqnarray}
\Delta\beta=\beta_2-\beta_1,\;\;\Delta V=V_2-V_1,\;\;\Delta
\xi=\xi_2-\xi_1,
\end{eqnarray}
the above requirements then amount to
\begin{eqnarray}
P_\beta>0,|\Delta\beta|\ll\beta,|\Delta V|\ll
1,\beta\Delta\xi\gg1\gg|\Delta\beta\Delta\xi|.\label{eqn:assu}
\end{eqnarray}
Here, $P_\beta \equiv dP/d\beta>0$ corresponds to the
Vakhitov-Kolokolov criterion for the linear stability of solitary
waves in Eq. (\ref{eqn:1}) \cite{Kivshar_Agrawal, VK}.

Below, we numerically study the weak interaction of solitary waves
in Eq. (\ref{eqn:1}). This equation is numerically integrated by the
pseudo-spectral method coupled with the fourth-order Runge-Kutta
integration along the time direction. Since each solitary wave has
four parameters, we have eight parameters in the initial conditions.
Due to the phase, translation and Galilean invariances of Eq.
(\ref{eqn:1}), we can fix $\sigma_{0,1}=0$, $x_{0,1}+x_{0,2}=0$ and
$V_0\equiv (V_{0,1}+V_{0,2})/2=0$ without any loss of generality.
Also, for simplicity, we take $\Delta V_0=V_{0,2}-V_{0,1}=0$ in all
our simulations of this section, i.e. the two solitary waves are
initially at rest. This leaves four free parameters in the initial
conditions (\ref{pdeinitial}): $\Delta x_{0} \equiv x_{0,2}-x_{0,1},
\Delta\phi_0\equiv \phi_{0,2}-\phi_{0,1}=-\sigma_{0,2},
\beta_{0}\equiv (\beta_{0,1}+\beta_{0,2})/2$, and $\Delta \beta_0
\equiv \beta_{0,2}-\beta_{0,1}$.
We define the exit velocity $\Delta V_\infty \equiv
{\displaystyle\lim_{t\rightarrow +\infty}} \Delta V$. We also define
the collision time $\tilde{t}$ as the time when the two solitary
waves are the closest (i.e. the separation distance between peaks of
the two solitary waves the smallest) during interactions. The life
time of interaction is defined as the time length from the beginning
$(t=0$) to the collision time $\tilde{t}$, which is equal to the
collision time in value. Thus $\tilde{t}$ will be used to denote the
life time as well. Of the four parameters in initial conditions, we
will fix $\beta_0, \Delta \beta_0$ and $\Delta x_0$, and use
$\Delta\phi_0$ as the control parameter and vary it continuously
between 0 and $2\pi$. At each $\Delta\phi_0$ value, we simulate the
evolution of the two solitary waves and record the exit velocity and
the life time. Numerically, the exit velocity is determined as
follows. We let the solitary waves propagate for a long time. If
they still do not separate, we assign the exit velocity as zero. If
they do separate, we wait till they have separated far apart and
their velocities stabilized. Then we locate the positions of maximum
solitary wave amplitudes at serval different time values. The
average separation velocity of the two solitary waves in these time
intervals is assigned as the exit velocity. The numerical life time
is simply the time when the two solitary waves are the closest in
the simulations. Below, we carry out numerical studies of weak
interactions as described above on two different nonlinearities: the
cubic-quintic and exponential nonlinearities.

\subsection{Weak interactions for the cubic-quintic nonlinearity}
Our first example of nonlinearity is the cubic-quintic nonlinearity
(\ref{cqNL}), which arises in a wide array of physical systems such
as optics \cite{Kivshar_Agrawal} and boson condensates
\cite{Kovalev,Barashenkov}. In this nonlinearity, we set
\begin{eqnarray} \label{alphagamma}
\alpha=1, \quad \gamma=0.04.
\end{eqnarray}
It is easy to verify that all solitary waves (\ref{cqNLs}) in this
case are linearly stable using the Vakhitov-Kolokolov criterion. In
our simulations, we set $\Delta x_0=10$ and $\beta_0=1$. The $x$
interval is 70 units wide, discretized by 512 grid points; and the
time step size is 0.004.
\begin{figure}
\includegraphics[width=60mm,height=55mm]{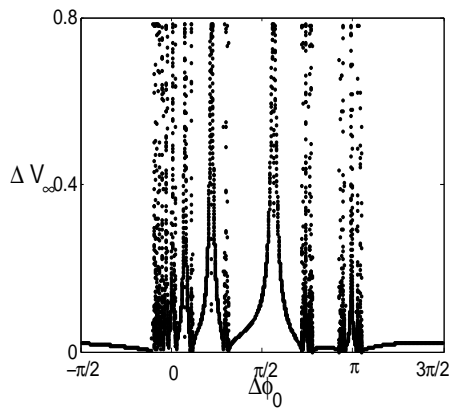}
\caption{\label{cqnonequal}~ The graph of exit velocity $\Delta
V_\infty$ versus the initial phase difference $\Delta\phi_0$ in the
non-equal initial amplitude case of the cubic-quintic NLS equation
(\ref{eqn:1}), (\ref{cqNL}). The cubic and quintic nonlinearity
coefficients are given in Eq. (\ref{alphagamma}), and the other
(fixed) initial parameters are $\Delta x_0=10$, $\beta_0=1$,
$\Delta\beta_0=-0.065$, and $\Delta V_0=0$. }
\end{figure}
\begin{figure*}
\includegraphics[width=80mm,height=70mm]{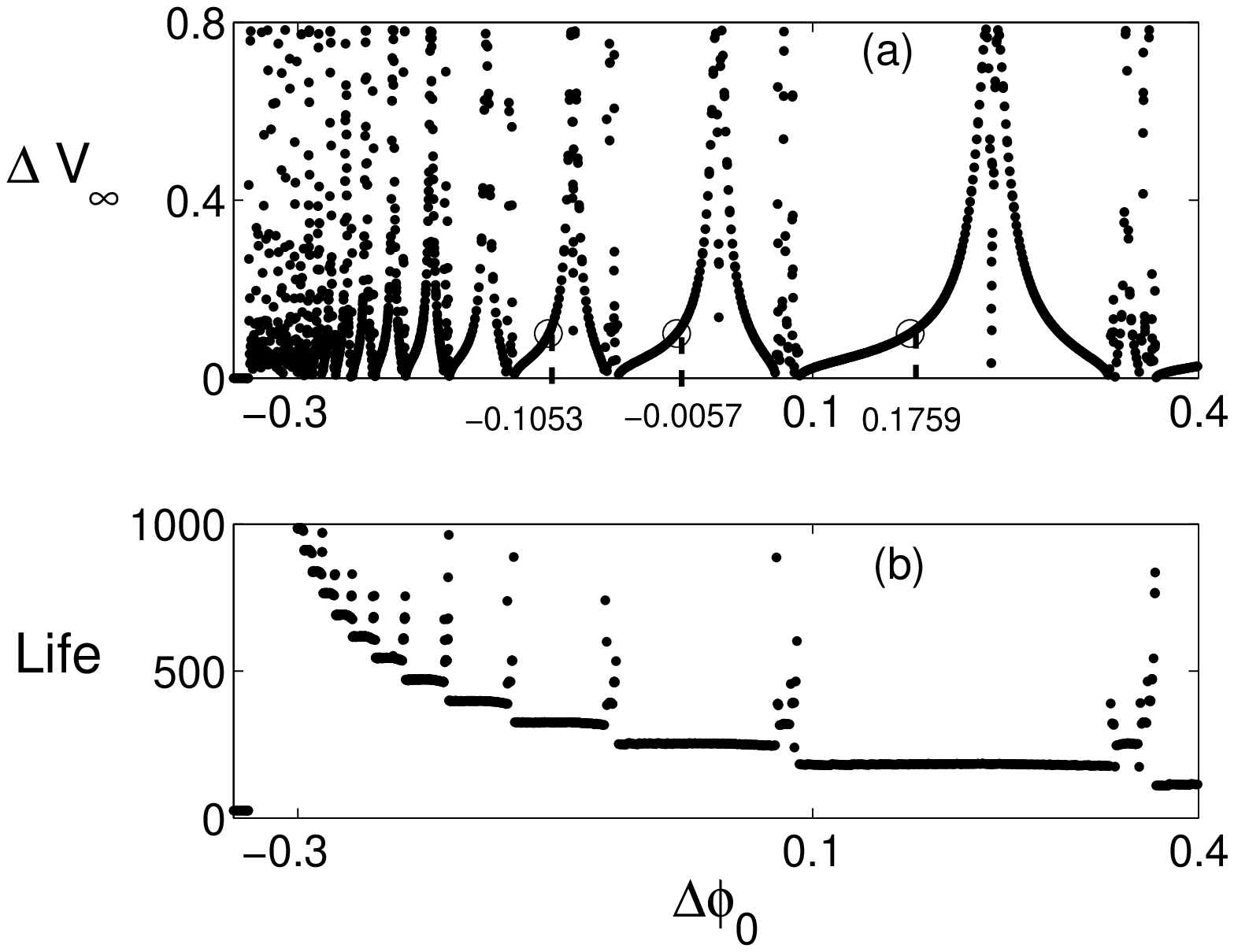} \hspace{0.4cm}
\includegraphics[width=80mm,height=70mm]{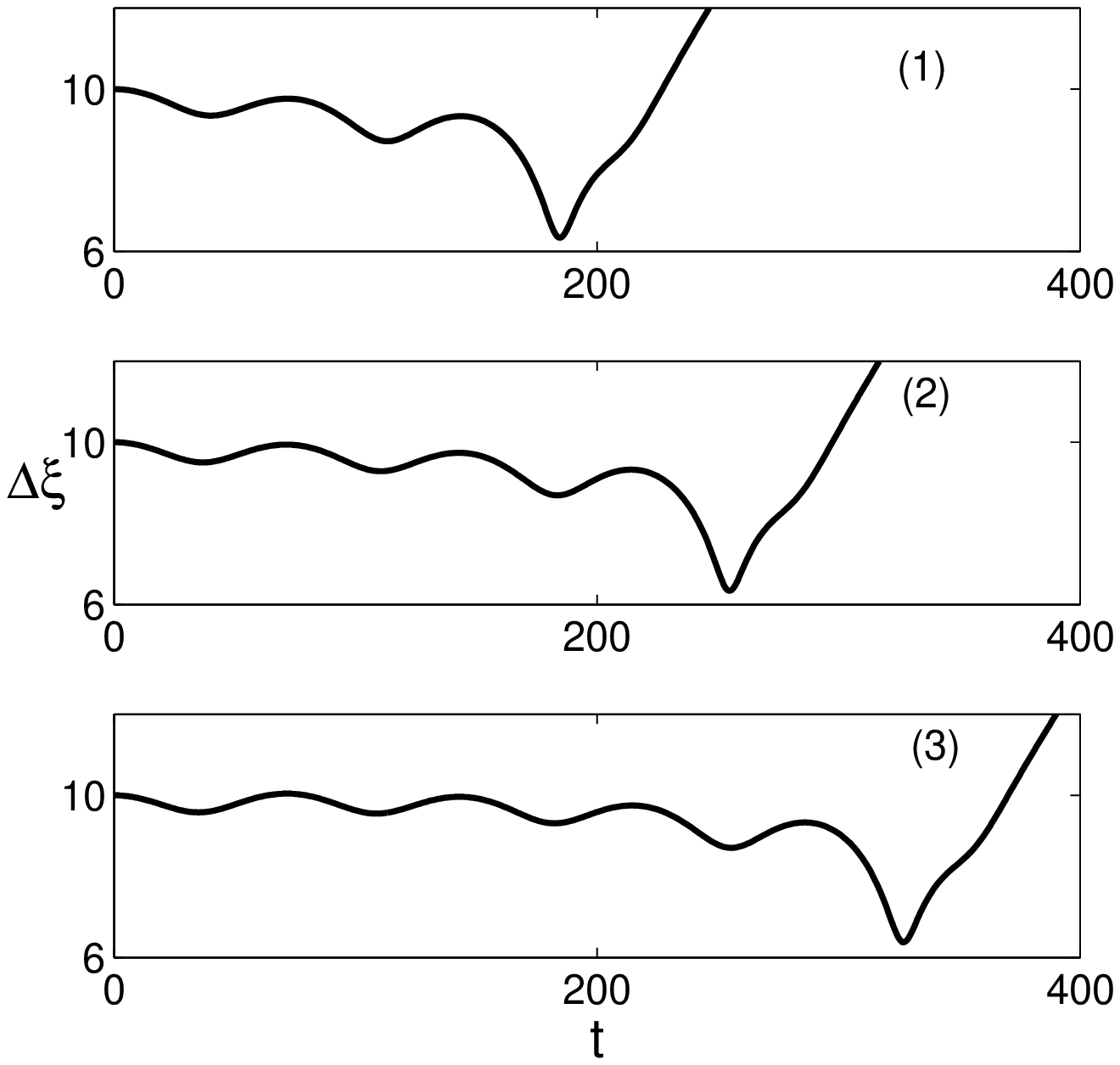}
\caption{\label{pdelevel1}~ (a) The exit velocity versus initial
phase difference graph of Fig. 1 re-plotted near the accumulation
point of the primary hill sequence; (b) the life time versus initial
phase difference graph; (1)-(3): separation versus time diagrams of
solitary wave interactions at three values of $\Delta\phi_0$ marked
by circles in (a): (1) 0.1759; (2) $-0.0057$; (3) $-0.1053$.}
\end{figure*}

\begin{figure*}
\includegraphics[width=63mm,height=47mm]{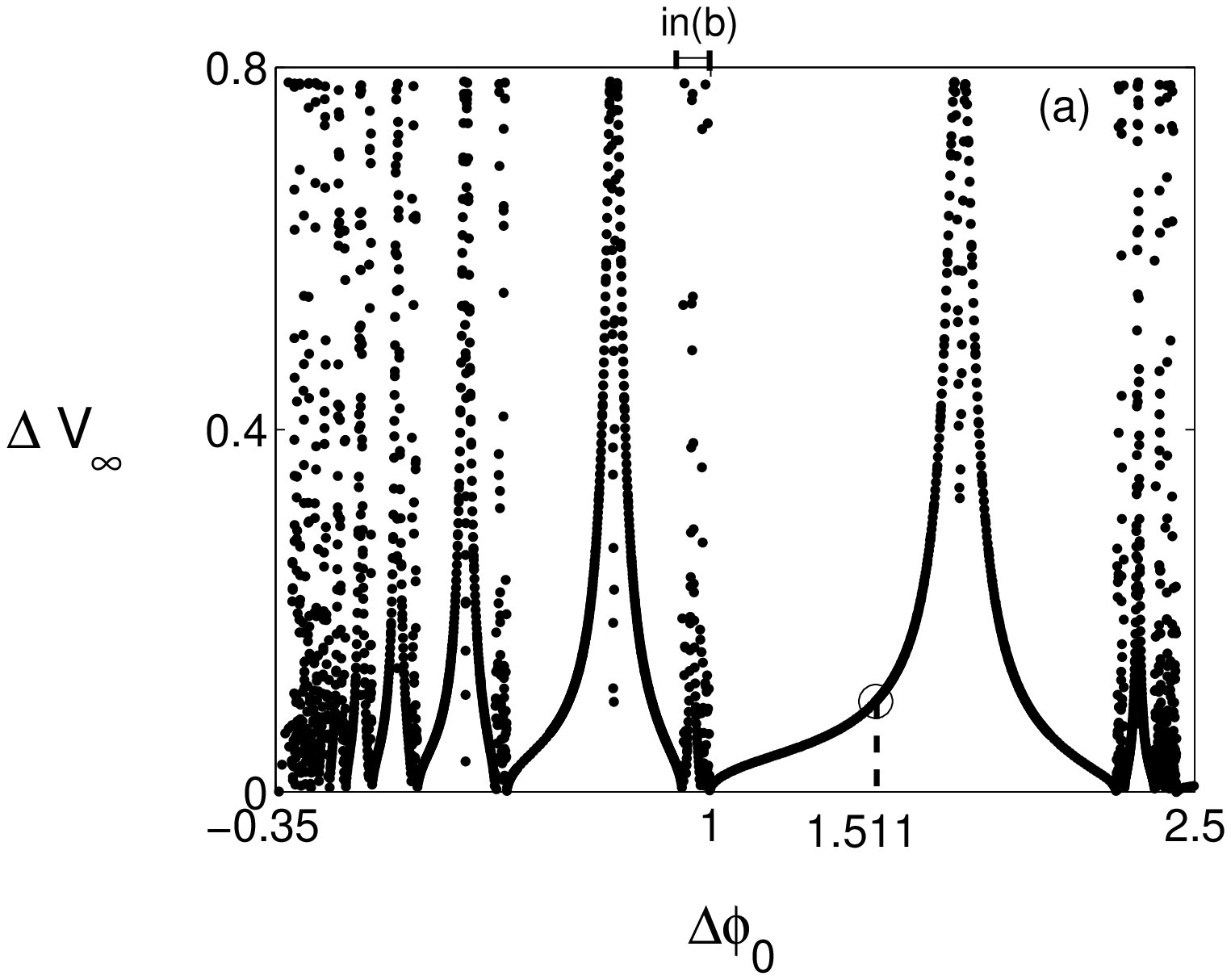}
\includegraphics[width=53mm,height=47mm]{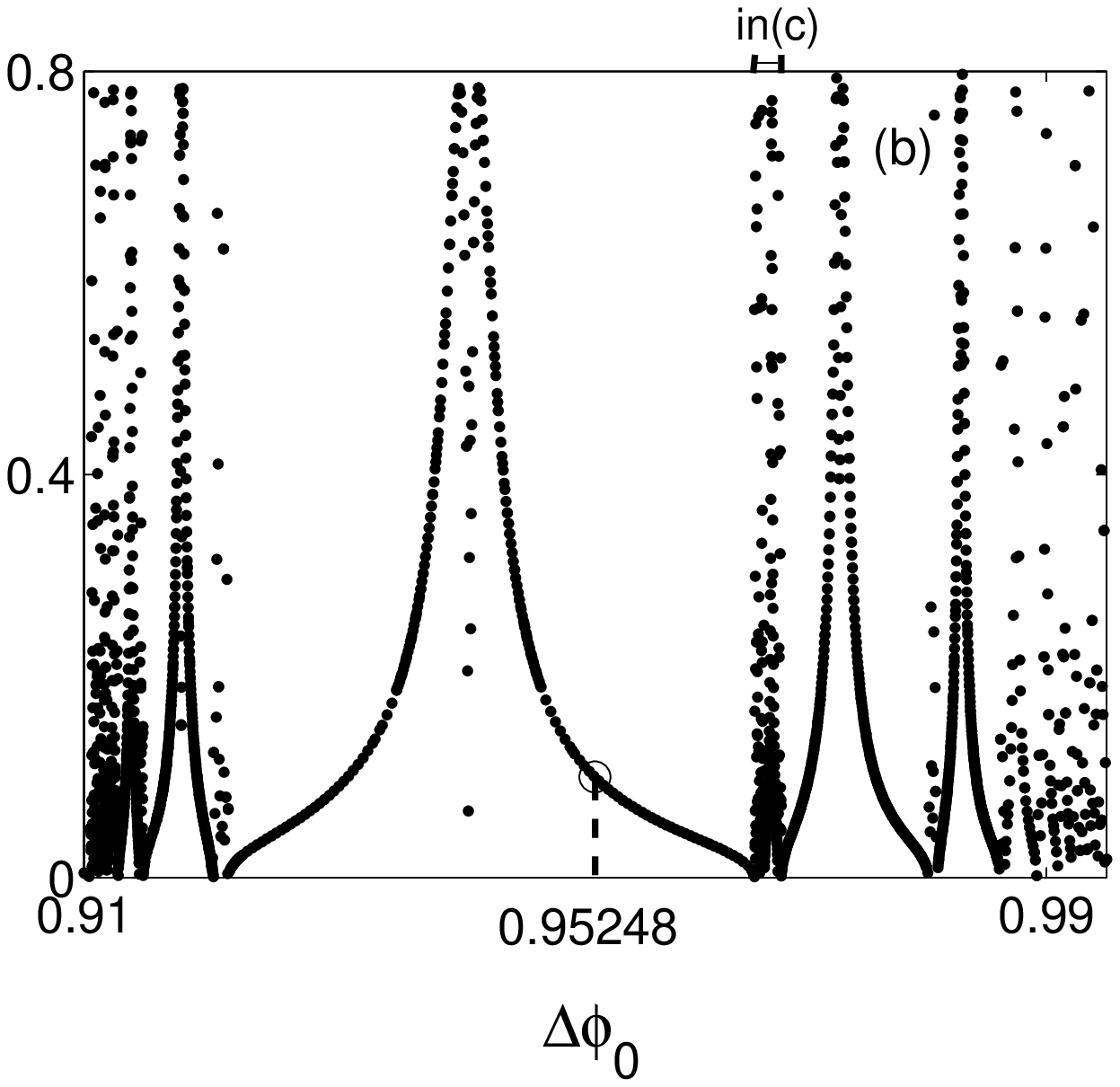}
\includegraphics[width=53mm,height=45mm]{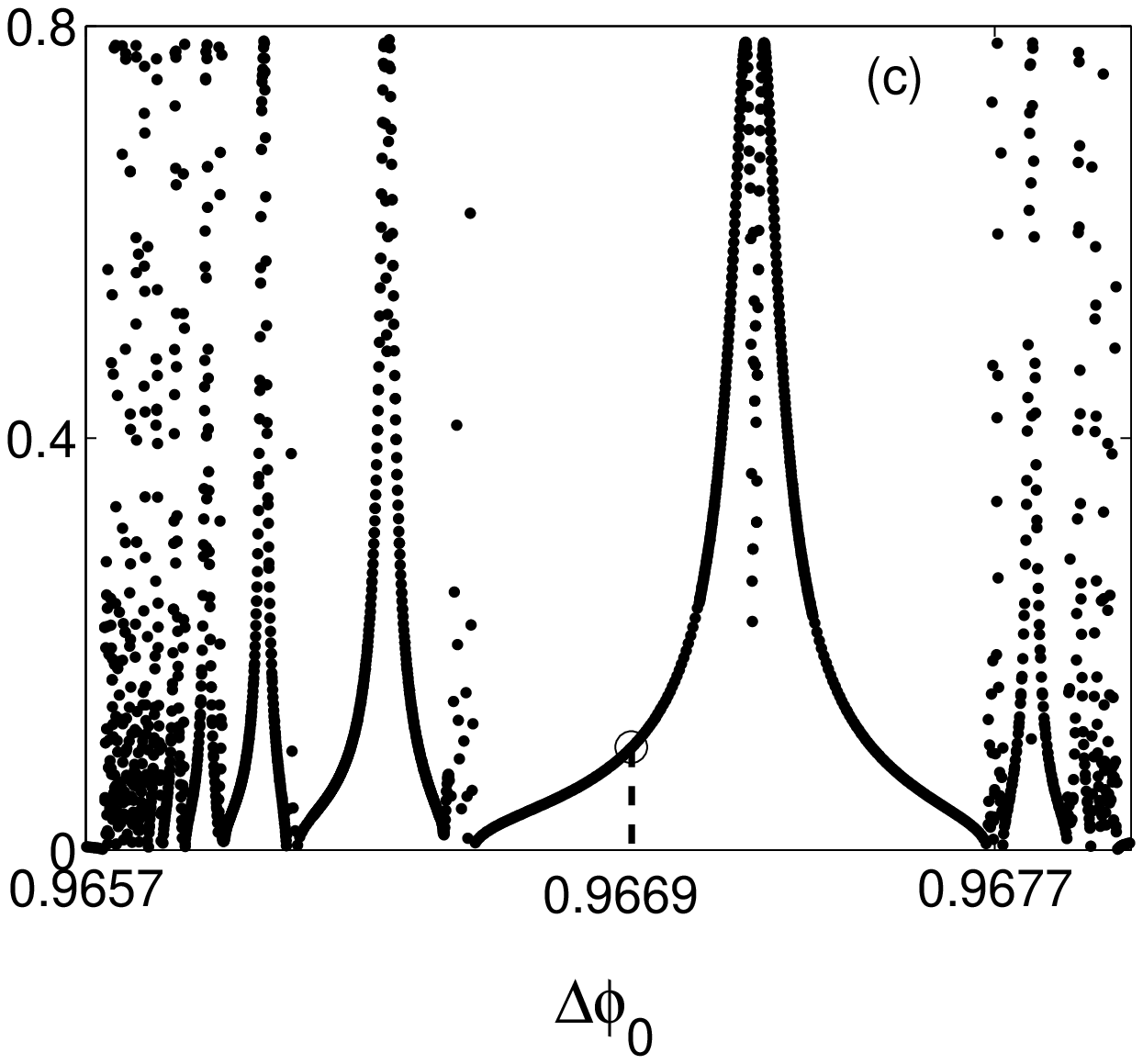}\\
\includegraphics[width=63mm,height=40mm]{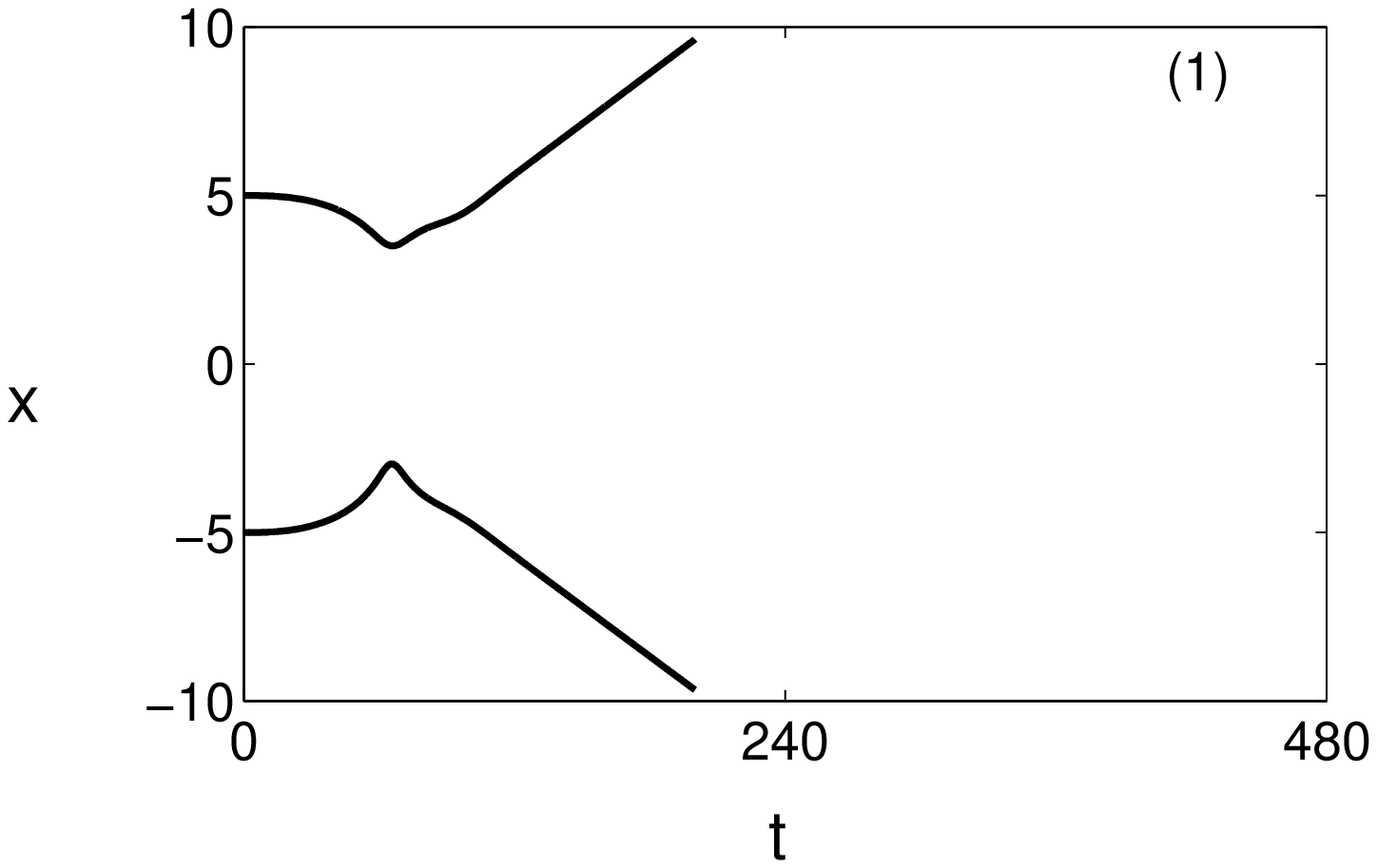}
\includegraphics[width=53mm,height=40mm]{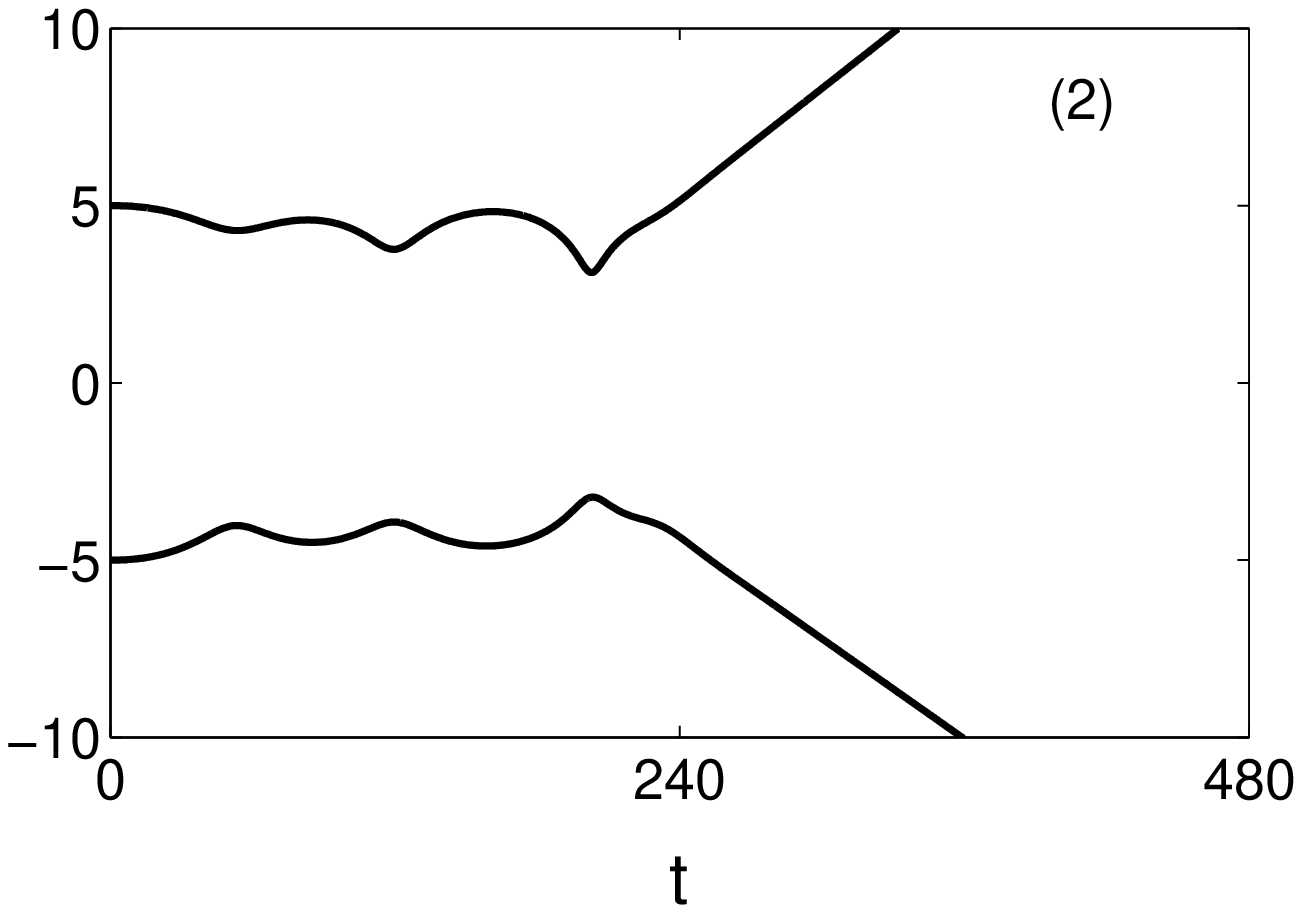}
\includegraphics[width=53mm,height=40mm]{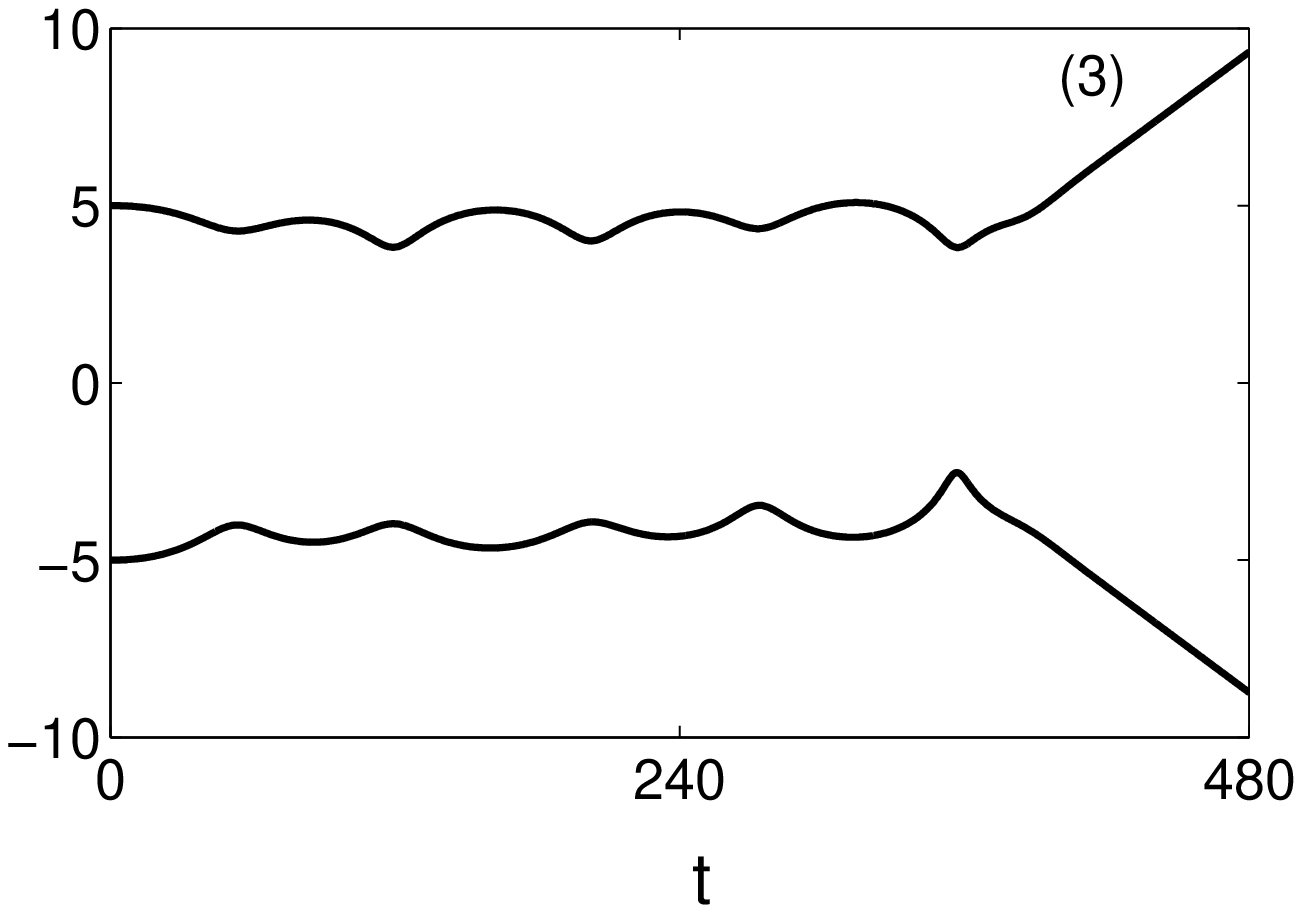}
\caption{\label{fig:PDEzoom}~ Top: the exit velocity versus initial
phase difference graph of Fig. 1 and its two zoomed-in structures;
bottom: soliton-positions versus time diagrams at three values of
$\Delta\phi_0 $ marked by circles in the top panel: (1) 1.511; (2)
0.95248; (3) 0.9669. }
\end{figure*}

We first study the nonequal-amplitude case, and take
$\Delta\beta_0=-0.065$. The $\Delta V_\infty$ versus $\Delta\phi_0$
diagram is shown in Fig.\ref{cqnonequal}. The prominent structures
in this graph can be split into two regions: one region is
$-0.34<\Delta\phi_0<2.5$, and the other region is
$2.9<\Delta\phi_0<3.3$. The structures in these two regions turn out
to be quite similar (except a horizontal reflection with respect to
a vertical axis), thus we focus on the larger region
$-0.34<\Delta\phi_0<2.5$ below. The main structure in this region
forms a sequence of hills; their widths get smaller from the right
to the left, and their heights are about the same. These hills will
be called the primary hills. This primary-hill sequence converges to
the accumulation point $\Delta\phi_{0c}=-0.339$. In order to see
this hill sequence near the accumulation point $\Delta\phi_{0c}$
more clearly, we zoom in the region $[-0.35,0.4]$, and the zoomed-in
diagram is shown in Fig.\ref{pdelevel1}. In this figure, the
cascading sequence can be seen very clearly (see
Fig.\ref{pdelevel1}(a)). In Fig.\ref{pdelevel1}(b), the
corresponding life-time diagram is displayed. We can see that on the
same hill, interactions have roughly the same life time. On
different hills, life times are different: hills closer to the
accumulation point $\Delta\phi_{0c}$ have longer life times. Between
hills, even longer life times can be seen, suggesting more complex
dynamics there. To explore differences in interaction dynamics on
different hills, we select three points
$\Delta\phi_0=0.1759,-0.0057,-0.1053$ (marked in
Fig.\ref{pdelevel1}(a) by circles) on three adjacent primary hills.
These points are at the same relative positions (roughly halfway
between the peak and bottom) of the respective hills. At these
points, the interaction dynamics is plotted in
Fig.\ref{pdelevel1}(1-3). Here only the separation distance
$\Delta\xi$ versus time $t$ graphs are shown. We find that these
three dynamical processes are similar, except that the oscillation
times before final separation differ by one from one hill to the
next. The life times $\tilde{t}_n$ of interactions on this primary
hill sequence are found to be an almost perfect linear function of
the hill index $n$ as
\begin{equation}
\omega \tilde{t}_n=2n\pi+\delta,   \label{time_PDE}
\end{equation}
where the least-square linear fit gives
\begin{equation} \label{PDEvalue}
\omega=0.08605, \quad \delta=2.8897.
\end{equation}
Here the life time of each primary hill is measured numerically at
the relative location of that hill shown in Fig.\ref{pdelevel1}(a)
by circles. This life-time formula has the same form as those for
all window sequences reported before
\cite{Campbell2,Campbell3,Kivshar1,YangTan,Goodman_Haberman2}.
%

In addition to the primary hill sequence as described above,
Fig.\ref{cqnonequal} also possesses higher-order structures between
primary hills. To demonstrate, we first isolate the long interval
$[-0.35, 2.5]$ in Fig. \ref{cqnonequal} and re-plot that part of the
graph in Fig.\ref{fig:PDEzoom}(a). Then we zoom into its
sub-interval $[0.91, \;0.995]$, which is between the two largest
primary hills in Fig.\ref{fig:PDEzoom}(a). The zoomed-in graph is
shown in Fig.\ref{fig:PDEzoom}(b). We see that the zoomed-in graph
is similar to Fig.\ref{fig:PDEzoom}(a), but the cascading direction
has reversed. This behavior is analogous to that reported in
\cite{anninos, YangTan2} for the $\phi^4$ model and the coupled NLS
equations. The main structure in this zoomed-in window is again two
sequences of hills, accumulating to the left and right respectively.
We call them secondary hills. Between secondary hills, we see even
higher-order structures. To see these structures more clearly, we
zoom into the sub-interval $[0.9657, \;0.96785]$, which is between
the two largest secondary hills in Fig.\ref{fig:PDEzoom}(b). The
zoomed-in graph is shown in Fig.\ref{fig:PDEzoom}(c). We see that it
is again similar to Fig. \ref{fig:PDEzoom}(b) but with a reversed
cascading direction. One can zoom into the regions between these
tertiary hills in Fig.\ref{fig:PDEzoom}(c) further, and will get
even higher order structures which are similar to the ones shown in
Fig.\ref{fig:PDEzoom}. Thus Fig.\ref{fig:PDEzoom}(a) is a fractal
structure! We have also explored the interaction dynamics on this
fractal. To demonstrate, we pick three $\Delta \phi_0$ values
~$1.511,\;0.95248,\;0.9669$ which are at the same relative positions
of the fractal (roughly halfway between the peak and bottom of the
widest hills) in Fig.\ref{fig:PDEzoom}(a)-(c) (marked by circles).
The interaction dynamics at these three points are displayed in
Fig.\ref{fig:PDEzoom}(1-3) respectively. Here the positions of
maximum amplitudes of the interacting waves are plotted against
time. We see that these dynamical patterns are clearly similar,
except that the numbers of oscillations before final separation are
different.

\begin{figure}
\includegraphics[width=70mm,height=60mm]{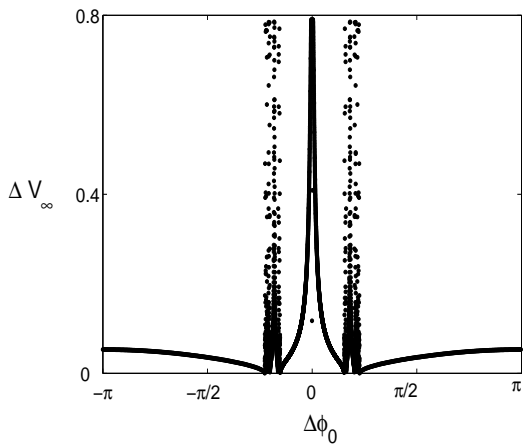}
\caption{\label{fig:cqequal}~ The exit velocity versus initial phase
difference graph in the equal initial amplitude case of the
cubic-quintic NLS equation \ref{eqn:1}), (\ref{cqNL}). The cubic and
quintic nonlinearity coefficients as well as the initial conditions
are the same as in Fig. 1, except that $\Delta\beta_0=0$ now.}
\end{figure}

In the above numerical simulations, the two solitary waves have
different initial amplitudes ($\Delta\beta_0=-0.065$). We have also
studied interactions of equal-amplitude solitary waves, i.e. with
$\Delta\beta_0=0$, while keeping the other parameters the same. In
this case, the graph of exit velocity $\Delta V_\infty$ versus
initial phase difference $\Delta \phi_0$ is shown in
Fig.\ref{fig:cqequal}. This graph is symmetric with respect to
$\Delta \phi_0$ for obvious reasons. Examination of this graph shows
that it is also a fractal. Thus fractal dependence arises in weak
interactions of both equal and non-equal amplitude solitary waves.

\subsection{Weak interactions with exponential nonlinearity}
\begin{figure}
\includegraphics[width=70mm,height=60mm]{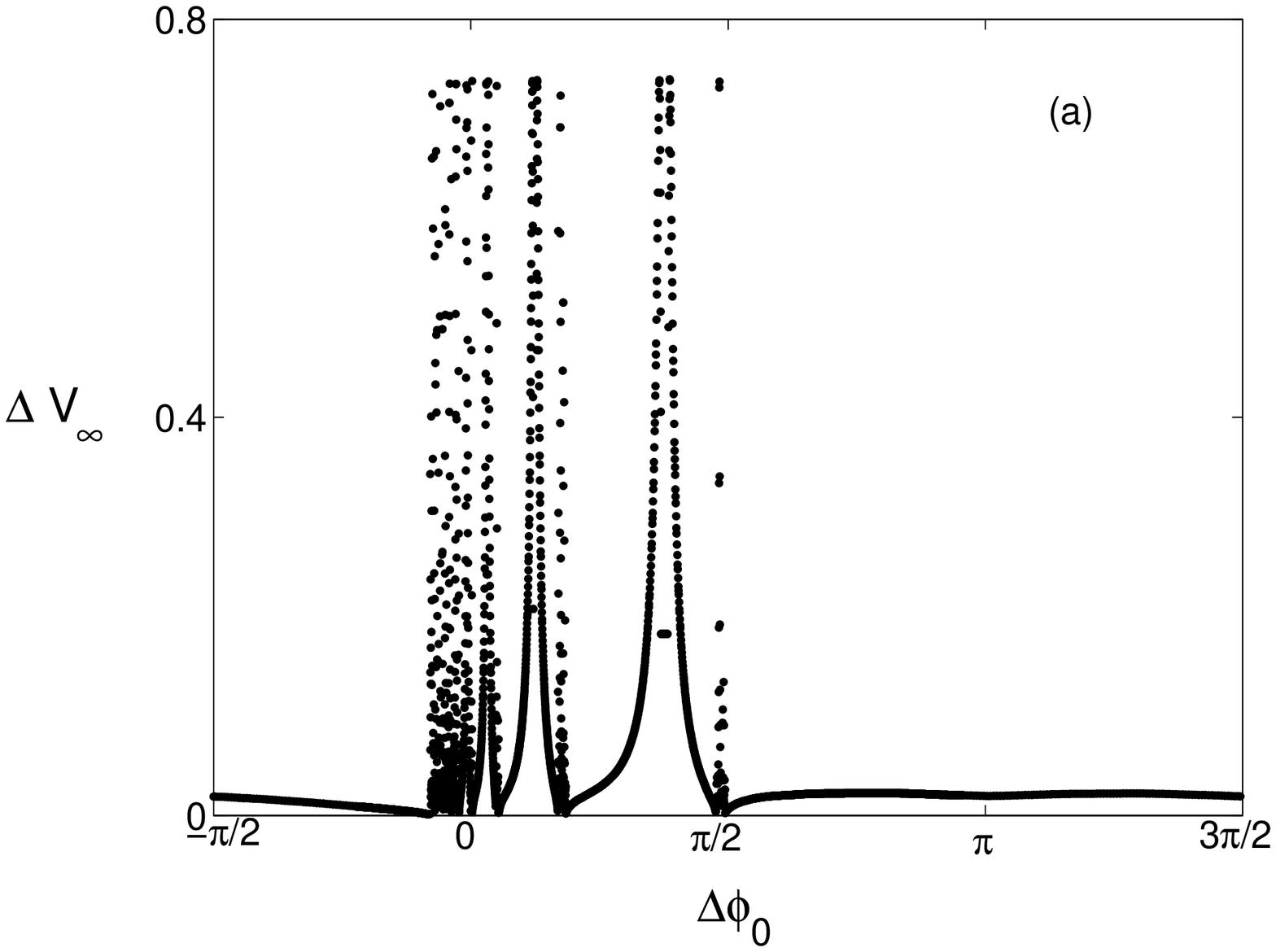}
\includegraphics[width=70mm,height=60mm]{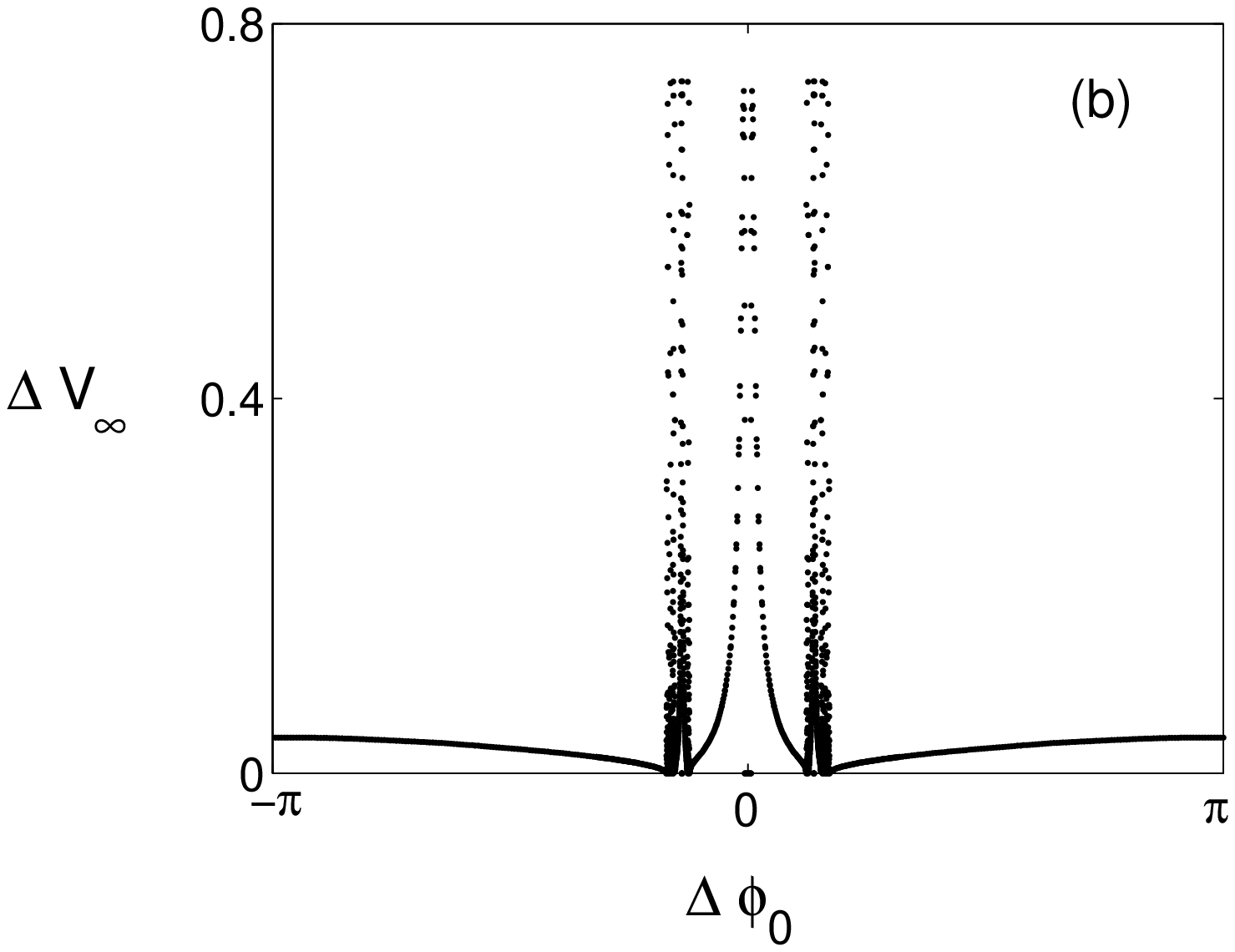}
\caption{\label{fig:exPDE} The exit velocity versus initial phase
difference graphs for the exponential nonlinearity (\ref{exp_non}):
(a) the non-equal initial amplitude case with
$\Delta\beta_0=-0.045$; (b) the equal initial amplitude case with
$\Delta\beta_0=0$. The other (fixed) initial parameters are
$\beta_0=2.3$, $\Delta x_0=8$, and $\Delta V_0=0$. }
\end{figure}
To explore whether the above fractal structures for weak
interactions persist or not with other types of nonlinearities, we
consider in this subsection a different type of nonlinearity --- the
exponential nonlinearity, with
\begin{equation} \label{exp_non}
F(|U|^2)=e^{|U|}-1.
\end{equation}
Here, $-1$ is introduced into this function to meet the condition
$F(0)=0$. Note that this nonlinearity does not have any parameters.
Throughout this subsection, we set the initial separation $\Delta
x_0=8$, and average propagation constant $\beta_0=2.3$. We study two
cases, one for non-equal amplitudes with $\Delta\beta_0=-0.045$, and
the other for equal amplitudes with $\Delta\beta_0=0$. For both
cases, the control parameter is $\Delta \phi_0$ as before. In our
simulations, the $x$ interval was 70 units wide, discretized by 512
grid points. The time step size was 0.002. The $\Delta V_\infty$
verse $\Delta\phi_0$ graphs for both cases are plotted in Fig.
\ref{fig:exPDE}. We have verified that both graphs in this figure
are fractals. Comparing these fractals with those in Figs.
\ref{cqnonequal} and \ref{fig:cqequal} of the cubic-quintic
nonlinearity, we see that the fractal structures for these two
different nonlinearities are very similar. The only major difference
between them is in the non-equal amplitude case, where there is only
one primary hill sequence (accumulating toward the left) for the
exponential nonlinearity, while there are two primary hill sequences
for the cubic-quintic nonlinearity. It is remarkable that two very
different nonlinearities exhibit quite similar fractal dependence on
initial conditions. Thus fractal scattering appears to be a
universal feature in weak interactions of Eq. (\ref{eqn:1}) rather
than an accident. This leads us to the following questions: how can
we analytically establish the universal nature of fractal
scatterings for Eq. (\ref{eqn:1}) with general nonlinearities? how
can we analytically explain the major differences of fractals for
different nonlinearities? These questions will be answered in the
following sections.

\section{\label{sec:level3}Dynamical Equations\setcounter{equation}{0}}
\quad To study weak interactions analytically, we use the
Karpman-Solov'ev method \cite{Karpman_Solovev} by treating the
interference as a small perturbation to each solitary wave (see also
\cite{Gorshkov}). This method has been successfully used before on
the NLS equation \cite{Hasegawa_Kodama, Karpman_Solovev, Gorshkov,
Gerdjikov_PRL}, the modified NLS equation \cite{Gerdjikov_Yang}, the
Manakov equations \cite{Yang_Manakov}, as well as the
(non-integrable) coupled NLS equations \cite{YangPRE01}. To proceed,
we first need to consider the evolution of a single solitary wave in
the perturbed generalized NLS equation
\begin{equation}
iU_t+U_{xx}+F(|U|^2)U=\epsilon G,  \label{eqn:3}
\end{equation}
where function $G$ is a perturbation term, and $\epsilon$ is a small
parameter. Without perturbations ($\epsilon=0$), the solitary wave
(\ref{eqn:solution}) is an exact solution of Eq.(\ref{eqn:3}), and
its internal parameters $V,\beta,\sigma_0,x_0$ are time-independent.
When the perturbation is turned on, these internal parameters of the
solitary wave will evolve slowly on the time scale $T=\epsilon t$.
The multiple-scale perturbation theory for this slow evolution is
well known \cite{YangKaup, YangPRE01}. We write the perturbed
solution as
\begin{equation}
U=\hat{\Phi}(\theta,t,T)e^{iV\theta/2+i\sigma}, \label{eqn:4}
\end{equation}
where
\begin{equation}
\theta=x-\int^t_0Vdt-x_0,\quad\sigma=\int^t_0(\beta+V^2/4)dt-\sigma_0.
\end{equation}
Here $V(T),\beta(T),\sigma_0(T),x_0(T)$ are all functions of slow
time $T$. Next, we will derive the dynamical equations (ODEs) for
the slow-time evolution of these parameters. Substituting
(\ref{eqn:4}) into (\ref{eqn:3}), we get the equation for
$\hat{\Phi}$ as
\begin{eqnarray}
i\hat{\Phi}_t+\hat{\Phi}_{\theta\theta}-\beta\hat{\Phi}+F(\hat{\Phi}^2)\hat{\Phi}=\nonumber
\\\epsilon
Ge^{-i\phi}
-\epsilon\left[i\hat{\Phi}_{\beta}\beta_T-i\hat{\Phi}_{\theta}x_{0T}\right]\nonumber\\
-\epsilon \left(Vx_{0T}/2-V_T\theta/2+\sigma_{0T} \right)\hat{\Phi},
\end{eqnarray}
where $\phi$ is defined in Eq. (\ref{def_phi}). We expand the
amplitude function $\hat{\Phi}$ into a perturbation series
\begin{equation}
\hat{\Phi}=\Phi(\theta;\beta)+\epsilon\tilde{\Phi}+O(\epsilon^2).
\end{equation}
The equation at order $\epsilon^0$ is satisfied automatically since
$\Phi$ satisfies Eq.(\ref{eqn:2}). At order $\epsilon$, the equation
for $\tilde{\Phi}$ can be written as
\begin{equation}
i\Psi_t+\mathcal{L}\Psi=H,   \label{eqn:5}
\end{equation}
where
\begin{eqnarray}
\Psi=\left(\begin{array}{c}\tilde{\Phi}+\tilde{\Phi}^*\\\tilde{\Phi}^*-\tilde{\Phi}\end{array}\right),
\quad
\mathcal{L}=\left(\begin{array}{c}0\quad\mathcal{L}_0\\\mathcal{L}_1\quad0\end{array}\right),
\end{eqnarray}
\begin{eqnarray}
&&\mathcal{L}_0=-\partial_{\theta\theta}+\beta-F(\Phi^2),\nonumber\\
&&\mathcal{L}_1=-\partial_{\theta\theta}+\beta-F(\Phi^2)-2\Phi^2F'(\Phi^2),
\end{eqnarray}
and
\begin{eqnarray}
H=\left[\begin{array}{l}-G^*e^{i\phi}+Ge^{-i\phi}-2i\Phi_{\beta}\beta_T+2i\Phi_{\theta}x_{0T}\\
-G^*e^{i\phi}- Ge^{-i\phi}+(Vx_{0T}-\theta
V_T+2\sigma_{0T})\Phi\end{array}\right].
\end{eqnarray}
Here the superscript "*" represents complex conjugation. Operator
$\mathcal{L}$ has two eigenfunctions and two generalized
eigenfunctions associated with the zero eigenvalue,
\begin{eqnarray}
\Psi_1=\left(\begin{array}{c}\Phi_{\theta}\\0\end{array}\right),\;\;\quad\quad\Psi_2=\left(\begin{array}{c}0\\\Phi\end{array}\right),\nonumber\\
\tilde{\Psi}_1=\left(\begin{array}{c}0\\-\theta\Phi/2\end{array}\right),\quad\tilde{\Psi}_2=\left(\begin{array}{c}-\Phi_{\beta}\\0\end{array}\right),
\end{eqnarray}
with the relations
\begin{eqnarray}
\mathcal{L}\Psi_k=0,  \hspace{0.5cm}
\mathcal{L}\tilde{\Psi}_k=\Psi_k, \quad k=1,2.
\end{eqnarray}
In order for the inhomogeneous solution $\Psi$ of the first-order
equation (\ref{eqn:5}) to be non-secular at large time, the
inhomogeneous term in Eq.(\ref{eqn:5}) must be orthogonal to the
above eigenfunctions and generalized eigenfunctions of the zero
eigenvalue, i.e.,
\begin{eqnarray}
\left<H,\Psi_k\right>=\left<H,\tilde{\Psi}_k\right>=0,\quad
k=1,2,\label{eqn:6}
\end{eqnarray}
under the inner product defined as
\begin{eqnarray}
\left<F_1,F_2\right>=\int^{\infty}_{-\infty}F_1^{\dag}\left(\begin{array}{c}
0 \quad 1\\1\quad0\end{array}\right)F_2 d\theta.
\end{eqnarray}
Here $F_k^{\dag}$ is the Hermitian of $F_k$. Evaluating the four
integrals in Eq.(\ref{eqn:6}), the slow-time evolution equations for
parameters $V(T),\beta(T),\sigma_0(T),x_0(T)$ will be obtained.
These evolution equations can be written as
\begin{eqnarray}
P\frac{dV}{dT}=2\int^{\infty}_{-\infty}\Phi_\theta(G^*e^{i\phi}+
Ge^{-i\phi})d\theta, \label{eqn:dy1}
\end{eqnarray}
\begin{eqnarray}
P_\beta\frac{d\beta}{dT}=\frac{1}{i}\int^{\infty}_{-\infty}\Phi(Ge^{-i\phi}-
G^*e^{i\phi})d\theta,   \label{eqn:dy2}
\end{eqnarray}
\begin{eqnarray}
P\frac{dx_0}{dT}=\frac{1}{i}\int^{\infty}_{-\infty}\Phi\theta(Ge^{-i\phi}-
G^*e^{i\phi})d\theta,  \label{eqn:dy3}
\end{eqnarray}
\begin{eqnarray}
P_\beta(\frac{V}{2}\frac{dx_0}{dT}+\frac{d\sigma_0}{dT})=\int^{\infty}_{-\infty}\Phi_\beta(G^*e^{i\phi}+
Ge^{-i\phi})d\theta. \label{eqn:dy4}
\end{eqnarray}
These equations will be critical for the development of weak
interaction theory of solitary waves below.

Now, we consider the weak interaction of two solitary waves. Here
the tail overlapping can be considered as a small perturbation which
causes the internal parameters of each solitary wave to evolve on a
slow time scale $\epsilon t$. Here $\epsilon$ is the magnitude of
tail overlapping which is exponentially small with solitary wave
spacing $\Delta \xi$. We will not introduce $\epsilon$ explicitly in
the next analysis. To the leading order, the interacting solution is
simply a superposition of two solitary waves,
\begin{eqnarray}
U=U_{1}+U_{2}, \nonumber\\
U_{k}=\Phi_{k}e^{i\phi_{k}}, \quad k=1,2,
\end{eqnarray}
where all parameters slowly vary over time. Picking up the dominant
interference terms, each solitary wave is governed by the following
perturbed generalized NLS equations:
\begin{eqnarray}
iU_{k,t}+iU_{k,\theta\theta}+F(|U_k|^2)U_k=H_k, \label{eqn:7}
\end{eqnarray}
where
\begin{eqnarray}
H_k=&-&(F(|U_k|^2)+F'(|U_k|^2)|U_k|^2)U_{3-k}\nonumber\\&-&F'(|U_k|^2)U_k^2U_{3-k}^*.
\end{eqnarray}
In this paper, we only study the weak interaction, so conditions
(\ref{eqn:assu}) are assumed. Since $|\Delta V|\ll 1$, the phase
difference
\begin{eqnarray}
\Delta\phi=\phi_2-\phi_1\approx-V\Delta\xi/2+\Delta\sigma,
\end{eqnarray}
which is independent of $\theta$.

 Now We apply the above solitary wave perturbation theory to
 Eq.(\ref{eqn:7}). In this problem,
 \begin{eqnarray}
 \epsilon
 Ge^{-i\phi}=&-&(F(\Phi_k^2)\Phi_{3-k}+F'(\Phi_k^2)\Phi_k^2\Phi_{3-k})e^{(-1)^{k+1}i\Delta\phi}\nonumber\\
 &-&F'(\Phi_k^2)\Phi_k^2\Phi_{3-k}e^{(-1)^ki\Delta\phi}.
\label{eqn:8}
\end{eqnarray}
Substituting (\ref{eqn:8}) into Eqs.(\ref{eqn:dy1})-(\ref{eqn:dy4}),
we obtain the following dynamical equations
\begin{widetext}
\begin{eqnarray}
P_k\frac{dV_k}{dt}=-4\int^{\infty}_{-\infty}\Phi_{k,\theta}(F(\Phi_k^2)\Phi_{3-k}+2F'(\Phi_k^2)\Phi_k^2\Phi_{3-k})d\theta\cos(\Delta\phi),
\label{eqn:dy1-1}
\end{eqnarray}
\begin{eqnarray}
P_{k,\beta_k}\frac{d\beta_k}{dt}=(-1)^k2\int^{\infty}_{-\infty}\Phi_kF(\Phi_k^2)\Phi_{3-k}d\theta\sin(\Delta\phi),\label{eqn:dy2-1}
\end{eqnarray}
\begin{eqnarray}
P_k\frac{dx_{k,0}}{dt}=(-1)^k2\int^{\infty}_{-\infty}\Phi_k\theta
F(\Phi_k^2)\Phi_{3-k}d\theta\sin(\Delta\phi), \label{eqn:dy3-1}
\end{eqnarray}
\begin{eqnarray}
P_{k,\beta_k}(\frac{V_k}{2}\frac{dx_{k,0}}{dt}+\frac{d\sigma_{k,0}}{dt})=-2\int^{\infty}_{-\infty}\Phi_{k,\beta_k}
(F(\Phi_k^2)\Phi_{3-k}+2F'(\Phi_k^2)\Phi_k^2\Phi_{3-k})d\theta\cos(\Delta\phi),
\label{eqn:dy4-1}
\end{eqnarray}
\end{widetext}
where $P_k, k=1, 2$ are powers of the two individual solitary
waves. These equations can be simplified greatly. Due to
assumptions (\ref{eqn:assu}), and noticing that $\Phi(\theta)$ and
$\Phi_\beta(\theta)$ are even functions of $\theta$, the
leading-order terms of the above integrals can be explicitly
calculated. For instance,
\begin{eqnarray}
&&\int^{\infty}_{-\infty}\Phi_k
F(\Phi_k^2)\Phi_{3-k}d\theta\nonumber\\&=&\int^{\infty}_{-\infty}\Phi
F(\Phi^2)ce^{(-1)^{k+1}\sqrt{\beta}\theta}d\theta e^{-\sqrt{\beta}\Delta\xi} \nonumber\\
&=&\int^{\infty}_{-\infty}(\beta\Phi-\Phi_{\theta\theta})ce^{\sqrt{\beta}\theta}d\theta
e^{-\sqrt{\beta}\Delta\xi}\nonumber\\&=&2\sqrt{\beta}c^2e^{-\sqrt{\beta}\Delta\xi}.
\end{eqnarray}
Similarly,
\begin{eqnarray}
&&\int^{\infty}_{-\infty}\Phi_{k,\theta}(F(\Phi_k^2)\Phi_{3-k}+2F'(\Phi_k^2)\Phi_k^2\Phi_{3-k})d\theta\nonumber\\
&=&\int^{\infty}_{-\infty}\Phi_\theta(F(\Phi^2)+2F'(\Phi^2)\Phi^2)ce^{(-1)^{k+1}\sqrt{\beta}\theta}d\theta e^{-\sqrt{\beta}\Delta\xi}\nonumber\\
\nonumber\\&=&(-1)^k\sqrt{\beta}\int^{\infty}_{-\infty}\Phi
F(\Phi^2)ce^{\sqrt{\beta}\theta}d\theta e^{-\sqrt{\beta}\Delta\xi}
\nonumber\\&=&(-1)^k2\beta c^2e^{-\sqrt{\beta}\Delta\xi},
\end{eqnarray}
\begin{eqnarray}
&&\int^{\infty}_{-\infty}\Phi_k\theta F(\Phi_k^2)\Phi_{3-k}d\theta
\nonumber\\&=&\int^{\infty}_{-\infty}\Phi\theta
F(\Phi^2)ce^{(-1)^{k+1}\sqrt{\beta}\theta}d\theta e^{-\sqrt{\beta}\Delta\xi}\nonumber\\
&=&(-1)^{k+1}\int^{\infty}_{-\infty}\Phi\theta
F(\Phi^2)ce^{\sqrt{\beta}\theta}d\theta
e^{-\sqrt{\beta}\Delta\xi}\nonumber\\
&  \frac{\underline{\textit{\footnotesize def}}}
{\;}&(-1)^{k+1}D_1e^{-\sqrt{\beta}\Delta\xi}, \label{eqD1}
\end{eqnarray}
\begin{eqnarray}
&&\int^{\infty}_{-\infty}\Phi_{k,\beta_k}(F(\Phi_k^2)\Phi_{3-k}+2F'(\Phi_k^2)\Phi_k^2\Phi_{3-k})d\theta\nonumber\\
&=&\int^{\infty}_{-\infty}\Phi_\beta(F(\Phi^2)+2F'(\Phi^2)\Phi^2)ce^{(-1)^{k+1}\sqrt{\beta}\theta}d\theta e^{-\sqrt{\beta}\Delta\xi}\nonumber\\
&=&\int^{\infty}_{-\infty}\Phi_\beta(F(\Phi^2)+2F'(\Phi^2)\Phi^2)ce^{\sqrt{\beta}\theta}d\theta e^{-\sqrt{\beta}\Delta\xi}\nonumber\\
& \frac{\underline{\textit{\footnotesize
def}}}{\;}&D_2e^{-\sqrt{\beta}\Delta\xi}. \label{eqD2}
\end{eqnarray}
With the above simplifications, the dynamical equations reduce to
\begin{eqnarray}
P\frac{dV_k}{dt}=(-1)^{k+1}8\beta
c^2\cos(\Delta\phi)e^{-\sqrt{\beta}\Delta\xi}, \label{eqn:simpD1}
\end{eqnarray}
\begin{eqnarray}
P_\beta\frac{d\beta_k}{dt}=(-1)^k4\sqrt{\beta}c^2\sin(\Delta\phi)e^{-\sqrt{\beta}\Delta\xi},\label{eqn:simpD2}
\end{eqnarray}
\begin{eqnarray}
P\frac{dx_{k,0}}{dt}=-2D_1\sin(\Delta\phi)e^{-\sqrt{\beta}\Delta\xi},
\label{eqn:simpD3}
\end{eqnarray}
\begin{eqnarray}
P_\beta(\frac{V}{2}\frac{dx_{k,0}}{dt}+\frac{d\sigma_{k,0}}{dt})=-2D_2\cos(\Delta\phi)e^{-\sqrt{\beta}\Delta\xi}
,\label{eqn:simpD4}
\end{eqnarray}
where $P$ is the power of the solitary wave with propagation
constant $\beta$, and $D_1,D_2$ are defined in
(\ref{eqD1}),(\ref{eqD2}). From the above equations, we find that
\begin{eqnarray}
\beta_t=V_t=0, \label{Eq1}
\end{eqnarray}
\begin{eqnarray}
\Delta\xi_t=\Delta V,\label{Eq2}
\end{eqnarray}
\begin{eqnarray}
\Delta\phi_t=\Delta\beta,\label{Eq3}
\end{eqnarray}
\begin{eqnarray}
\Delta V_t=-\frac{16\beta
c^2}{P}\cos(\Delta\phi)e^{-\sqrt{\beta}\Delta\xi},\label{Eq4}
\end{eqnarray}
\begin{eqnarray}
\Delta\beta_t=\frac{8\sqrt{\beta}c^2}{P_\beta}\sin(\Delta\phi)
e^{-\sqrt{\beta}\Delta\xi}\label{Eq5}.
\end{eqnarray}
Equations (\ref{Eq1})-(\ref{Eq5}) are the key results in the weak
interaction theory of solitary waves.
These equations can be further simplified by variable rescalings.
Introducing notations
\begin{eqnarray} \label{scaling1}
\psi=\Delta\phi,\zeta=-\sqrt{\beta}\Delta\xi,f=\frac{16\beta^{3/2}
c^2}{P},g=\frac{8\sqrt{\beta}c^2}{P_\beta},\quad
\end{eqnarray}
and
\begin{equation}  \label{time_scale}
\tau=\sqrt{f}\: t, \quad\varepsilon=\frac{g}{f}-1=\frac{P}{2\beta
P_\beta}-1,
\end{equation}
then the dynamical equations (\ref{Eq2})-(\ref{Eq5}) reduce to
\begin{eqnarray}\{\begin{array}{l}
\zeta_{\tau\tau}=\cos{\psi}e^{\zeta}\\
\psi_{\tau\tau}=(1+\varepsilon)\sin{\psi}e^{\zeta}
\end{array}\quad. \label{Dyreduce}
\end{eqnarray}
Eq. (\ref{Dyreduce}) is the final dynamical system we obtained for
the analytical treatment of weak interactions in the generalized NLS
equations (\ref{eqn:1}). It is important to remark that Eq.
(\ref{Dyreduce}) is universal for the generalized NLS equations with
arbitrary nonlinearities. It contains only a single parameter
$\varepsilon$, which depends on the specific form of nonlinearity.

Eq. (\ref{Dyreduce}) has the following general properties. First, it
is Hamiltonian with the conserved Hamiltonian (energy) as
\begin{equation}  \label{Edef}
E=\frac{1}{2}(\dot{\zeta}^2-\dot{\psi}^2)-e^\zeta \cos\psi
+\frac{\varepsilon}{2(1+\varepsilon)} \dot{\psi}^2.
\end{equation}
Here $\dot{(\;)}\equiv d/d\tau$. Second, it has some symmetry
properties. One is that it is time-reversible, i.e., if
$[\zeta(\tau), \psi(\tau)]$ is a solution with initial conditions
$(\zeta_0, \dot{\zeta}_0, \psi_0, \dot{\psi}_0)$, then
$[\zeta(-\tau), \psi(-\tau)]$ is a solution with initial conditions
$(\zeta_0, -\dot{\zeta}_0, \psi_0, -\dot{\psi}_0)$. Another symmetry
is on phase flipping, i.e., if $[\zeta(\tau), \psi(\tau)]$ is a
solution with initial conditions $(\zeta_0, \dot{\zeta}_0, \psi_0,
\dot{\psi}_0)$, then $[\zeta(\tau), -\psi(\tau)]$ is a solution with
initial conditions $(\zeta_0, \dot{\zeta}_0, -\psi_0,
-\dot{\psi}_0)$. Physically, this latter symmetry corresponds to the
interchange of the left and right solitary waves in the PDE
evolutions, which of course does not change the interaction outcome.
Eq. (\ref{Dyreduce}) also has the property that if $\psi(\tau)$ is a
solution, so is $\psi(\tau)+2n\pi$ for any integer of $n$. This
reflects the fact that in the PDE system, solution evolution remains
the same if the phase difference between the solitary waves changes
by a multiple of $2\pi$.

\begin{figure*}
\includegraphics[width=80mm,height=65mm]{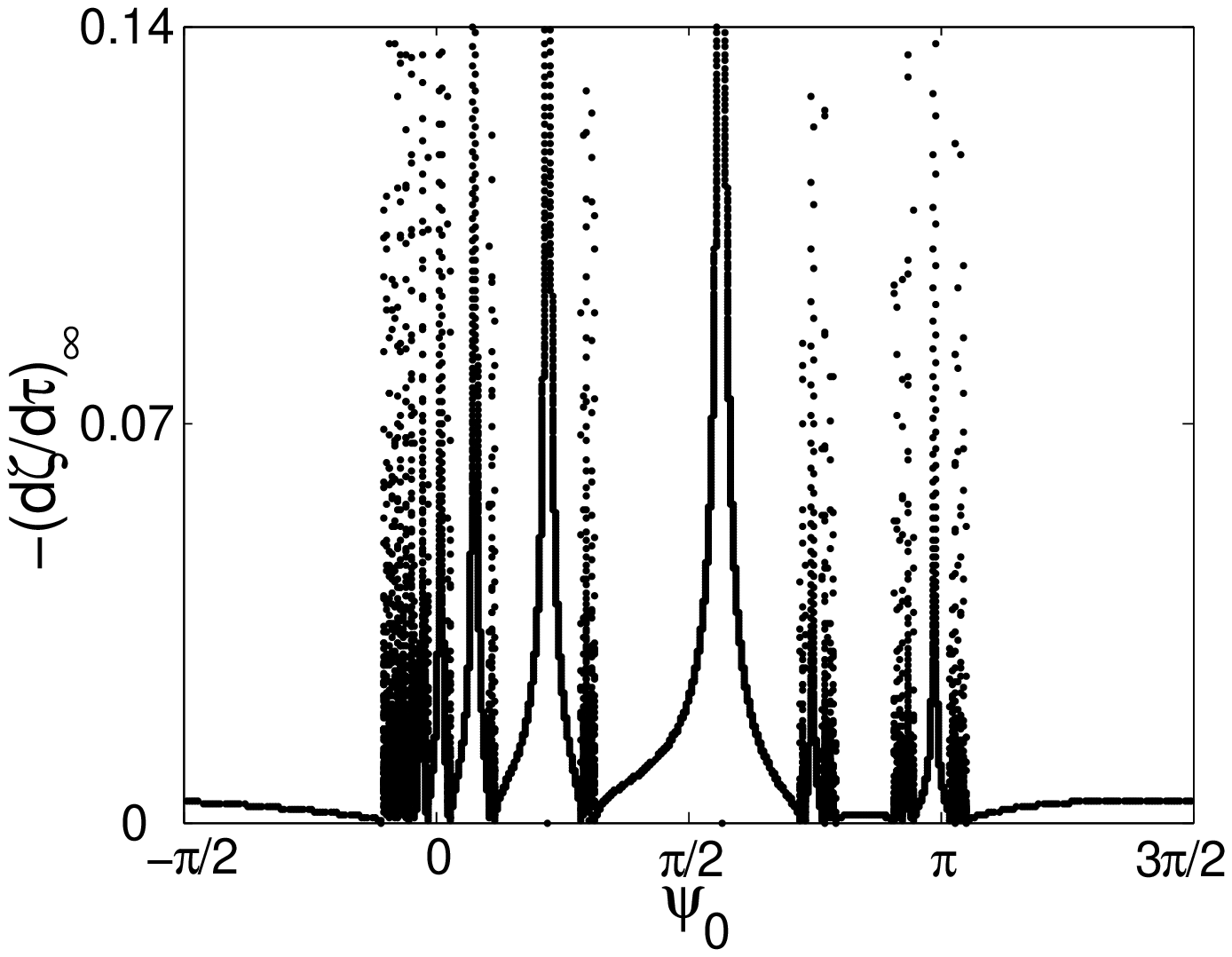}
\includegraphics[width=80mm,height=65mm]{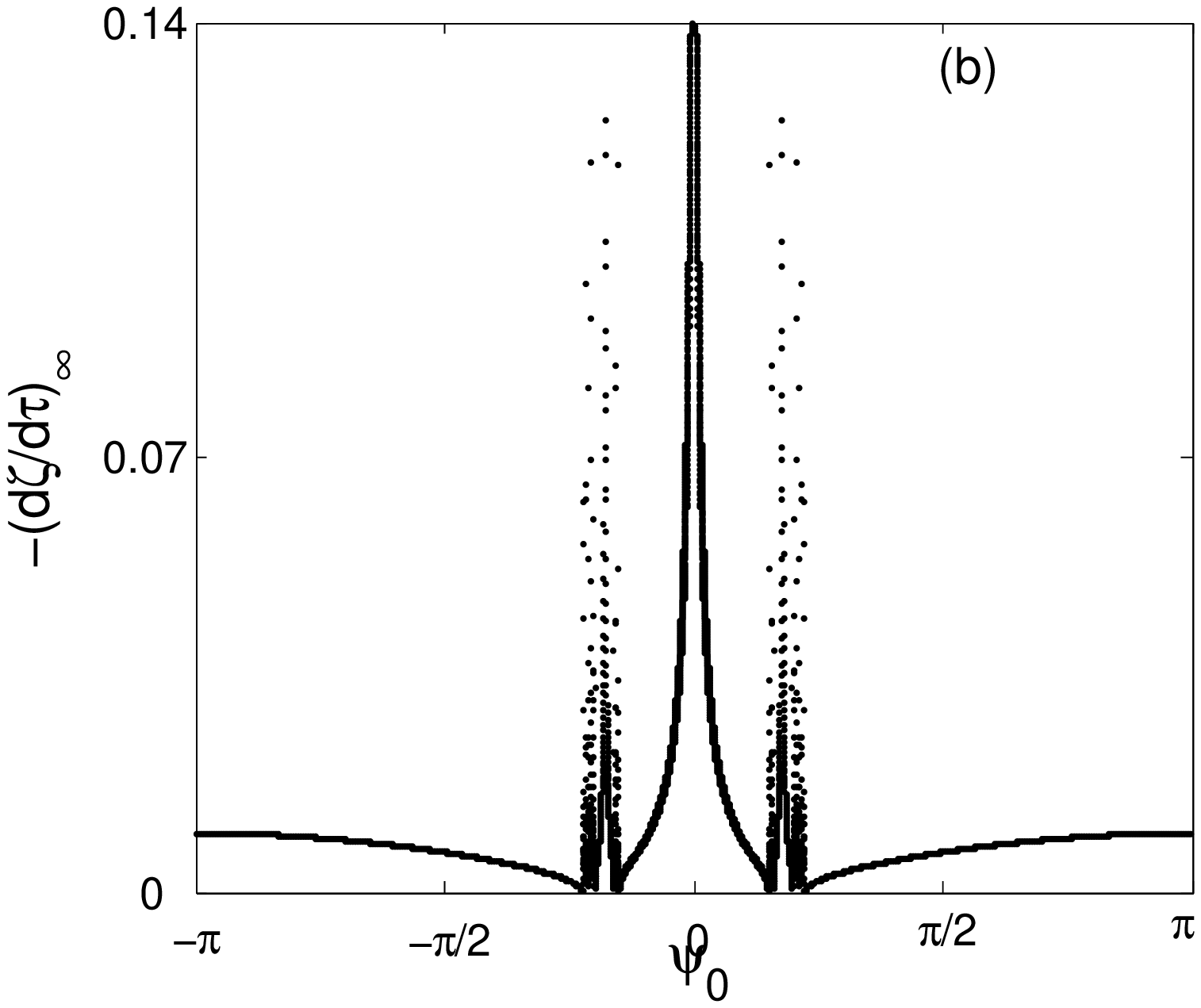}
\caption{\label{fig:ODEglobal} The exit velocity
($-\dot{\zeta}_\infty$) versus the initial phase difference
($\psi_0$) graphs from the ODE model (\ref{Dyreduce}). Here the
initial conditions for (a) and (b) are chosen corresponding to the
PDE simulation results in Figs. 1 and 4 respectively (see text). }
\end{figure*}

\begin{figure*}
\includegraphics[width=80mm,height=70mm]{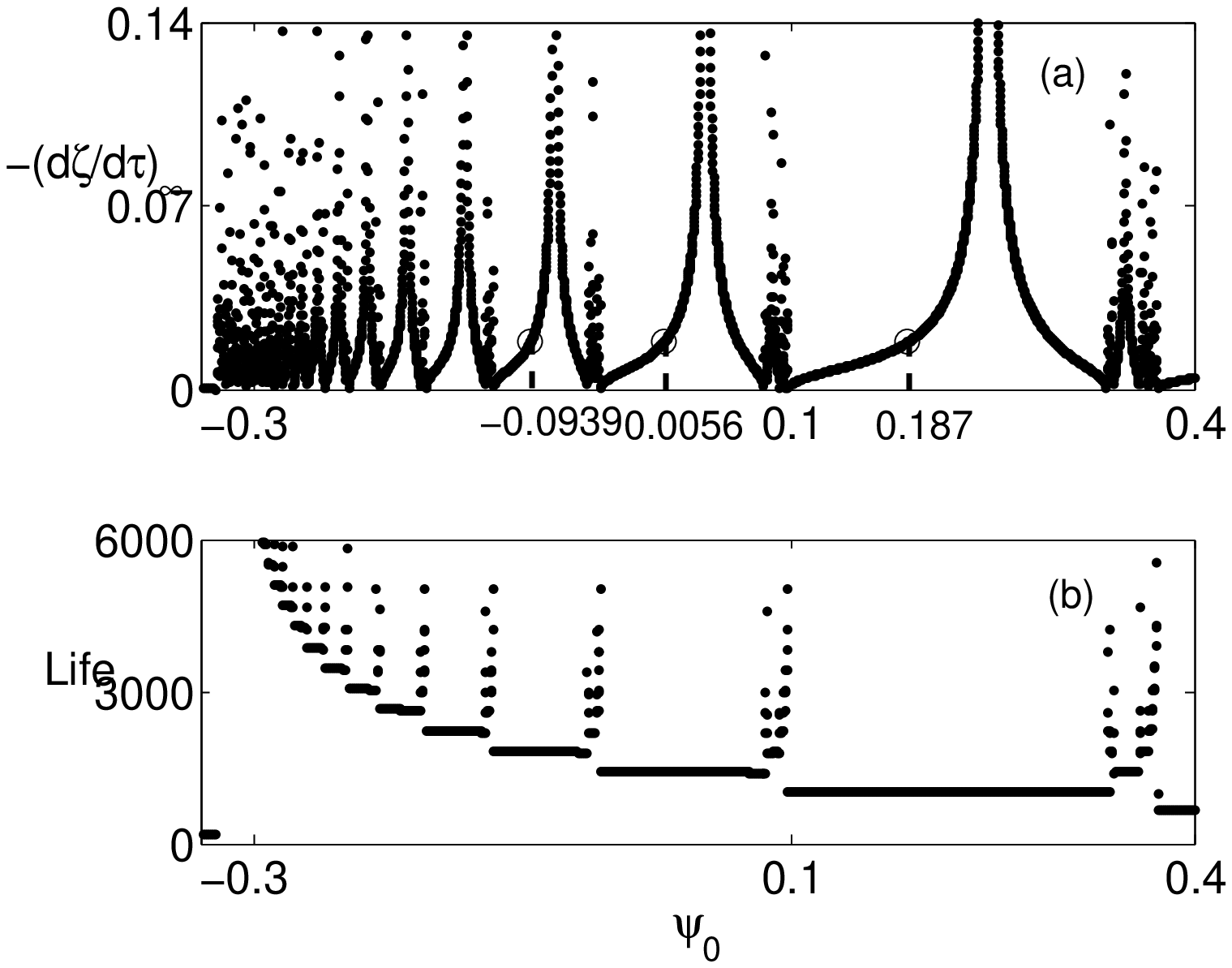}
\hspace{0.4cm}
\includegraphics[width=80mm,height=70mm]{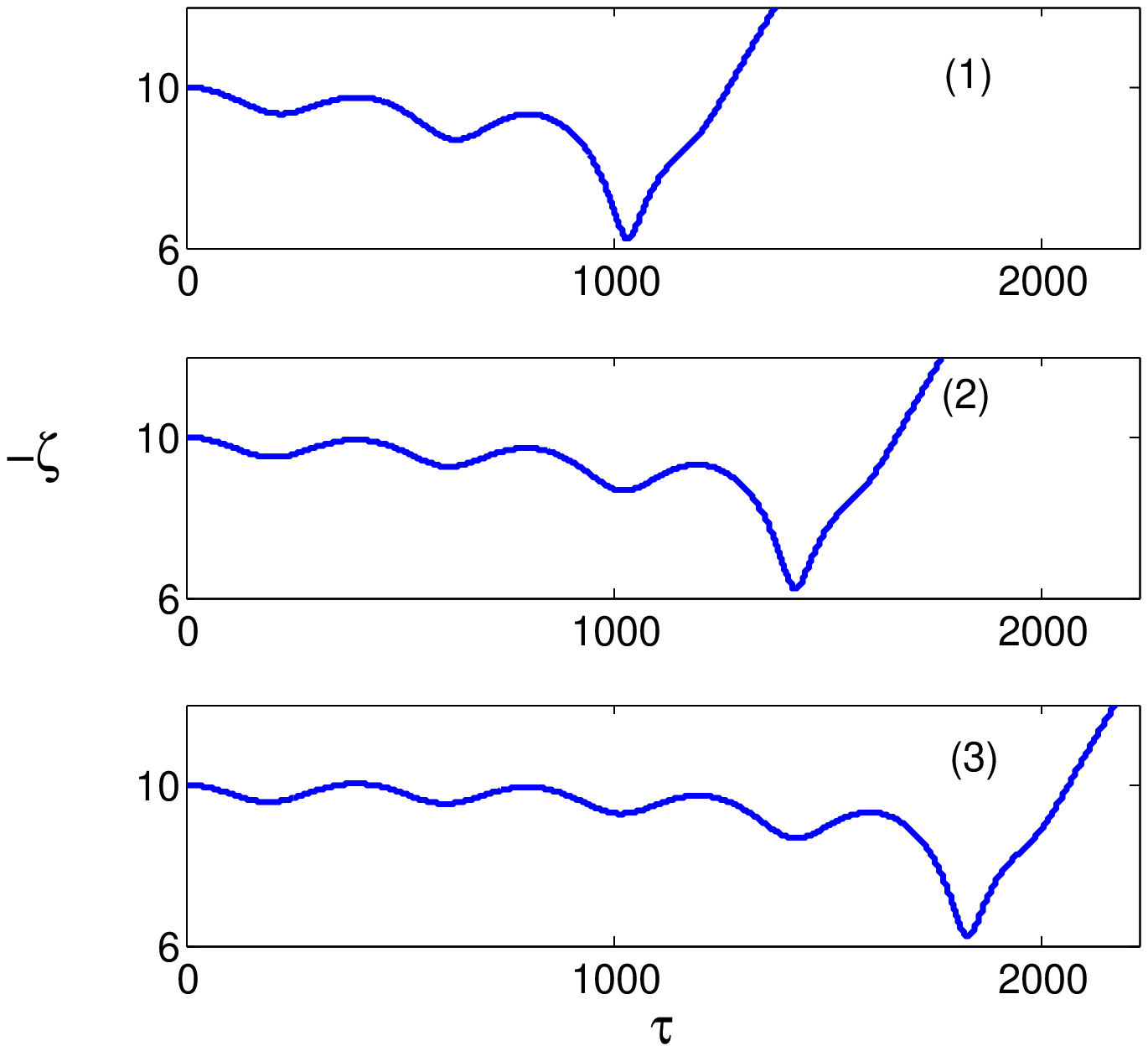}
%
\caption{\label{ODElevel1}~ (a) The exit velocity versus initial
phase difference graph of Fig. \ref{fig:ODEglobal}(a) re-plotted
near the accumulation point of the primary hill sequence; (b) the
life time versus initial phase difference graph; (1)-(3): separation
($-\zeta$) versus time ($\tau$) diagrams at three values of $\psi_0$
marked by circles in (a): (1) 0.187; (2) 0.0056; (3) $-0.0939$. All
these graphs are obtained from the ODE model (\ref{Dyreduce}), and
they should be compared to the corresponding PDE graphs in Fig.
\ref{pdelevel1}. }
\end{figure*}

\begin{figure*}
\includegraphics[width=63mm,height=47mm]{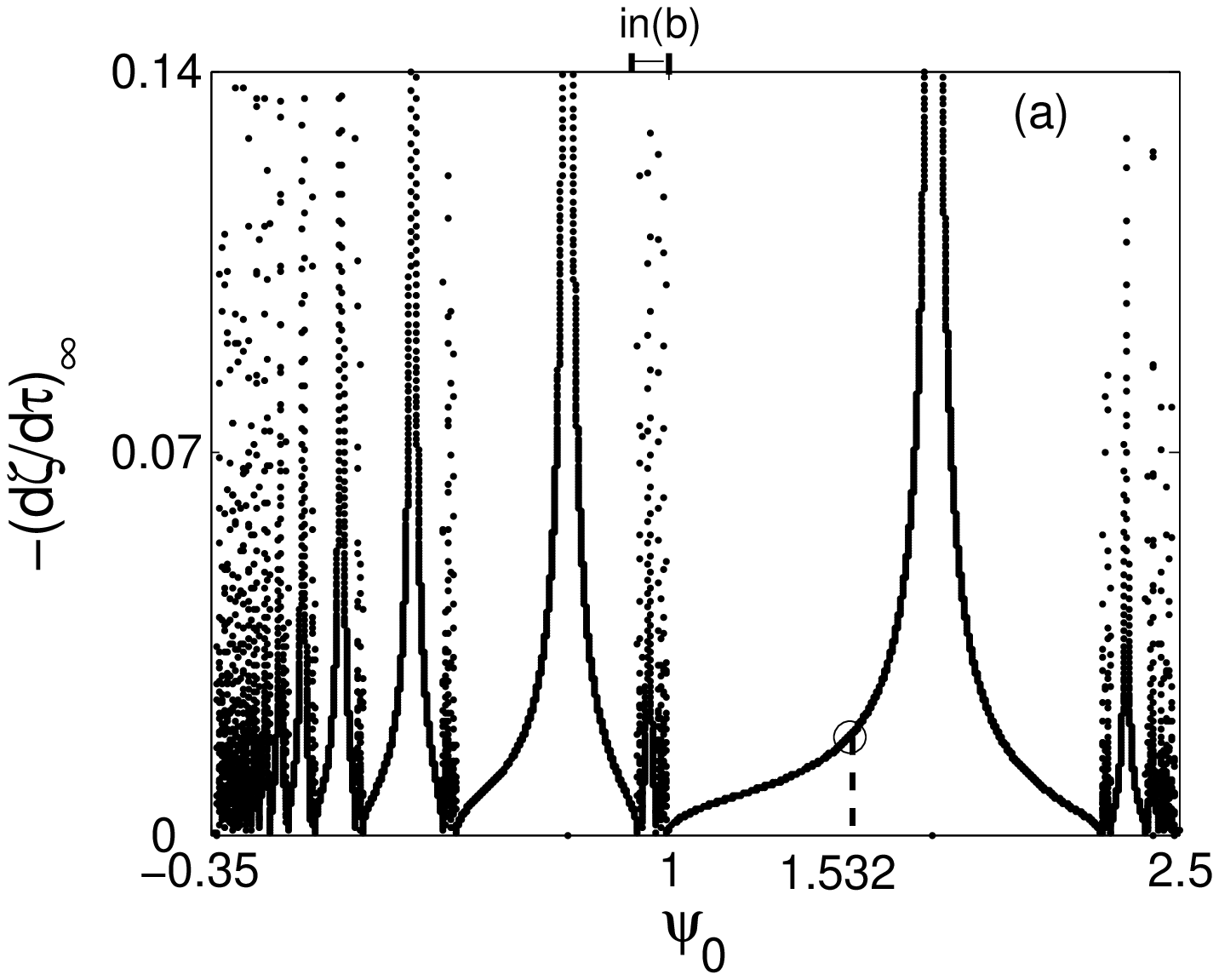}
\includegraphics[width=53mm,height=47mm]{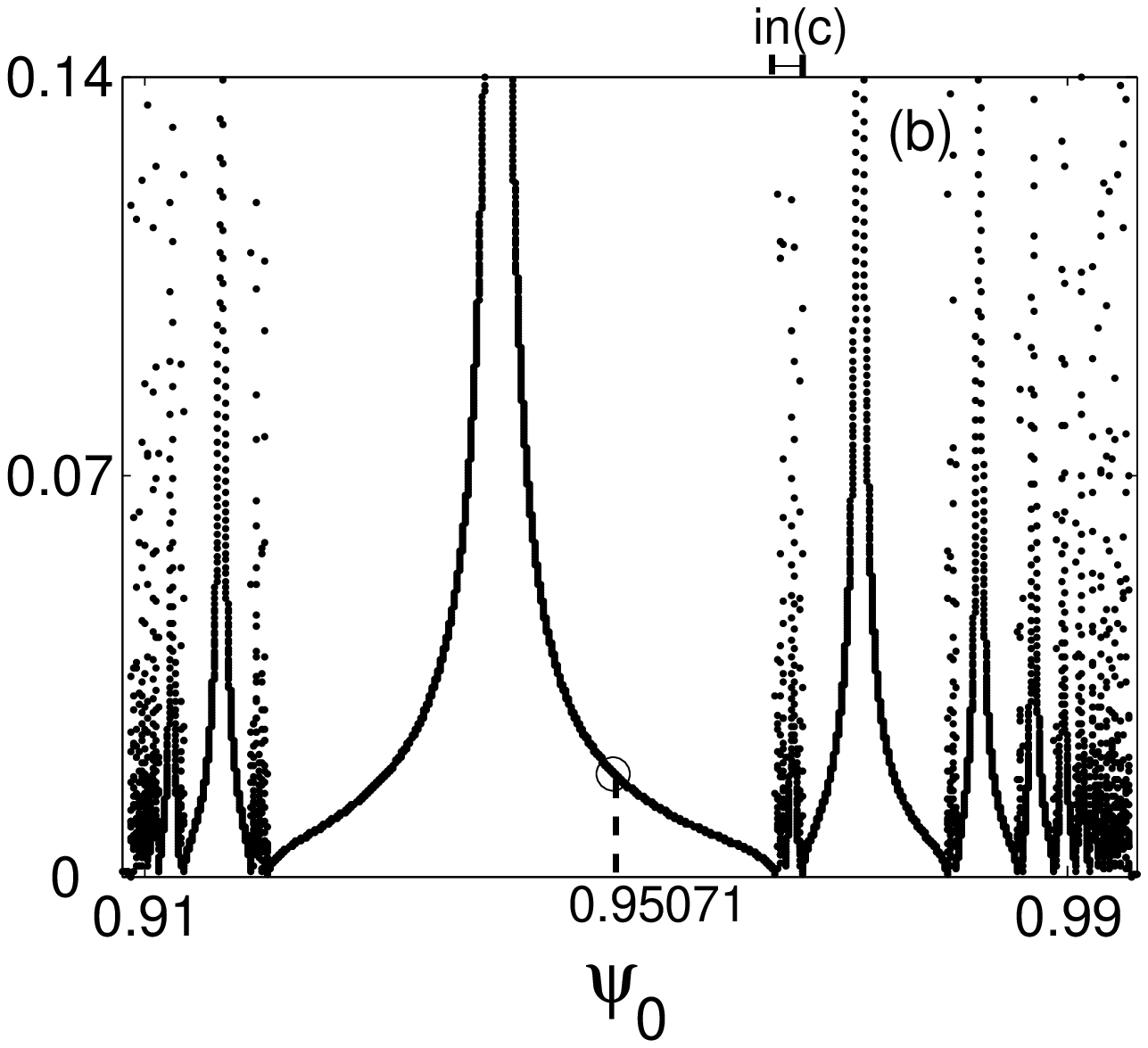}
\includegraphics[width=53mm,height=45mm]{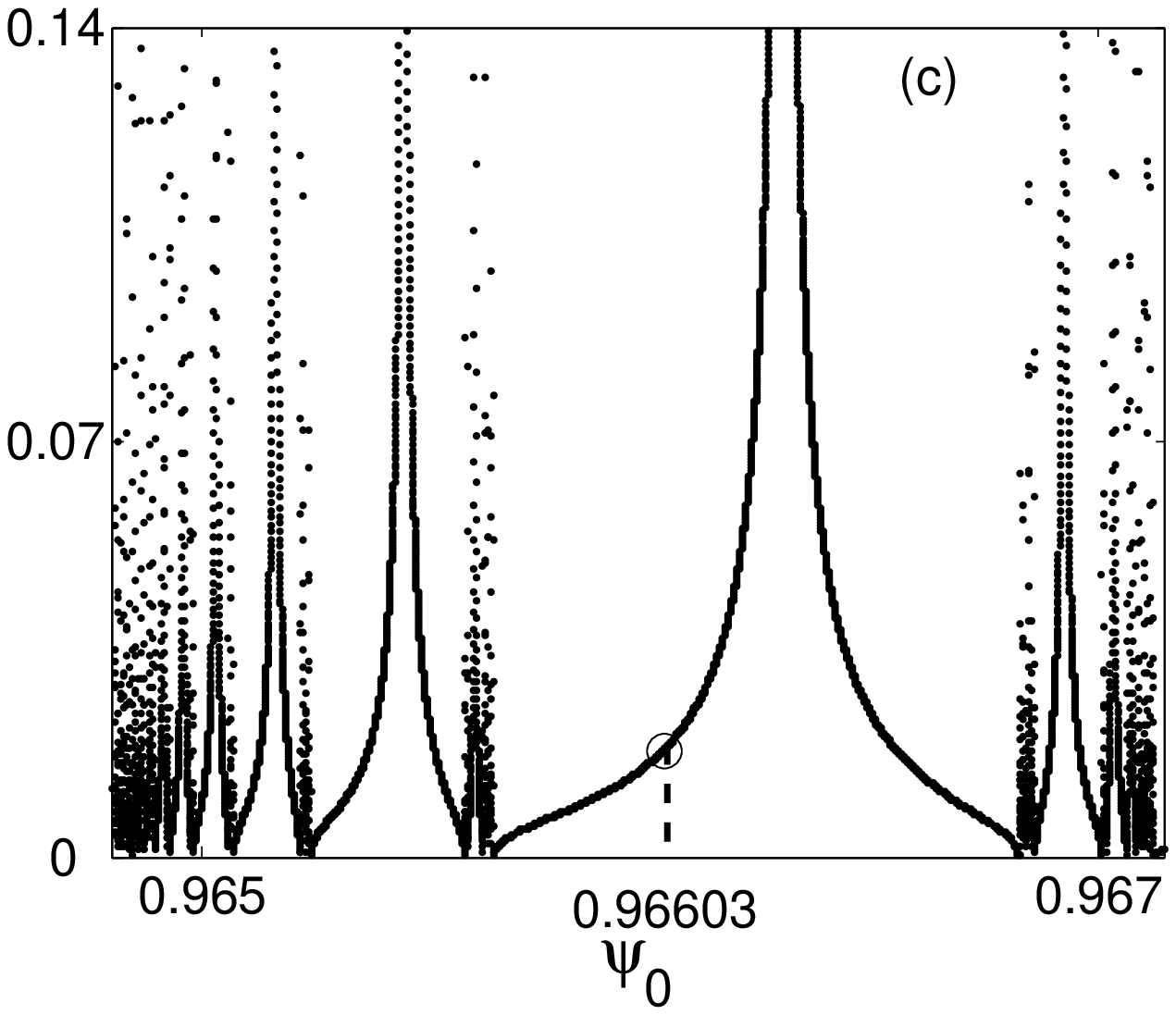}\\
\includegraphics[width=60mm,height=30mm]{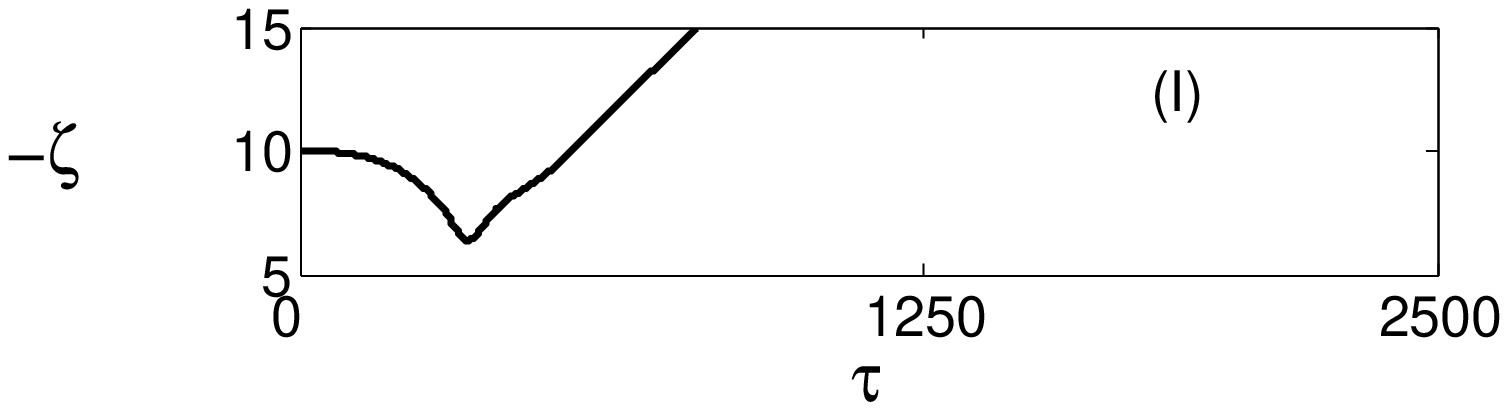}
\includegraphics[width=53mm,height=30mm]{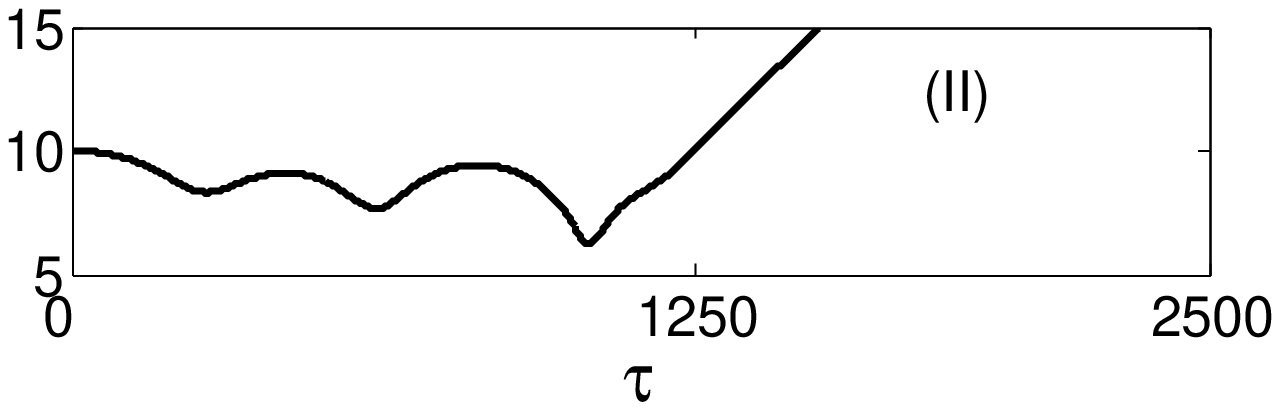}
\includegraphics[width=53mm,height=30mm]{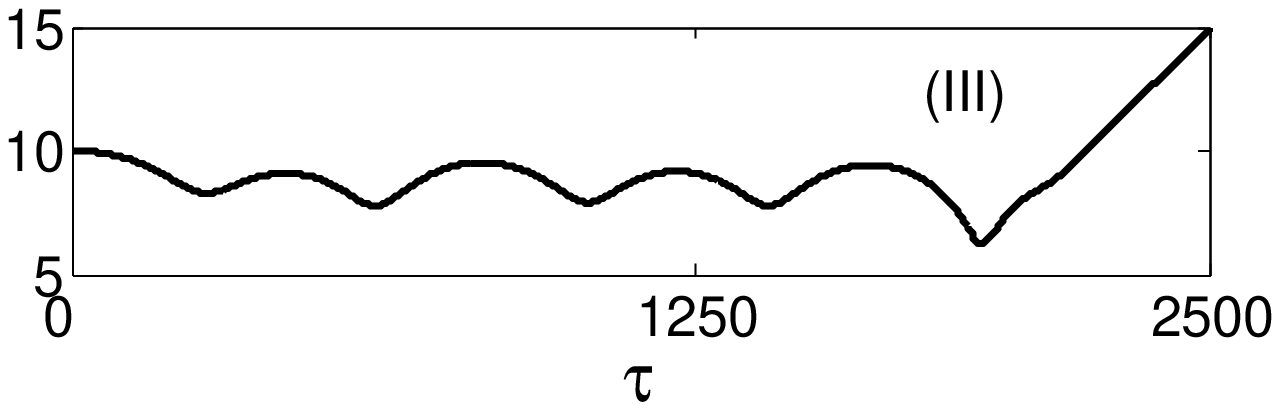}
\caption{\label{fig:ODE-zoom} Top: the exit velocity versus
initial phase difference graph of Fig. \ref{fig:ODEglobal}  and
its two zoomed-in structures; bottom: separation versus time
diagrams at three values of $\psi_0 $ marked by circles in the top
panel: (1) 1.532; (2) 0.95071; (3) 0.96603. These graphs from the
ODE model should be compared to the corresponding PDE graphs in
Fig. \ref{fig:PDEzoom}. }
\end{figure*}

The dynamical equations (\ref{Dyreduce}) are asymptotically accurate
in describing weak interactions in the PDE system (to the leading
order) when the spacing $\Delta \xi$ is large. Surprisingly, even
when the two solitary waves come close to each other, Eq.
(\ref{Dyreduce}) can still describe the interaction process very
well. This is analogous to weak interactions in the (integrable) NLS
equation \cite{Hasegawa_Kodama, Karpman_Solovev}. Below, we make
detailed comparisons between the ODE solutions of
Eq.(\ref{Dyreduce}) and the PDE solutions in Sec. 3(A) for the
cubic-quintic nonlinearity. Inserting the parameter values
$\alpha=1, \gamma=0.04$ and $\beta_0=1$ of PDE simulations into Eqs.
(\ref{Pformula}) and (\ref{cformula}), we get $P=3.74720$,
$P_{\beta}=1.64835$, and $c=2.69495$, thus $f=31.01080$,
$g=35.24845$, and $\varepsilon=0.13665$. Corresponding to the
initial conditions for PDE simulations in Sec. 3(A), the initial
conditions for the ODE system (\ref{Dyreduce}) with nonequal and
equal initial amplitudes are
\begin{equation} \label{ic_nonequal}
\zeta_0=-10, \; \dot{\zeta}_0=0, \; \dot{\psi}_0=-0.01167,
\end{equation}
and
\begin{equation} \label{ic_equal}
\zeta_0=-10, \; \dot{\zeta}_0=0, \; \dot{\psi}_0=0,
\end{equation}
respectively. In both cases, the initial phase difference $\psi_0$
is the control parameter as in PDE simulations. The ODE system
(\ref{Dyreduce}) is numerically solved by the fourth-order
Runge-Kutta method, with the time step set as $0.01$. The simulation
results on the exit velocity $-\dot{\zeta}_\infty$ versus $\psi_0$
graph are shown in Fig.\ref{fig:ODEglobal}.
Clearly, these graphs are very similar to Figs. \ref{cqnonequal} and
\ref{fig:cqequal}) from PDE simulations. We have also investigated
the detailed structures of Fig. \ref{fig:ODEglobal} in ways
analogous to what we did for Figs. 2 and 3. Specifically, we have
examined the primary hill sequence in Fig. \ref{fig:ODEglobal}(a),
and zoomed into regions between primary hills. The results are shown
in Figs.\ref{ODElevel1} and \ref{fig:ODE-zoom} respectively. Both
figures closely resemble Figs. 2 and 3 from the PDE simulations.

The agreement between the ODE model and the PDE simulations is not
only qualitative, but also quantitative. To demonstrate, we compare
the locations and life times of primary hill sequences in Figs. 2
and \ref{ODElevel1}. The comparison results are summarized in Table
1. Very good quantitative agreement between them can be seen. In the
ODE model, the life time is also an almost perfect linear function
of the hill index $n$ in the form (\ref{time_PDE}). When the time
rescaling (\ref{time_scale}) is recovered, the ODE model gives
\begin{equation} \label{ODEvalue}
\omega|_{ODE}=0.08570,\quad \delta|_{ODE}=2.9655,
\end{equation}
closely resembling the corresponding values (\ref{PDEvalue}) from
the PDE simulations.

\begin{table}
\caption{\label{tab:table1} Comparison on locations and life times
of primary hills in Figs. 2(a) and 7(a) from the PDE and ODE
simulations. }
\begin{ruledtabular}
\begin{tabular}{lcccc}
&\quad location (PDE) &location (ODE) & life (PDE) & life (ODE)\\
 \hline
 $n$ & $\Delta\phi_{0,n}$ & $\psi_{0,n}$ & $t_n$ & $\sqrt{f}\tau_n$\\
\hline
1&1.7735  & 1.7794  & 65 &68 \\
2&0.6985  & 0.7015  &117 &119\\
3&0.2430  & 0.2468  &183 &185\\
4&0.0280  & 0.0359  &253 &255\\
5&-0.0850 &-0.0763  &325 &327\\
6&-0.1530 &-0.1431  &398 &400\\
7&-0.1963 &-0.1863  &470 &473\\
8&-0.2258 &-0.2157  &544 &547\\
9&-0.2475 &-0.2367  &617 &620\\
10&-0.2630&-0.2523  &691 &693\\
\hline
$\infty$&-0.3392&-0.3280&$\infty$&$\infty$
\end{tabular}
\end{ruledtabular}
\end{table}


Above we have established that the reduced ODE system
(\ref{Dyreduce}) accurately describes weak interactions of the PDE
system. Since the ODE system (\ref{Dyreduce}) is universal for Eq.
(\ref{eqn:1}) regardless of details of its nonlinearities, we see
that the hill sequences and fractal structures in Eq.
(\ref{Dyreduce}) are universal for weak interactions of solitary
waves in the PDE system (\ref{eqn:1}), as Figs. \ref{cqnonequal},
\ref{fig:cqequal}, \ref{fig:exPDE}, and \ref{fig:ODEglobal} clearly
indicate.

Next we will turn our attention to the ODE system (\ref{Dyreduce}),
and analyze its solution dynamics in more detail. In particular, we
would like to understand why fractal structures arise in this
system, and how to analytically predict their locations and other
main features.

\section{Solutions of the integrable dynamical equations and their
singularity conditions
 \setcounter{equation}{0}}

Eq.(\ref{Dyreduce}) conserves energy (\ref{Edef}) for all values of
$\varepsilon$. When $\varepsilon=0$, it has another conserved
quantity,
\begin{eqnarray}
M=\dot{\zeta}\dot{\psi}-e^\zeta \sin\psi,
\end{eqnarray}
which can be called the momentum of this system.  In this particular
case, system (\ref{Dyreduce}) is an integrable Hamiltonian system
and can be analytically solved. Let $Y=\zeta+i\psi$,
Eq.(\ref{Dyreduce}) becomes
\begin{eqnarray}
Y_{\tau\tau}=e^Y.
\end{eqnarray}
The general solution of this equation is
\begin{eqnarray}
Y(\tau)=\ln\left[-2C^2_1\textrm{sech}^2\left(C_1\tau+C_2\right)\right],\label{GSolu}
\end{eqnarray}
where
\begin{equation}  \label{C1}
C_1=\frac{1}{2}\sqrt{{\dot{Y}_0}^2-2e^{Y_0}}=\frac{1}{\sqrt{2}}\sqrt{E+i
M},
\end{equation}
and
\begin{equation}  \label{C2}
C_2=-\textrm{arctanh}\left(\frac{\dot{Y}_0}{2C_1}\right).
\end{equation}
Here the branch of the square root function in (\ref{C1}) is chosen
such that $\rm{Re}(C_1)\ge 0$. It is noted that solutions $Y$ which
differ by a multiple of $2\pi i$ correspond to the same physical
solution, thus it does not matter which Riemann surface one takes
for the logarithmic function in Eq. (\ref{GSolu}). If $C_1=0$, i.e.,
$\dot{Y}_0=\pm \sqrt{2}\hspace{0.02cm} e^{Y_0/2}$, the solution
(\ref{GSolu}) degenerates to the form
\begin{eqnarray}
Y(\tau)=-2\ln\left(e^{-\frac{1}{2}Y_0}\mp
\frac{1}{\sqrt{2}}\;\tau\right).
\end{eqnarray}
The asymptotic behaviors of these solutions as
$\tau\rightarrow\infty$ can be easily determined. Let
\begin{equation} \label{C12abcd}
C_1=a+b \hspace{0.05cm}i, \quad \frac{C_2}{C1}=c+d \hspace{0.05cm}
i,
\end{equation}
where $a,b,c,d$ are real constants, then the following leading-order
asymptotic expressions for the solution can be obtained when
$\tau\rightarrow\infty$:
\begin{eqnarray}
&& (1)\; a \neq 0: \;\; Y(\tau) \rightarrow
-2|a|\tau-\mbox{sgn}(a) \hspace{0.02cm} 2b\tau \hspace{0.007cm} i;  \label{ane0} \\
&& (2)\;  a=0, \;  b\neq0: \nonumber \\
&& (2a)\; d=0: \; \nonumber\\
&& \quad Y(\tau) =\ln 2b^2-\ln\cos^2b(\tau+c);  \label{asym_dzero}\\
&&(2b)\;d\neq 0: \nonumber \\
&& \quad Y(\tau) = \ln
4b^2-\ln\left[\cosh 2bd+\cos 2b(\tau+c)\right] \nonumber \\
&&\quad\quad\quad\quad +
2i\arctan\left[\textrm{tanh} (bd)\hspace{0.01cm} \tan b(\tau+c)\right];  \\
&&(3)\; a=0,b=0: \;\; Y(\tau) \rightarrow -2\ln({\mp\tau}).
\end{eqnarray}

\begin{figure}
\includegraphics[width=80mm,height=60mm]{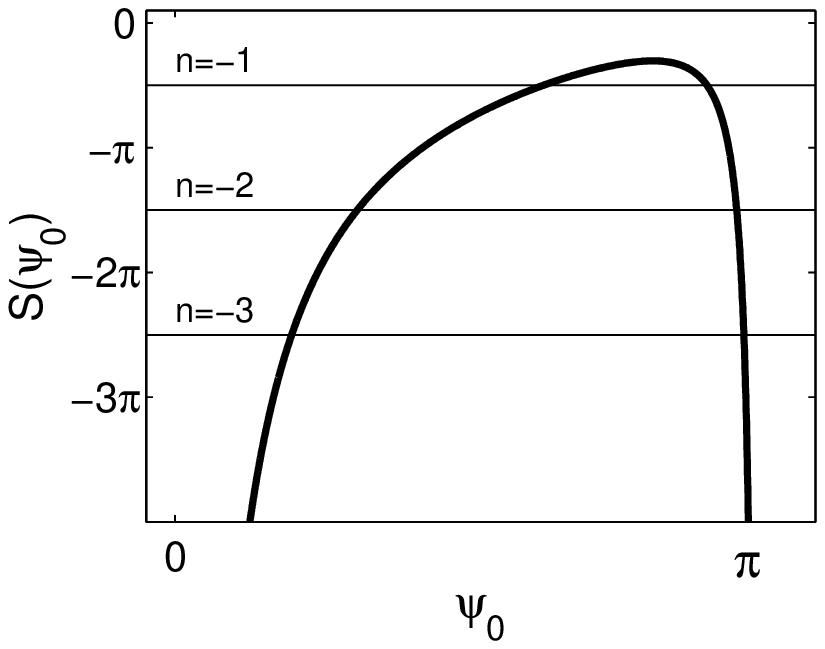}

\vspace{0.3cm}
\includegraphics[width=84mm,height=30mm]{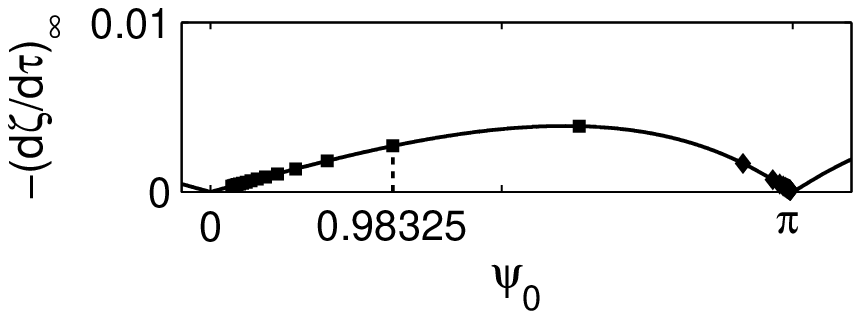}
\caption{Top: graph of the function $S(\psi_0)$ defined in Eq.
(\ref{condition2}) for the initial conditions (\ref{ic_nonequal}).
Intersections of the graph with horizontal lines are singularity
points. Bottom: exit velocity versus $\psi_0$ graph in the
integrable system (\ref{Dyreduce}). Both squares and diamonds are
singularity points. \label{figS}}
\end{figure}

From these asymptotic expressions, we see that when $a\ne 0$, the
two solitary waves eventually move away from each other with exit
velocity $2|a|$; when $a=0$ but $b\ne 0$, the solution is
time-periodic for both $d=0$ and $d\ne 0$, the difference being that
in the former case, the periodic solution exhibits finite-time
singularities (where $\zeta=\mbox{Re}(Y) \to \infty$), while in the
latter case, the solution has no singularities; when $a=b=0$, the
two solitary waves eventually separate logarithmically, and the exit
velocity is zero. As an example, we take the initial conditions
(\ref{ic_nonequal}). In this case, the graph of exit velocity
$-\dot{\zeta}_\infty (=2|a|)$ versus $\psi_0$ is plotted in Fig.
\ref{figS} (bottom panel). This graph is smooth everywhere, except
at $\psi_0=0,\pm\pi$ where it has a cusp (due to the absolute-value
function in $|a|$). The squares and diamonds on this graph will be
explained later. Clearly, this graph has no fractal structure
anywhere. Thus, fractal dependence is a signature of the dynamical
system (\ref{Dyreduce}) when it is non-integrable (with $\varepsilon
\ne 0$), not when it is integrable (with $\varepsilon=0$).

The above asymptotic states do not tell the full story about the
solution dynamics in the integrable system. For instance, for the
case of $a\ne 0$, even though the solution has a benign-looking
asymptotics (\ref{ane0}) as $\tau \to \infty$, the solution can
still develop a singularity (where the separation $\zeta \to
\infty$) at a finite time. These solutions with finite-time
singularities turn out to be critical for the appearance of fractal
structures in the non-integrable system, as our numerics in the next
section will indicate. Thus we analyze these singularity solutions
in more details below. The necessary and sufficient conditions for
singularities in solution (\ref{GSolu}) are that
\begin{eqnarray} \label{sing_cond}
\cosh(C_1\tilde{\tau}+C_2)=0,
\end{eqnarray}
and $\tilde{\tau}>0$, where $\tilde{\tau}$ is the time of
singularity. If $\tilde{\tau}<0$, i.e., singularities in the
solution occur at a negative time, such singularities are irrelevant
for the time evolution of Eq. (\ref{Dyreduce}) and need not be
considered. The solutions of Eq. (\ref{sing_cond}) are
\begin{eqnarray} \label{tildetau_form}
C_1\tilde{\tau}+C_2=\frac{1}{2}(2n+1)\pi i, \quad n=0, \pm 1, \pm 2,
\cdots.
\end{eqnarray}
This is a complex-valued relation, which gives two real relations on
$\tilde{\tau}$, $C_1$ and $C_2$. When $a\ne 0$, i.e., $C_1$ is not
purely imaginary, we find by separating the real and imaginary parts
of Eq. (\ref{tildetau_form}) that the solution (\ref{GSolu}) has a
single finite-time singularity of the type $\ln(\tau-\tilde{\tau})$
if the following conditions are satisfied:
\begin{equation}
S\equiv \frac{\mbox{Im}(C_1^*C_2)}{\mbox{Re}(C_1)}
=\frac{1}{2}\left(2n+1\right)\pi, \quad n=0, \pm 1, \pm 2, \cdots,
\label{condition2}
\end{equation}
\begin{equation}
\tilde{\tau}= -\frac{\rm{Re}(C_2)}{\rm{Re}(C_1)}> 0.
\label{condition1}
\end{equation}
Here Re$(\cdot)$ and Im$(\cdot)$ represent the real and imaginary
parts of a complex number. When $a=0 \; (b\ne 0)$, singularity
solutions exist if $d=0$. These solutions have an infinite number of
finite-time singularities of the type $\ln(\tau-\tilde{\tau})$, as
the formula (\ref{asym_dzero}) indicates. Physically, at the time of
singularity $\tilde{\tau}$, the two solitary waves strongly collide,
thus $\tilde{\tau}$ is the collision time. Whether conditions
(\ref{condition2}) and (\ref{condition1}) can be satisfied depends
on the initial conditions (which determine the $C_1$ and $C_2$
values, see (\ref{C1}), (\ref{C2})). In the text below, we will call
initial conditions $(\zeta_0, \dot{\zeta}_0, \psi_0, \dot{\psi}_0)$
which satisfy Eqs. (\ref{condition2}) and (\ref{condition1}) as
singularity points. At singularity points, solutions of the
integrable dynamical system (\ref{Dyreduce}) develop finite-time
singularities.

To demonstrate how to determine singularity points in the
initial-condition space, we take initial conditions
(\ref{ic_nonequal}) of Fig. \ref{fig:ODEglobal}(a) as an example.
Here $\psi_0$ is a control parameter. With these initial conditions,
the graph of function $S(\psi_0)$ is plotted in Fig. \ref{figS} (top
panel). This graph has a maximum 0.96. As $\psi_0 \to 0^+$ or
$\pi^-$, $S(\psi_0) \to -\infty$.
As we can see from this graph, for any value of $n\le -1$, Eq.
(\ref{condition2}) has two roots, $\psi_{0, n}^{(1)}$ and $\psi_{0,
n}^{(2)}$. We have checked that these roots satisfy the other
singularity condition (\ref{condition1}), thus these $\psi_{0,
n}^{(1)}$ and $\psi_{0, n}^{(2)}$ values are singularity points. It
is noted that the graph of function $S(\psi_0)$ also has another
piece in the interval $\pi < \psi_0 < 2\pi$, which is the mirror
image of that shown in Fig. \ref{figS} around the point
$\psi_0=\pi$. But in that interval, $\tilde{\tau}<0$, not satisfying
the second singularity condition (\ref{condition1}), thus we did not
plot that piece of the graph in Fig. \ref{figS}.

Next, we examine these singularity points $\psi_{0, n}^{(1)}$ and
$\psi_{0, n}^{(2)}$ in more detail. These points form two infinite
sequences with $n=-1, -2, \dots$, which accumulate at $\psi_0=0^+$
and $\pi^-$ respectively. In Fig. \ref{figS} (bottom panel), these
two sequences are marked by squares and diamonds on the exit
velocity versus $\psi_0$ graph. Calculating the asymptotics of $C_1$
from Eq. (\ref{condition2}) and substituting it into Eq.
(\ref{condition1}), we find that the collision times
$\tilde{\tau}_n$ of both sequences have the following asymptotic
expressions:
\begin{equation} \label{collide_time}
\omega \tilde{\tau}_n= 2|n|\pi + \pi, \quad n \to -\infty,
\end{equation}
where $\omega=2\mbox{Im}(C_1|_{\psi_0=0})$ and
$2\mbox{Im}(C_1|_{\psi_0=\pi})$ for the left and right sequences
respectively. The asymptotic formulas for the locations of these two
singularity sequences \{$\psi_{0, n}^{(1)}$\} and \{$\psi_{0,
n}^{(2)}$\} can also be calculated. We find from (\ref{condition2})
that
\begin{equation} \label{location_sing}
\psi_{0, n}^{(1)} \to \frac{A_1}{(2n+1)\pi}, \quad  n \to -\infty,
\end{equation}
and
\begin{equation} \label{location_sing2}
\pi-\psi_{0, n}^{(2)} \to \frac{A_2}{(2n+1)\pi}, \quad  n \to
-\infty,
\end{equation}
where
\[A_1=\left.8e^{-\zeta_0}\rm{Re}(C_2)\rm{Im}^2(C_1)\right|_{\psi_0=0},
\]
and
\[A_2=\left.8e^{-\zeta_0}\rm{Re}(C_2)\rm{Im}^2(C_1)\right|_{\psi_0=\pi}.
\]

\begin{figure}
\includegraphics[width=80mm,height=65mm]{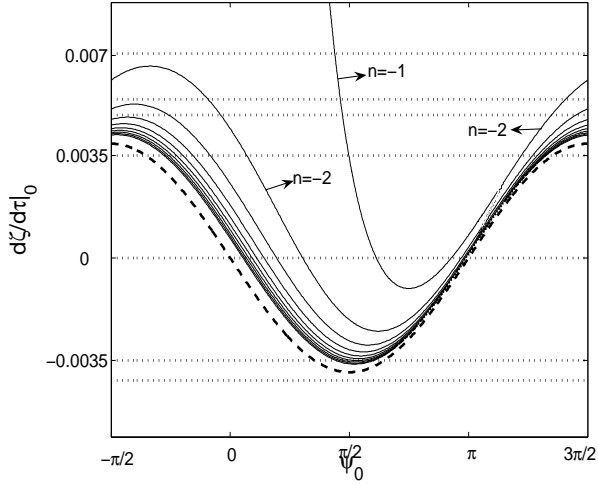}
\centering \caption{\label{singularpoints}Singularity points
satisfying conditions (\ref{condition2})-(\ref{condition1}) in the
($\psi_0$, $\dot{\zeta}_0$) plane. The dashed curve is the
accumulation curve. Here $\zeta_0=-10, \dot{\psi}_0=-0.01167$. }
\end{figure}

The above detailed analysis on singularity points was performed for
the particular initial conditions (\ref{ic_nonequal}) where the two
solitary waves are initially stationary ($\dot{\zeta}_0=0$). What
will happen if $\dot{\zeta}_0 \neq 0$? To answer this question, we
fix $\zeta_0$ and $\dot{\psi}_0$ as in Eq. (\ref{ic_nonequal}), vary
$\dot{\zeta}_0$, and examine how singularity points move in the
($\psi_0$, $\dot{\zeta}_0$) plane. The results are shown in
Fig.\ref{singularpoints}. The top curve corresponds to $n=-1$,  the
next curve corresponding to $n=-2$, and so on. All curves are
bounded from both above and below except the top one (with $n=-1$).
When $n\rightarrow -\infty$, these curves approach the accumulation
curve plotted by the dashed line in Fig.\ref{singularpoints}. Below
this accumulation curve, there are no singularity points. The
analytical formula for this accumulation curve can be easily
derived. On this accumulation curve, $C_1$ must be pure imaginary,
thus
\begin{eqnarray}
&M=\dot{\zeta}_{0c}\dot{\psi}_0-e^{\zeta_0}\sin{\psi_{0}}=0,\label{accumulation1}\\
&E=\frac{1}{2}(\dot{\zeta}_{0c}^2-\dot{\psi}_0^2)-e^{\zeta_0}\cos{\psi_{0}}<0.
\label{accumulation2}
\end{eqnarray}
Here $(\dot{\zeta}_{0c}, \psi_{0})$ is an accumulation point. From
Eq. (\ref{accumulation1}), we see that the function of the
accumulation curve is
\begin{equation} \label{sine_curve}
\dot{\zeta}_{0c}=\frac{e^{\zeta_0}\sin{\psi_{0}}}{\dot{\psi}_0}.
\end{equation}
The maximum and minimum of this curve are
\begin{equation} \label{threshold}
\dot{\zeta}_{0c, \rm{min}}=-\dot{\zeta}_{0c, \rm{max}}=
-\left|\frac{e^{\zeta_0}}{\dot{\psi}_0}\right|.
\end{equation}
For the $\zeta_0$ and $\dot{\psi}_0$ values of Fig.
\ref{singularpoints}, we get $\dot{\zeta}_{0c, \rm{min}}=-0.00389$.
If $\dot{\zeta}_0 < \dot{\zeta}_{0c, \rm{min}}$, there are no
singularity solutions for any value of $\psi_0$. When
$\dot{\zeta}_{0c, \rm{min}}< \dot{\zeta}_0 < \dot{\zeta}_{0c,
\rm{max}}$, two infinite sequences of singularity points can be
found. When $\dot{\zeta}_0>\dot{\zeta}_{0c, \rm{max}}$, however, the
number of singularity points becomes finite; this number gradually
decreases (down to one) as $\dot{\zeta}_0$ increases.

The above calculations of singularity points and their accumulation
curves were made for special choices of initial conditions
$\zeta_0=-10$ and $\dot{\psi}_0=-0.01167$ (see Fig.
\ref{singularpoints}). In view of the importance of singularity
points for fractal structures which we will reveal in the next
section, we would like to discuss singularity points and their
accumulation curves further for {\it general} initial conditions
below.

First, we examine the accumulation curve in the ($\psi_0$,
$\dot{\zeta}_0$) plane for general initial conditions $\zeta_0$ and
$\dot{\psi}_0$. In this general case, the accumulation curve (if it
exists) is necessarily given by Eq. (\ref{sine_curve}). But the
curve (\ref{sine_curve}) (or portions of it) may not satisfy
condition (\ref{accumulation2}), thus may not actually be the
accumulation curve. Below we determine what portions of the curve
(\ref{sine_curve}) are the accumulation curve. Before we do so, let
us first point out that conditions (\ref{accumulation1}) and
(\ref{accumulation2}) are not only the necessary, but also
sufficient conditions for the accumulation curve. In addition, the
accumulation of singularity points toward the accumulation curve is
always from the upper side, not lower side. To show these, we only
need to prove that condition (\ref{condition1}) holds only on the
upper edge of the curve (\ref{sine_curve}), but not the lower edge
of it. On the upper edge of (\ref{sine_curve}), $C_1$ is purely
imaginary, and $\rm{sgn}(M)=\rm{sgn}(\dot{\psi}_{0})$. Thus
$\rm{sgn}(\rm{Im}(C_1))=\rm{sgn}(M)=\rm{sgn}(\dot{\psi}_{0})$.
Consequently, $\rm{Re}(\dot{Y}_0/2C_1)>0$. Notice that for any
complex number $z$, $\mbox{Re}[\mbox{tanh}(z)]$ and $\mbox{Re}(z)$
have the same sign, hence $\rm{Re}(C_2)<0$. Then due to
$\rm{Re}(C_1)>0$, condition (\ref{condition1}) thus holds. By
similar reasoning, we can show that on the lower edge of the curve
(\ref{sine_curve}), condition (\ref{condition1}) does not hold. Thus
singularity points accumulate toward (\ref{sine_curve}) only from
above, not below.

Now we turn to equations (\ref{accumulation1}) and
(\ref{accumulation2}), and use them to determine the accumulation
curve for the general case.  Substituting Eq. (\ref{sine_curve})
into inequality (\ref{accumulation2}) and simplifying, this
inequality becomes
\begin{equation}
-\frac{1}{\dot{\psi}_0^2}\left(\dot{\psi}_0^2+e^{\zeta_0}(1+\cos\psi_{0})\right)
\left(\dot{\psi}_0^2-e^{\zeta_0}(1-\cos\psi_{0})\right)<0,
\end{equation}
which is equivalent to
\begin{equation} \label{cospsi2}
\cos\psi_{0} > 1-e^{-\zeta_0}\dot{\psi}_0^2.
\end{equation}
Thus the accumulation curve is the parts of curve (\ref{sine_curve})
where $\psi_{0}$ satisfies the constraint (\ref{cospsi2}). If
$\dot{\psi}_0^2>2e^{\zeta_0}$, condition (\ref{cospsi2}) is
satisfied for all values of $\psi_0$, hence the entire curve
(\ref{sine_curve}) is the accumulation curve. If $0<\dot{\psi}_0^2
\le 2e^{\zeta_0}$, portions of the curve (\ref{sine_curve}) centered
at $\psi_0=\pi$ do not satisfy condition (\ref{cospsi2}), thus do
not belong to the accumulation curve. The rest of the curve
(\ref{sine_curve}) does satisfy condition (\ref{cospsi2}), thus is
the accumulation curve. If $\dot{\psi}_0=0$ (equal initial amplitude
case), no value of $\psi_0$ satisfies condition (\ref{cospsi2}),
thus accumulation points do not exist.

Next, we derive two general properties about singularity points in
the $(\psi_0, \dot{\zeta}_0)$ plane for general initial conditions
$\zeta_0$ and $\dot{\psi}_0$. One property is that, if
$\dot{\zeta}_{0c}$ is on the accumulation curve, then for any
$\dot{\zeta}_0 <\dot{\zeta}_{0c}$, singularity points can not exist.
We will prove this by showing that $\tilde{\tau}<0$ for
$\dot{\zeta}_0 <\dot{\zeta}_{0c}$. To show $\tilde{\tau}<0$, we only
need to show $\rm{Re}(\dot{Y}_0/2C_1)<0$ (see above). Without loss
of generality, we only show this for $\dot{\psi}_0<0$; the proof for
$\dot{\psi}_0>0$ is similar (in fact, as has been pointed out
before, flipping the sign of $\dot{\psi}_0$ physically amounts to
interchanging the positions of the left and right solitary waves and
thus does not affect the interaction outcome). For $\dot{\psi}_0<0$
and $\dot{\zeta}_0 <\dot{\zeta}_{0c}$, $C_1$ is in the first
quadrant (as $M>0$). If $\dot{\zeta}_0<0$, then $\dot{Y}_0$ is in
the third quadrant, thus $\rm{Re}(\dot{Y}_0/2C_1)<0$ holds. Now we
consider $0<\dot{\zeta}_0<\dot{\zeta}_{0c}$. In this case,
$\dot{Y}_0$ is in the fourth quadrant, hence $i\dot{Y}_0$ lies in
the first quadrant (like $C_1$). To show
$\rm{Re}(\dot{Y}_0/2C_1)<0$, we only need to show
$\rm{arg}(i\dot{Y}_0)<\rm{arg}(2C_1)$. Since both $i\dot{Y}_0$ and
$C_1$ are in the first quadrant, we only need to show
$\rm{arg}(-\dot{Y}_0^2)<\rm{arg}(4C_1^2)$. Notice that
\begin{equation}
-\dot{Y}_0^2+4C_1^2=-2e^{Y_0},
\end{equation}
which is independent of $\dot{\zeta}_0$. In addition, the angle of
$-2e^{Y_0}$ falls in between those of $-\dot{Y}_0^2$ and $4C_1^2$.
Thus to show $\rm{arg}(-\dot{Y}_0^2)<\rm{arg}(4C_1^2)$, we only need
to show $\rm{arg}(-\dot{Y}_0^2)<\rm{arg}(-2e^{Y_0})$. Note that
\begin{equation}
-\dot{Y}_0^2=\dot{\psi}_0^2-\dot{\zeta}_0^2-2i\dot{\psi}_0\dot{\zeta}_0,
\end{equation}
whose angle is an increasing function of $\dot{\zeta}_0$ when
$\dot{\psi}_0<0$, thus for $\dot{\zeta}_0 <\dot{\zeta}_{0c}$,
\begin{equation} \label{argY}
\rm{arg}(-\dot{Y}_0^2)<\rm{arg}(\dot{\psi}_0^2-\dot{\zeta}_{0c}^2-2i\dot{\psi}_0\dot{\zeta}_{0c}).
\end{equation}
Now recall that $\dot{\zeta}_{0c}$ lies on the accumulation curve,
thus it satisfies the conditions (\ref{accumulation1}) and
(\ref{accumulation2}). Substituting these conditions into
(\ref{argY}), and recalling our assumptions of $\dot{\psi}_0<0$ and
$0<\dot{\zeta}_0 <\dot{\zeta}_{0c}$, we find that the right hand
side of (\ref{argY}) is less than $\rm{arg}(-2e^{Y_0})$, thus
inequality $\rm{arg}(-\dot{Y}_0^2)<\rm{arg}(-2e^{Y_0})$ is proved.
Summarizing the above arguments, we conclude that for any
$\dot{\zeta}_0$ below the accumulation curve, singularity points do
not exist in the $(\psi_0, \dot{\zeta}_0)$ plane.

Another general property about singularity points is that, at
sufficiently large values of $\dot{\zeta}_0$, there is a unique
singularity point in the $\psi_0$ interval. The proof is as follows.
It is easy to check that when $\dot{\zeta}_0^2+\dot{\psi}_0^2 \gg
e^{\zeta_0}$ and $|\dot{\zeta}_0|\ _{\sim }^{>}\ |\dot{\psi}_0|$,
functions $S$ and $\tilde{\tau}$ have the following leading-order
asymptotic expressions,
\begin{equation} \label{tauasym}
\tilde{\tau} \to \frac{1}{\dot{\zeta}_0}\mbox{ln}
\frac{2(\dot{\zeta}_0^2+\dot{\psi}_0^2)}{e^{\zeta_0}},
\end{equation}
\begin{equation} \label{Sasym}
S \to \frac{1}{2}\mbox{sgn}(\dot{\zeta}_0)\psi_0+S_A,
\end{equation}
where
\begin{equation}
S_A=\frac{\dot{\psi}_0}{2|\dot{\zeta}_0|}\ln
\frac{2(\dot{\zeta}_0^2+\dot{\psi}_0^2)}{e^{\zeta_0}}-
\mbox{arctan}\frac{\dot{\psi}_0}{|\dot{\zeta}_0|}-\frac{1}{2}\pi,
\end{equation}
and the relative errors are
$O\left(e^{\zeta_0}/(\dot{\zeta}_0^2+\dot{\psi}_0^2)\right)$. From
Eq. (\ref{Sasym}), we see that the rise of $S$ value over the
interval $0\le \psi_0 \le 2\pi$ is $\pi$, which guarantees that Eq.
(\ref{condition2}) has a single solution in the $\psi_0$ interval
for a single value of $n$. From Eq. (\ref{tauasym}), we see that
when $\dot{\zeta}_0>0$ is sufficiently large, $\tilde{\tau}>0$ over
the entire $\psi_0$ interval. Thus singularity conditions
(\ref{condition2}) and (\ref{condition1}) admit a unique singularity
point. We note by passing that when $\dot{\zeta}_0$ is sufficiently
large negative, $\tilde{\tau}<0$ over the entire $\psi_0$ interval,
thus there can not be any singularity points. This is consistent
with the previous general property proved above.

To summarize the above results on singularity points and
accumulation points and slightly extend them, we present the
following classifications on singularity solutions in the integrable
system (\ref{Dyreduce}):
\begin{enumerate}
\item $\dot{\psi}_0=0$ (equal initial amplitudes):
\begin{itemize}
\item If $\dot{\zeta}_0 > -\sqrt{2}e^{\zeta_0/2}$, a singularity solution exists
at the single singularity point $\psi_0=0$. Here $M=0$, and $E>0$
for $\dot{\zeta}_0 > \sqrt{2}e^{\zeta_0/2}$ and $E<0$ otherwise;
\item If $\dot{\zeta}_0 < -\sqrt{2}e^{\zeta_0/2}$, there are no
singularity solutions for any ${\psi_0}$;
\end{itemize}

\item $\dot{\psi}_0\ne 0$, $\dot{\zeta}_0=0$
(non-equal initial amplitudes, zero initial velocities):
\begin{itemize}
\item If $\dot{\psi}_0^2 > 2e^{\zeta_0}$, singularity solutions exist
at two infinite sequences of $\psi_0$ values, accumulating at
$\psi_0=\{0^+, \pi^-\}$, or $\{\pi^+, 2\pi^-\}$, for
$\dot{\psi}_0<0$ and $\dot{\psi}_0>0$ respectively;
\item If $0<\dot{\psi}_0^2 < 2e^{\zeta_0}$: singularity solutions exist at
one infinite sequence of $\psi_0$ values, accumulating at
$\psi_0=0^+$ or $2\pi^-$ for $\dot{\psi}_0<0$ and $\dot{\psi}_0>0$
respectively;
\end{itemize}
On these sequences of singularity points, $M\ne 0$ and $E\ne 0$
generically (at the accumulation points, $M=0$, and $E<0$);

\item $\dot{\psi}_0\ne 0$ (the general non-equal initial amplitude
case):

In this case, the accumulation curve is the parts of curve
(\ref{sine_curve}) where $\psi_0$ satisfies the constraint
(\ref{cospsi2}). When $\dot{\psi}_0^2>2e^{\zeta_0}$, the entire
curve (\ref{sine_curve}) is the accumulation curve. When
$0<\dot{\psi}_0^2 \le 2e^{\zeta_0}$, the accumulation curve is
(\ref{sine_curve}) except portions of it which are centered at
$\psi_0=\pi$.

For $\dot{\zeta}_0$ below the accumulation curve, there are no
singularity points; at sufficiently large $\dot{\zeta}_0$ values,
there is a single singularity point.

At all these singularity values, $E$ and $M$ are non-zero
generically (except the accumulation points where $M=0$).
\end{enumerate}
It is noted that in the above classifications, case (2) is just a
special case of case (3), and can be readily deduced from (3). Case
(1) can be deduced from (3) as well under the limit $\dot{\psi}_0
\to 0$. But cases (1) and (2) are important special cases, hence we
listed them out separately.

\section{Origins of fractal structures in the non-integrable dynamical equations
 \setcounter{equation}{0}}

We have known from Figs. \ref{fig:ODEglobal}, \ref{ODElevel1} and
\ref{fig:ODE-zoom} that the non-integrable ODE system
(\ref{Dyreduce}) exhibits hill sequences and fractal structures
which coincide with those in the PDE simulations, but such
structures do not exist when this ODE system becomes integrable. The
natural question then is: where do the fractal structures in the
non-integrable system (\ref{Dyreduce}) come from? In this section,
we will establish through careful numerics that these fractal
structures bifurcate from singularity points of the integrable
system.
\begin{figure}
\includegraphics[width=80mm,height=65mm]{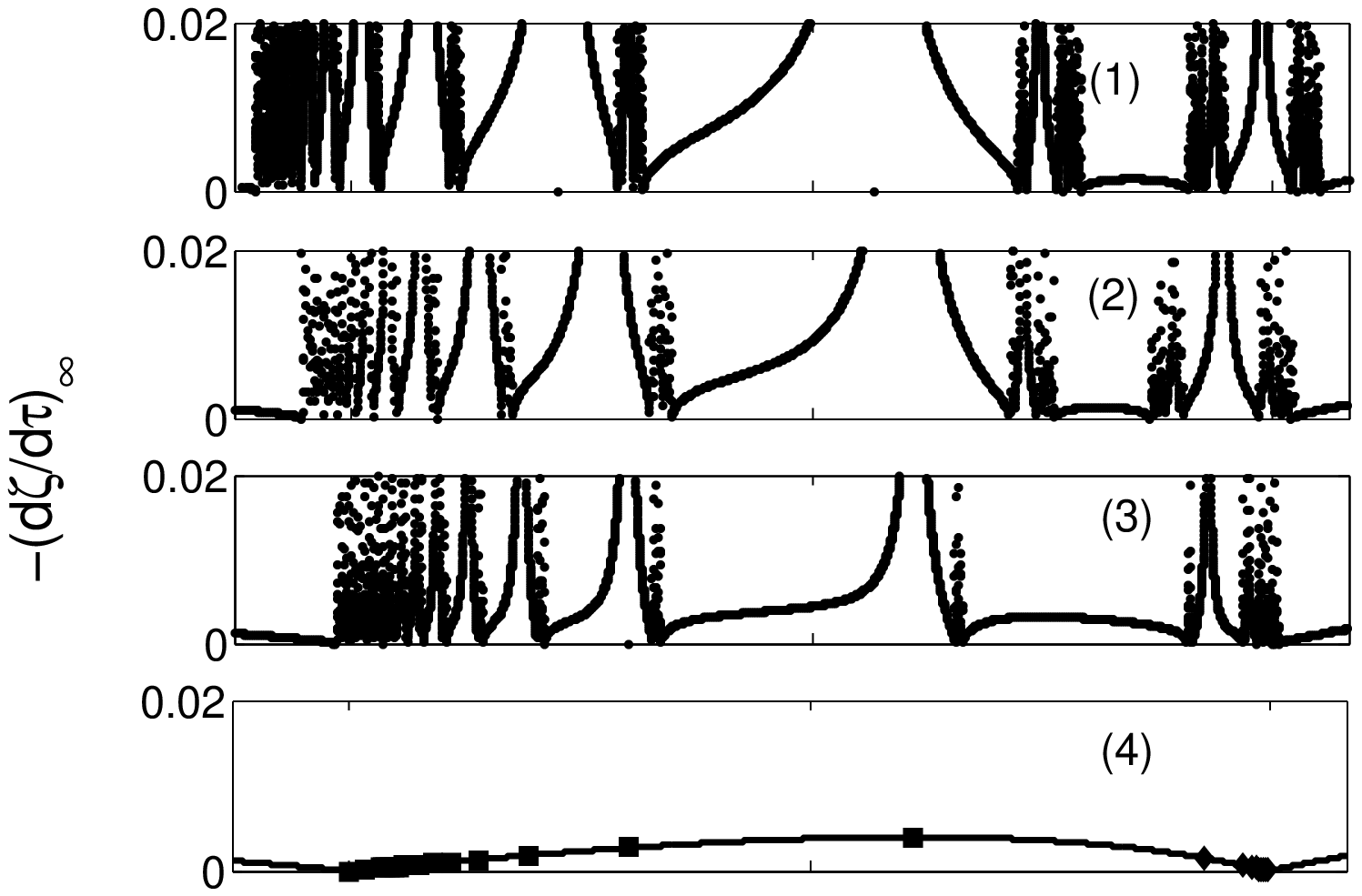}\\
\includegraphics[width=80mm,height=40mm]{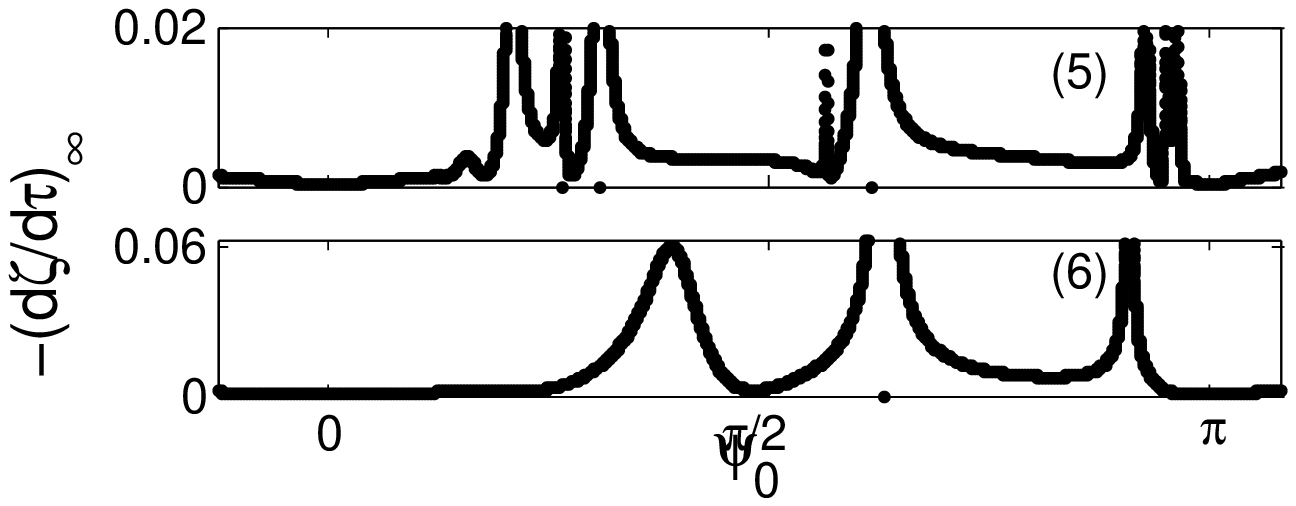}
\centering \caption{\label{ga2zero}~ The exit velocity versus
initial phase difference graphs in the ODE model (\ref{Dyreduce}) at
various values of $\varepsilon$: (1) 0.13665; (2) 0.036; (3) 0.0036;
(4) 0; (5) $-0.0036$; (6) $-0.036$. The initial conditions are given
in (\ref{ic_nonequal}). The squares and diamonds in (4) are
singularity points of the integrable system (see Fig. 9, bottom). }
\end{figure}

\begin{figure}
\includegraphics[width=80mm,height=50mm]{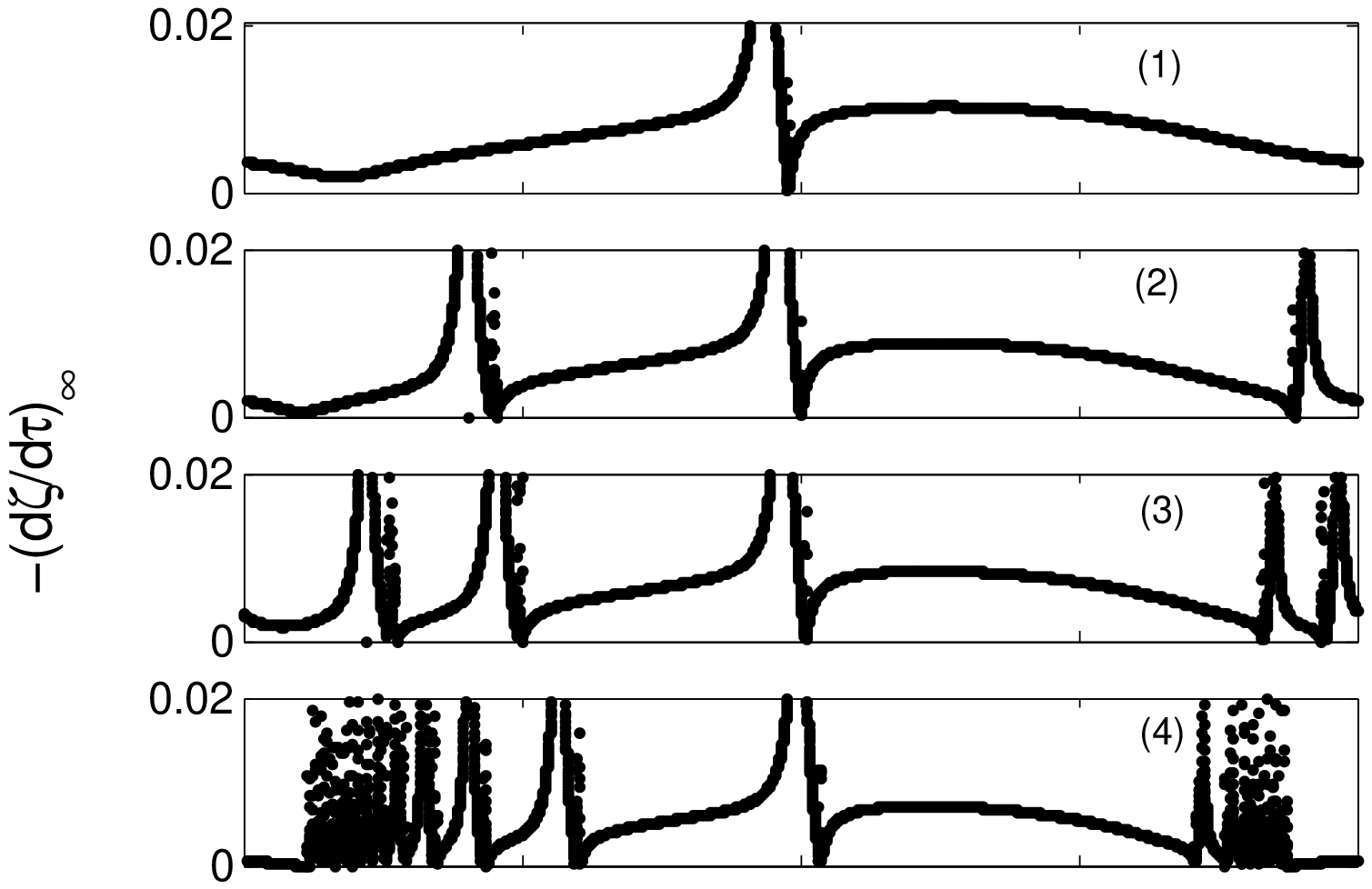}\\
\includegraphics[width=80mm,height=50mm]{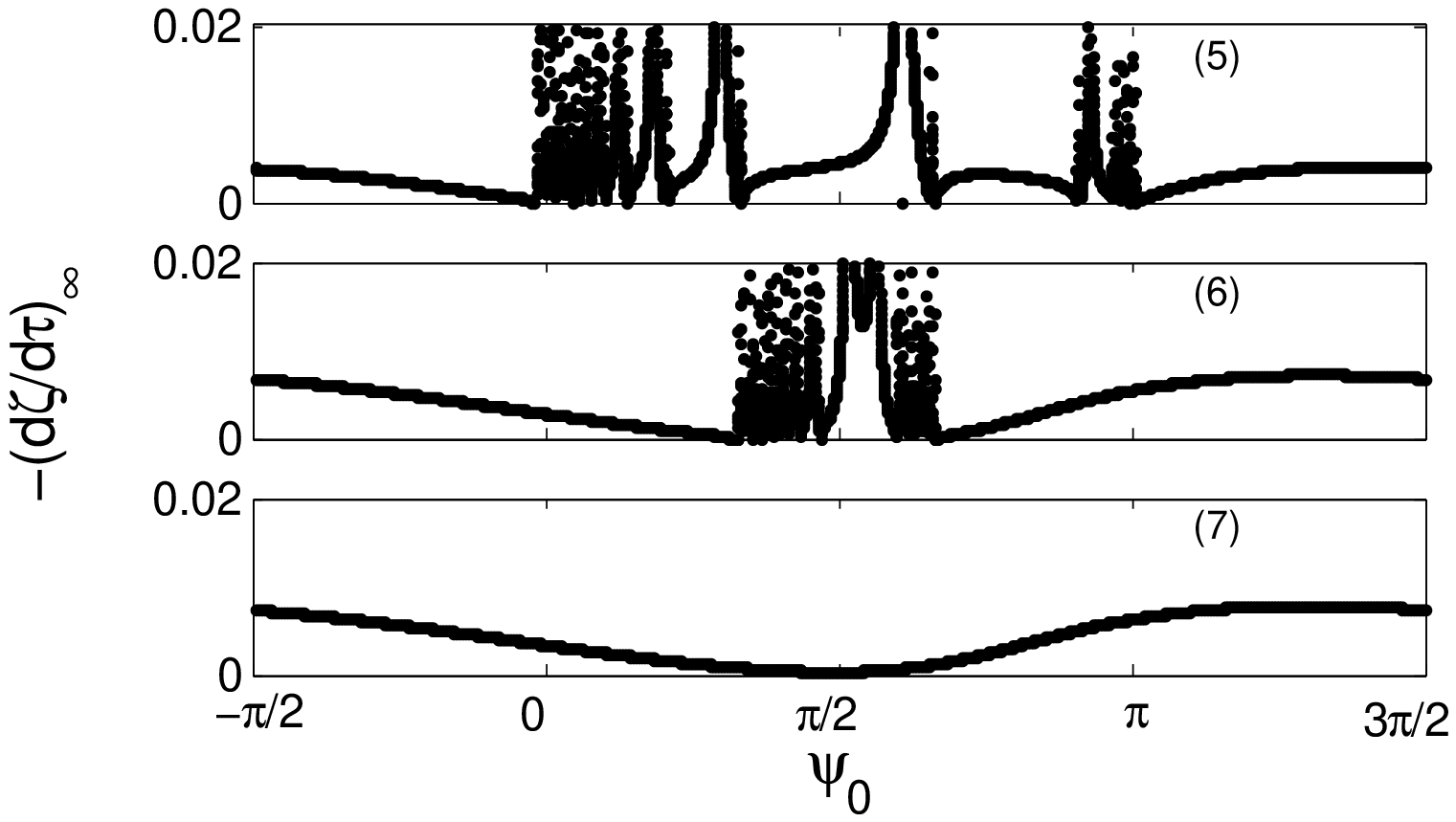}
\centering \caption{\label{evolution} The exit velocity versus
initial phase difference graphs in the ODE model (\ref{Dyreduce})
at various values of $\dot{\zeta}_0$: (1) 0.00707; (2) 0.00548;
(3) 0.00495; (4) 0.00350; (5) 0; (6) $-0.00350$; (7) $-0.00424$.
Here $\zeta_0=-10, \dot{\psi}_0=-0.01167$, and
$\varepsilon=0.0036$. }
\end{figure}

To determine the origin of these fractals, we take the same
initial conditions (\ref{ic_nonequal}) as in Fig.
\ref{fig:ODEglobal}(a), but gradually decrease the value of
$\varepsilon$ from 0.13665 of Fig. \ref{fig:ODEglobal}(a) down to
zero (the integrable case), then down further to negative values.
In this process, we closely monitor how the fractal structure of
Fig. \ref{fig:ODEglobal}(a) changes as $\varepsilon$ decreases.
The result is shown in Fig. \ref{ga2zero}. Here, the
$-\dot{\zeta}_\infty$ verse $\psi_0$ graphs are plotted at six
decreasing $\varepsilon$ values: $\varepsilon=0.13665, 0.036,
%
%
0.0036, 0, -0.0036$ and $-0.036$. We see that as $\varepsilon$
decreases from 0.13665 but above zero, primary hill sequences and
the fractal regions between them persist and are clearly visible
in Fig. \ref{ga2zero}(1, 2, 3). Indeed, we have zoomed into the
sensitive regions between primary hills in each of Figs.
\ref{ga2zero}(1, 2, 3), and obtained higher order structures which
look very similar to those shown in Fig. \ref{fig:ODE-zoom}. As
$\varepsilon \to 0^+$, our key observation is that, the peaks of
individual primary hills as well as the nearby fractal regions
collapse to sequences of points on the smooth
$-\dot{\zeta}_\infty$ curve of the integrable system (see Figs.
\ref{ga2zero}(3, 4)). Closer examination tells us that, these
sequences of points in Fig. \ref{ga2zero}(4) are nothing but the
two sequences of singularity points of the integrable system which
we plotted in Fig. \ref{figS}! In other words, hill sequences and
fractal structures in the non-integrable system bifurcate from the
singularity points of the integrable system! However, this
bifurcation is one-sided: as $\varepsilon$ decreases below zero,
no fractal regions appear, see Fig. \ref{ga2zero}(5). A finite
number of primary hills, reminiscent of primary hill sequences for
positive $\varepsilon$ values, do exist. But the whole graph is
smooth, and it has no fractal structures inside (even the
spike-looking parts of the graph in Fig. \ref{ga2zero}(5) turn out
to be smooth upon closer examination). Furthermore, as
$\varepsilon$ decreases further below zero, the number of primary
hills keeps decreasing, and the graph becomes more smooth, see
Fig. \ref{ga2zero}(6). Thus, fractal structures are a signature of
the non-integrable system (\ref{Dyreduce}) only for positive
values of $\varepsilon$, not negative values of $\varepsilon$.

To further substantiate our claim on fractal structures of the
non-integrable system bifurcating from singularity points of the
integrable system, we tune initial conditions so that singularity
points in the integrable system gradually disappear in the
$\psi_0$ interval, and check if fractal structures in the
non-integrable system disappear as well (for small $\varepsilon$).
Specifically, we fix the $\zeta_0$ and $\dot{\psi}_0$ values in
Eq. (\ref{ic_nonequal}) and tune the $\dot{\zeta}_0$ value, as we
did in Fig. \ref{singularpoints}. The $\varepsilon$ value in Eq.
(\ref{Dyreduce}) is taken as $\varepsilon=0.0036$, which is very
small. Thus, the non-integrable system is weakly perturbed from
the integrable one. For the above initial conditions, singularity
points of the integrable system have been displayed in Fig.
\ref{singularpoints} in the $(\psi_0, \dot{\zeta}_0)$ plane. We
gradually decrease the $\dot{\zeta}_0$ value. For each
$\dot{\zeta}_0$, we numerically compute the exit velocity versus
$\psi_0$ graph in the perturbed (non-integrable) system
(\ref{Dyreduce}), and compare how this graph relates to
singularity points of the integrable system in Fig.
\ref{singularpoints}. To illustrate, we pick seven representative
$\dot{\zeta}_0$ values, which are 0.00707, 0.00548, 0.00495,
0.00350, 0, $-0.00350$ and $-0.00424$ in decreasing order. These
seven $\dot{\zeta}_0$ values are marked by horizontal dashed lines
in Fig. \ref{singularpoints}. As we can see from that figure, at
these seven $\dot{\zeta}_0$ values, the numbers of singularity
points in the $\psi_0$ interval are 1, 3, 5, $\infty$, $\infty$,
$\infty$, and 0 respectively. For each of these seven
$\dot{\zeta}_0$ values, the corresponding exit velocity versus
$\psi_0$ graph in the perturbed system (\ref{Dyreduce}) is shown
in Fig. \ref{evolution}. We notice from this figure that the
numbers of primary hills and fractal regions near these hills at
these $\dot{\zeta}_0$ values are equal to 1, 3, 5, $\infty$,
$\infty$, $\infty$, and 0 respectively
--- exactly like singularity points in the integrable system! In
particular, when singularity points in the integrable system
disappear, so do primary hills and fractal structures in the weakly
perturbed non-integrable system. Furthermore, the locations of
primary hills and fractal regions closely follow those of
singularity points of the integrable system. Thus, the connections
between them are unmistakable. Fig. \ref{evolution}, together with
Fig. \ref{ga2zero}, establishes beyond doubt that primary hills and
fractal structures in the non-integrable system (\ref{Dyreduce})
indeed bifurcate from singularity points of the integrable system.

\begin{figure}
\includegraphics[width=70mm,height=55mm]{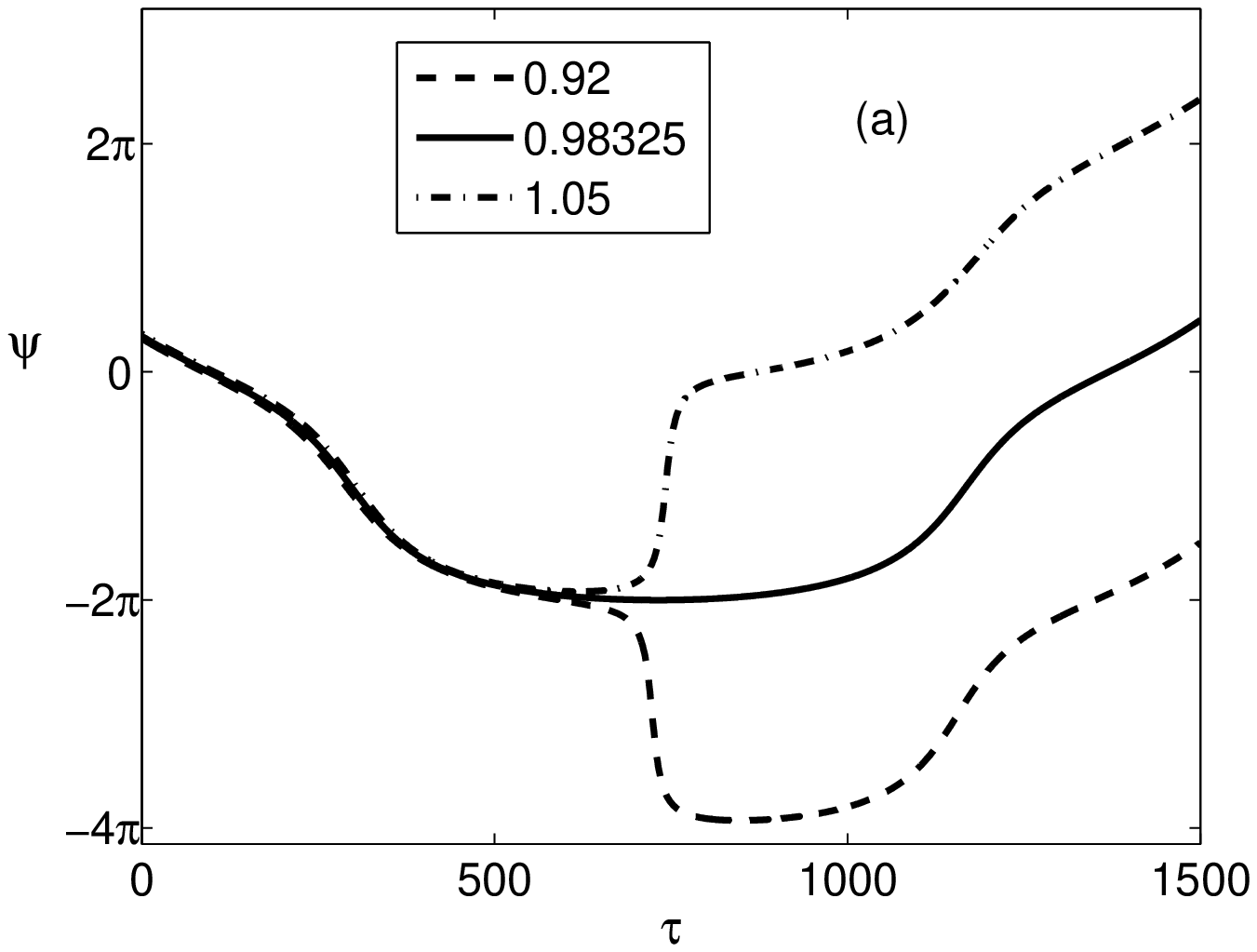}
\includegraphics[width=70mm,height=55mm]{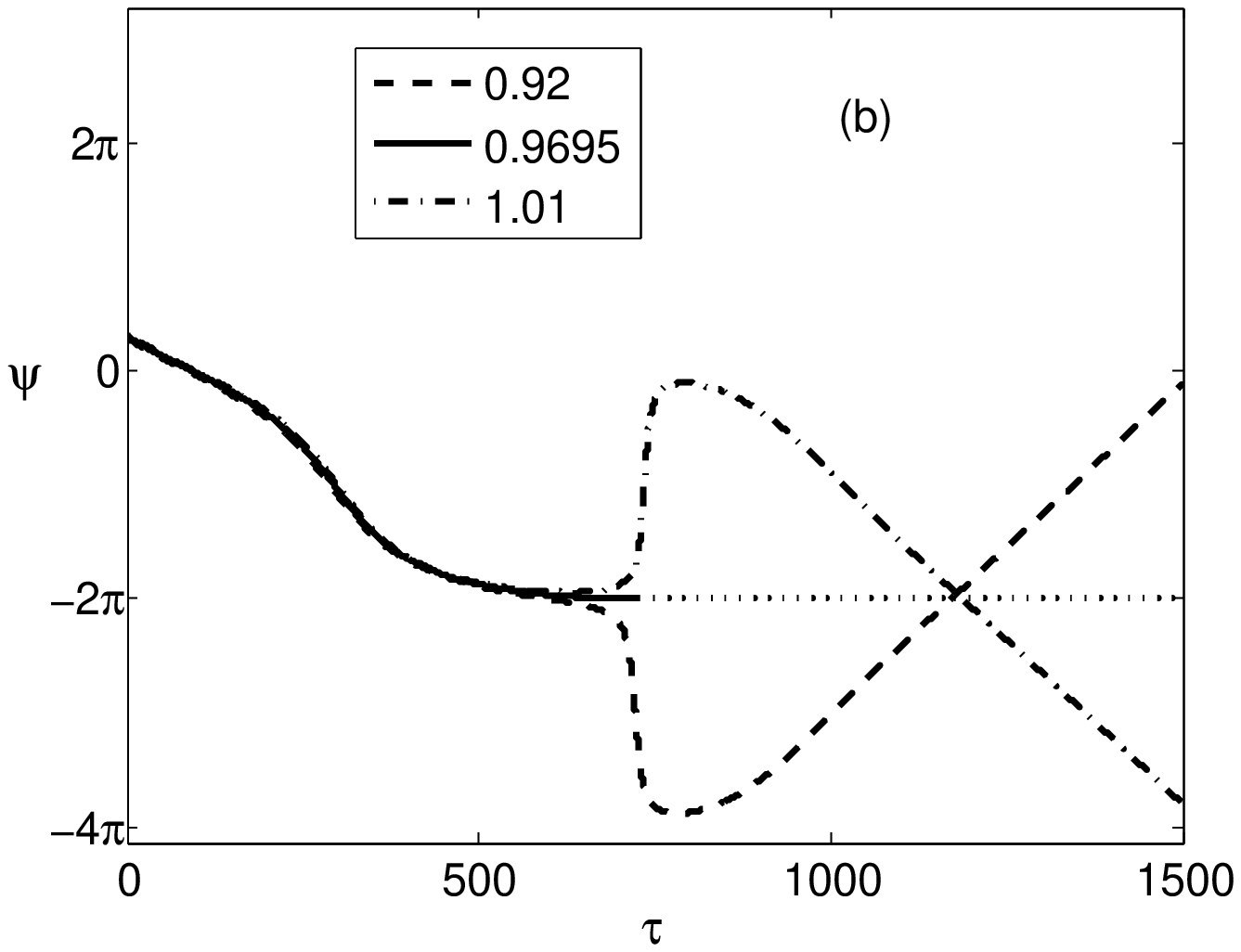}
\caption{\label{senstive} ~Evolutions of $\psi$ versus $\tau$ at a
singularity point $\psi_0$ and its left and right neighbor points in
the dynamical system (\ref{Dyreduce}) with initial conditions
(\ref{ic_nonequal}). (a) $\varepsilon=0$ (the integrable case); here
$\psi_0=0.98325$ is the singularity point which is marked in Fig.
\ref{figS} (bottom panel); the left and right neighbor points are
taken as $\psi_0=0.92$ and 1.05; (b) $\varepsilon=0.0036$ (the
positively perturbed case); here $\psi_0=0.9695$ is the singularity
point; its left and right neighbor points are taken as $\psi_0=0.92$
and 1.01. }
\end{figure}

The bifurcation of fractal structures from singularity points of the
integrable system indicates that near such points, the solutions of
the perturbed system (\ref{Dyreduce}) are very sensitive to initial
conditions. To shed light on why this sensitivity occurs, we present
some numerical results below. First, we look at the integrable
system (with $\varepsilon=0$). Taking the initial conditions as
(\ref{ic_nonequal}), evolutions of $\psi$ versus $\tau$ at the
singularity point $\psi_0=0.98325$ (marked in Fig. \ref{figS},
bottom panel) and its left and right near neighbors $\psi_0=0.92$
and 1.05 are plotted in Fig. \ref{senstive}(a) using the solution
formula (\ref{GSolu}). An interesting feature about these evolutions
is that for initial $\psi_0$ values at the two sides of the
singularity point, the phase functions $\psi(\tau)$ have drastically
different trajectories as they go through the time $\tau\approx 700$
where the two solitary waves interact strongly (this time is the
singularity time of the singular solution at $\psi_0=0.98325$). For
$\psi_0$ below the singularity point, the phase sharply (but
continuously) decreases by $2\pi$, while for $\psi_0$ above the
singularity point, the phase sharply (but continuously) increases by
$2\pi$. Recall that $\dot{\psi}\propto \Delta \beta$ and it
determines the relative energy (amplitude) distributions among the
two solitary waves [see Eqs. (\ref{Eq3}), (\ref{scaling1}) and
(\ref{time_scale})], we know that on the two sides of the
singularity point, the energies have opposite distributions among
the two solitary waves during their strong interactions. However,
after the interaction is completed, the asymptotic slopes of the
three $\psi(\tau)$ trajectories in Fig. \ref{senstive}(a) are almost
the same, signaling that the interaction outcome is actually
insensitive to the $\psi_0$ values (the roughly $2\pi$ difference
between these phase trajectories does not affect the physical
solutions). This is why outcomes of weak interactions in the
integrable system (\ref{Dyreduce}) do not exhibit sensitive
dependence on initial conditions (see Fig. \ref{figS} bottom panel).
However, when system (\ref{Dyreduce}) is positively perturbed, the
results are completely different. To demonstrate, we take
$\varepsilon=0.0036$ now, while the initial conditions
(\ref{ic_nonequal}) remain the same. In this slightly perturbed
system, the solution develops finite-time singularity at the
singularity point $\psi_0=0.9695$, which is the counterpart of the
singularity point mentioned above in the integrable system. The
phase function at this singularity point is plotted in Fig.
\ref{senstive}(b) (solid line). (It is noted that in this perturbed
case, we do not have exact solution formulas, hence this solution
was obtained by numerically integrating Eq. (\ref{Dyreduce}). Due to
the finite-time singularity in the solution, our numerical
integration can not go beyond the singularity time $\tau \approx
700$. The solution beyond the singularity time, shown in Fig.
\ref{senstive}(b) as dotted lines, was inferred from our numerics at
nearby $\psi_0$ values.) On the two sides of the singularity point,
we select two nearby values $\psi_0=0.92$ and 1.01. The phase
trajectories at these $\phi_0$ values are also plotted in Fig.
\ref{senstive}(b). We see that as these trajectories go through the
time $\tau \approx 700$, one sharply decreases by $2\pi$, while the
other sharply increases by $2\pi$, similar to what happens in the
integrable case (see Fig. \ref{senstive}(a)). However, after these
sharp decreases/increases, the trajectories turn around and start to
move in the opposite direction. Eventually, these trajectories
approach drastically different asymptotic slopes (one positive and
the other one negative in fact), indicating that the interaction
outcomes are very different for these slight changes in the $\psi_0$
values. This is the phenomenon of sensitive dependence on initial
conditions which occurs in the perturbed system (\ref{Dyreduce})
(with $\varepsilon>0$), but not the integrable system (with
$\varepsilon=0$).

\begin{figure}
\includegraphics[width=85mm,height=40mm]{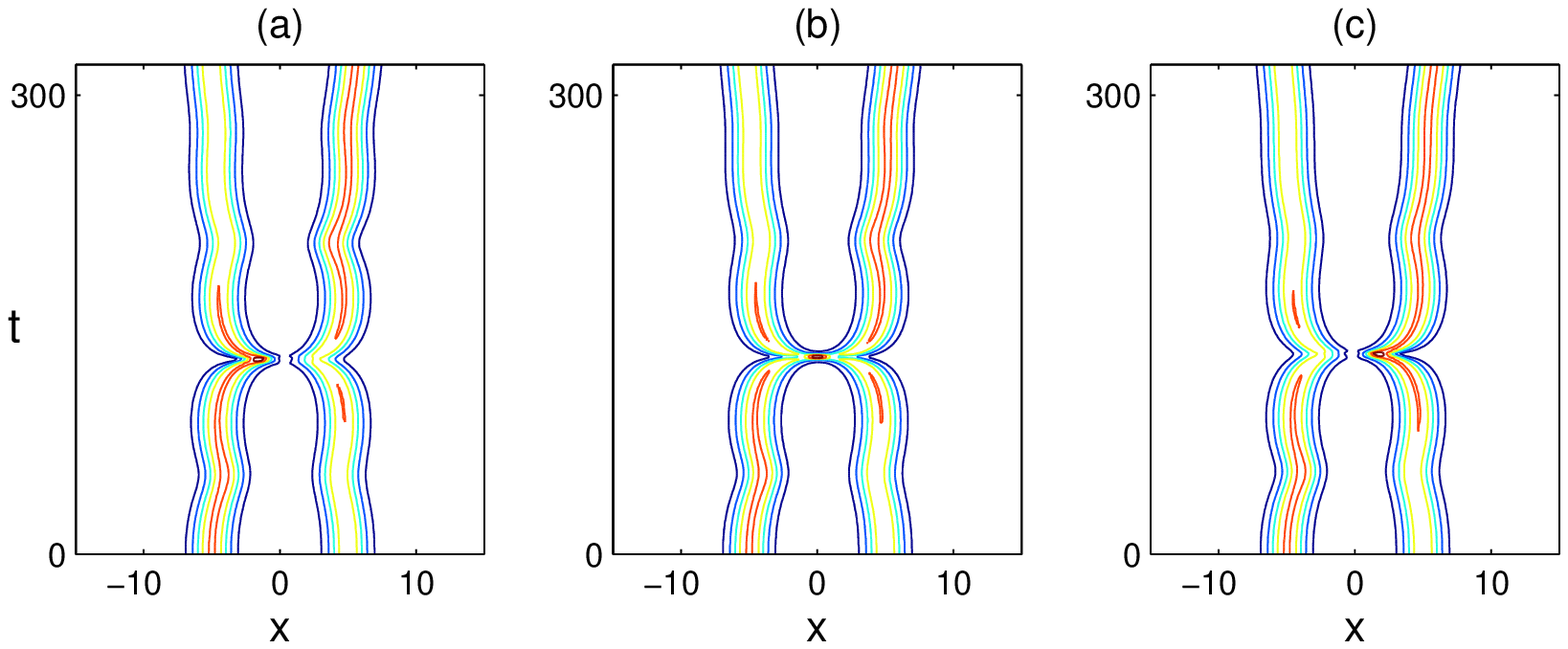}

\includegraphics[width=85mm,height=40mm]{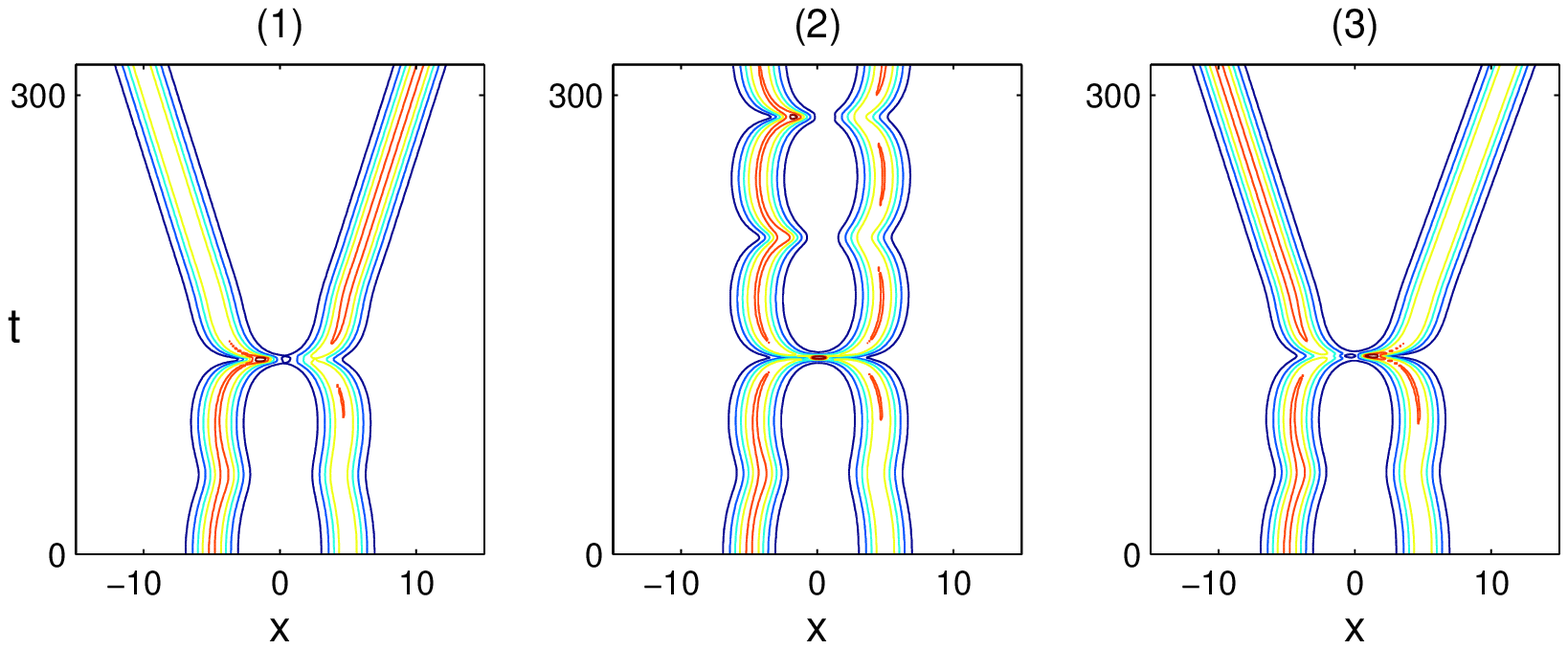}
\caption{Evolutions of $|U(x, t)|$ in the PDE (\ref{eqn:1}) with
cubic-quintic nonlinearity (\ref{cqNL}), corresponding to the ODE
simulations of Fig. \ref{senstive}. Top row: the integrable (NLS)
equation ($\gamma=0$); (a) $\Delta\phi_0=0.92$, (b)
$\Delta\phi_0=0.98325$, (c) $\Delta\phi_0=1.05$. Bottom row: the
perturbed case (with $\gamma=0.0010$); (1) $\Delta\phi_0=0.92$; (2)
$\Delta\phi_0= 0.972$; (3) $\Delta\phi_0=1.01$. The values of other
initial parameters are given in the text.  \label{contour}}
\end{figure}

It is also enlightening to look at this sensitive dependence on
initial conditions from the viewpoint of PDE evolutions. To
illustrate, we take the cubic-quintic nonlinearity (\ref{cqNL}) in
Eq. (\ref{eqn:1}), and take $\alpha=1$, $\beta_0=1$. Then for the
$\varepsilon$ values and initial conditions used in the ODE
simulations of Fig. \ref{senstive}, and in view of the variable
rescalings (\ref{scaling1}) and (\ref{time_scale}), the
corresponding PDE parameters for Fig. \ref{senstive}(a) (the
integrable case) are $\gamma=0$, $\Delta \beta_0=-0.066016$, $\Delta
V_0=0$, $\Delta x_0=10$, and the corresponding PDE parameters for
Fig. \ref{senstive}(b) (the perturbed case) are $\gamma=0.0010$,
$\Delta \beta_0=-0.066$, $\Delta V_0=0$, $\Delta x_0=10$. For these
PDE parameters, the PDE evolution results (in the form of contour
plots) at three $\Delta \phi_0$ values corresponding to those in the
ODE simulations of Fig. \ref{senstive} are displayed in Fig.
\ref{contour}. In the integrable (NLS) case (top row of Fig.
\ref{contour}), we take the three $\Delta \phi_0$ values exactly the
same as those in Fig. \ref{senstive}(a), i.e. $\Delta \phi_0=0.92,
0.98325, 1.05$. In this case, at the lower $\Delta \phi_0$ value,
the left solitary wave retains its higher energy at the collision
time; at the singularity point of $\Delta \phi_0$, the two waves
completely coalesce at the collision time, signaling the singularity
formation in the ODE system; at the higher $\Delta \phi_0$ value,
the right solitary wave gets higher energy at the collision time.
However, after interaction, the two waves always separate, and the
right wave always gets higher energy, in all three cases. Recall
that before interaction, the left wave has higher energy, thus we
can call these interaction outcomes transmission. In these
interactions, even though the intermediate process (especially the
collision segment) rather strongly depends on the initial phase
difference $\Delta \phi_0$, the interaction outcome is insensitive
to it. These PDE evolution results completely resemble the ODE
simulations in Fig. \ref{senstive}(a). In the perturbed
(non-integrable) case, the PDE simulation results are quite
different from the integrable ones (as in the ODE simulations). In
the perturbed case, we take the three $\Delta \phi_0$ values to be
$0.92, 0.972$ and $1.01$. Notice that the first and third of these
$\Delta \phi_0$ values are exactly the same as those in the ODE
simulations of Fig. \ref{senstive}(b), while the middle $\Delta
\phi_0$ value of $0.972$ is slightly different from the
corresponding ODE value of $0.9695$. This slight difference in the
middle $\Delta \phi_0$ value is necessary in order for the
corresponding ODE and PDE simulations to exhibit the same behaviors,
and this difference is due to the modeling error of the PDE
evolutions by the ODE system (\ref{Dyreduce}). At $\Delta
\phi_0=0.92$ (see Fig. \ref{senstive}(1)), the interaction outcome
is similar to the integrable ones (top row of Fig. \ref{senstive})
in that it is also transmission. But at $\Delta \phi_0=0.972$ (Fig.
\ref{senstive}(2)), the two waves strongly coalesce, then form an
oscillating bound state. At $\Delta \phi_0=1.01$, on the other hand,
the two exiting waves have opposite energy distributions from Fig.
\ref{senstive}(1); this interaction outcome can be called
reflection. Thus, in the perturbed case, the interaction outcome is
sensitive to initial conditions, which distinctively contrasts the
integrable case. Again, these PDE evolution results for the
perturbed case completely resemble the ODE simulations in Fig.
\ref{senstive}(b).

The above ODE and PDE simulations corroborate the fact that the
source of this sensitive dependence on initial conditions in the
perturbed system lies in the finite-time singularities of solutions
in the integrable dynamical system (\ref{Dyreduce}). From the PDE
point of view, the origin of this sensitive dependence can be traced
to the coalescing of the two solitary waves in the integrable PDE
system. At the moment, our understanding on this sensitive
dependence and fractal structures in the perturbed system is still
very limited. For instance, we can not yet explain any quantitative
details inside these fractal structures, nor can we explain why this
sensitive dependence occurs only for one-sided perturbations of the
integrable system (with $\varepsilon>0$). These are non-trivial
questions which merit further analysis, but they are beyond the
scope of the present article.


The fact of primary hills and fractal structures in the
non-integrable system (\ref{Dyreduce}) bifurcating from singularity
points of the integrable system has far reaching consequences. One
important consequence is that, the main features of primary hill
sequences shown in Figs. \ref{pdelevel1}(a) and \ref{ODElevel1}(a)
for PDEs and ODEs can now be analytically explained. For instance,
the life-time formula (\ref{time_PDE}) for primary hill sequences in
the weakly perturbed system (\ref{Dyreduce}) is nothing but the
analogous collision-time (singularity time) formula
(\ref{collide_time}) for sequences of singularity points in the
integrable system. To make a quantitative comparison between these
formulas, we take the initial conditions (\ref{ic_nonequal}) which
was used in the PDE and ODE simulations of Figs. \ref{pdelevel1}(a)
and \ref{ODElevel1}(a). When the time rescaling (\ref{time_scale})
is recovered, the collision-time formula (\ref{collide_time}) of the
integrable system becomes
\begin{equation} \label{time_anal}
0.0839\tilde{t}_n=2n\pi +\pi,
\end{equation}
which compares very favorably with the life-time formulae
(\ref{time_PDE}), (\ref{PDEvalue}) and (\ref{ODEvalue}) in direct
PDE and ODE simulations. The small differences in the $\omega$ and
$\delta$ values between the analytical formula (\ref{collide_time})
and the PDE/ODE ones (\ref{time_PDE}) are caused by the not-so-small
value of $\varepsilon=0.13665$. As $\varepsilon \to 0$, these
quantitative differences will vanish. Regarding the locations of
individual hills in the primary-hill sequence, they are described by
the formula (\ref{location_sing}) for singularity-point locations of
the integrable system when $\varepsilon \ll 1$. Note that the form
of this formula is different from all previous ones on window
sequences in solitary-wave collisions
\cite{Campbell2,Campbell3,Kivshar1,TanYang,Goodman_Haberman2}.

For $\varepsilon>0$, each primary hill is paired with a sensitive
(fractal) region at its foot (see Figs. \ref{ga2zero} and
\ref{evolution}). Similar to primary hill sequences, the locations
of these fractal regions are described by the same formula
(\ref{location_sing}) in the limit $\varepsilon \to 0^+$.

The fact of primary hills and fractal structures bifurcating from
singularity points of the integrable system also explains major
features of interaction results in Fig. \ref{fig:exPDE}(a) for the
exponential nonlinearity (\ref{exp_non}). We have noticed that,
unlike Fig. \ref{cqnonequal}, this graph has only one infinite
sequence of primary hills accumulating toward the left (the right
sequence of Fig. \ref{cqnonequal} is absent). This phenomenon is due
to the fact that for the choices of initial conditions for Fig.
\ref{fig:exPDE}(a), there is only one infinite sequence of
singularity points in the integrable system. To see it, we first
calculate the $f, g$ and $\varepsilon$ values for Fig.
\ref{fig:exPDE}(a), which are found to be
\begin{equation}
f=228.8211, \quad g=231.91770, \quad \varepsilon=0.01353.
\end{equation}
Thus in the scaled dynamical equation (\ref{Dyreduce}), the initial
conditions corresponding to those for Fig. \ref{fig:exPDE}(a) are
\begin{equation}
\zeta_0=-12.13260, \quad \dot{\zeta}_0=0, \quad
\dot{\psi}_0=-0.00297.
\end{equation}
Notice that $\dot{\psi}_0^2 < 2e^{\zeta_0}$, thus according to the
classifications of singularity points in the end of the previous
section (case 2), the integrable equation (\ref{Dyreduce}) with the
above initial conditions has only one sequence of singularity points
in the $\psi_0$ interval, accumulating to the left toward
$\psi_0=0^+$. This is in perfect agreement with the primary-hill
sequence of Fig. \ref{fig:exPDE}(a) from direct PDE simulations.

In many of the interaction results presented in this paper, the exit
velocity versus $\psi_0$ graphs have infinite sequences of primary
hills (see Figs. \ref{cqnonequal} and \ref{fig:exPDE} for instance);
when zooming into the sensitive regions between primary hills, one
gets infinite sequences of secondary hills. It is important to
understand that these two infinite sequences are pure coincidence,
and are totally un-related. Each primary hill corresponds to a
particular singularity point of the integrable system, thus the
number of primary hills is equal to the number of singularity points
in the integrable system. This number can be either infinite or
finite, depending on the choices of initial conditions. For
instance, Figs. \ref{evolution}(1, 2, 3) have 1, 3 and 5 primary
hills, corresponding to the same numbers of singularity points on
the top three dashed lines of Fig. \ref{singularpoints}. On the
other hand, at the foot of each primary hill, there is always an
{\it infinite} sequence of secondary hills (when $\varepsilon>0$).
In other words, secondary hills always exist as an infinite, not
finite, sequence. For example, if one zooms into each of the three
sensitive regions at the foot of the three primary hills in Fig.
\ref{evolution}(2), one always gets an infinite sequence of
secondary hills. Thus secondary-hill structures are un-related to
primary-hill structures. If we zoom into the sensitive regions
between secondary hills, we always get infinite sequences of
tertiary hills which are very similar to the sequences of secondary
hills both qualitatively and quantitatively (see Figs.
\ref{fig:PDEzoom}(b, c)). This process can continue indefinitely.
Thus, our conclusion is that sensitive regions between primary hills
are fractal structures (in the sense that portions of these
structures, when amplified, are the same as the strcutures
themselves). But the whole graph with primary hills is not a
fractal.

\section{Applications to the generalized NLS equations with various nonlinearities \setcounter{equation}{0}}
In previous sections, we have shown that for the cubic-quintic and
exponential nonlinearities at selected parameters ($\alpha=1,
\gamma=0.04, \beta_0=1$ for the former, and $\beta_0=2.3$ for the
latter), weak interactions of solitary waves exhibit hill sequences
and fractal structures for a wide range of initial conditions, and
the reduced ODE model (\ref{Dyreduce}) accurately captures these
interaction dynamics both qualitatively and quantitatively. In this
section, we consider a larger question: for a given form of
nonlinearity in the PDE (\ref{eqn:1}), can it exhibit fractal
structures? For example, with the cubic-quintic nonlinearity
(\ref{cqNL}), for what parameters $\alpha$ and $\gamma$ can one
possibly find fractal structures? This question can be answered by
applying our previous results on the ODE model (\ref{Dyreduce}). For
demonstration purpose, we will do so for three forms of
nonlinearity: cubic-quintic, exponential, and saturable
nonlinearities.

From the analysis of the ODE system (\ref{Dyreduce}) in the previous
section, we have found that fractal structures in weak interactions
can only occur for $\varepsilon>0$, not for $\varepsilon<0$. Thus,
once we have obtained the functional dependence of $\varepsilon$ on
system parameters, it will quickly become clear when fractal
structures can arise. The analytical expression for $\varepsilon$ is
given in Eq. (\ref{time_scale}). Notice that due to the
Vakhitov-Kolokolov stability criterion \cite{Kivshar_Agrawal,VK},
the solitary wave is linearly stable only when $P_\beta >0$, i.e.
$\varepsilon >-1$. Below, we will use the $\varepsilon$ formula in
(\ref{time_scale}) to calculate $\varepsilon$ for general system
parameters in the three nonlinearities mentioned above.

\begin{figure*}
\includegraphics[width=55mm,height=45mm]{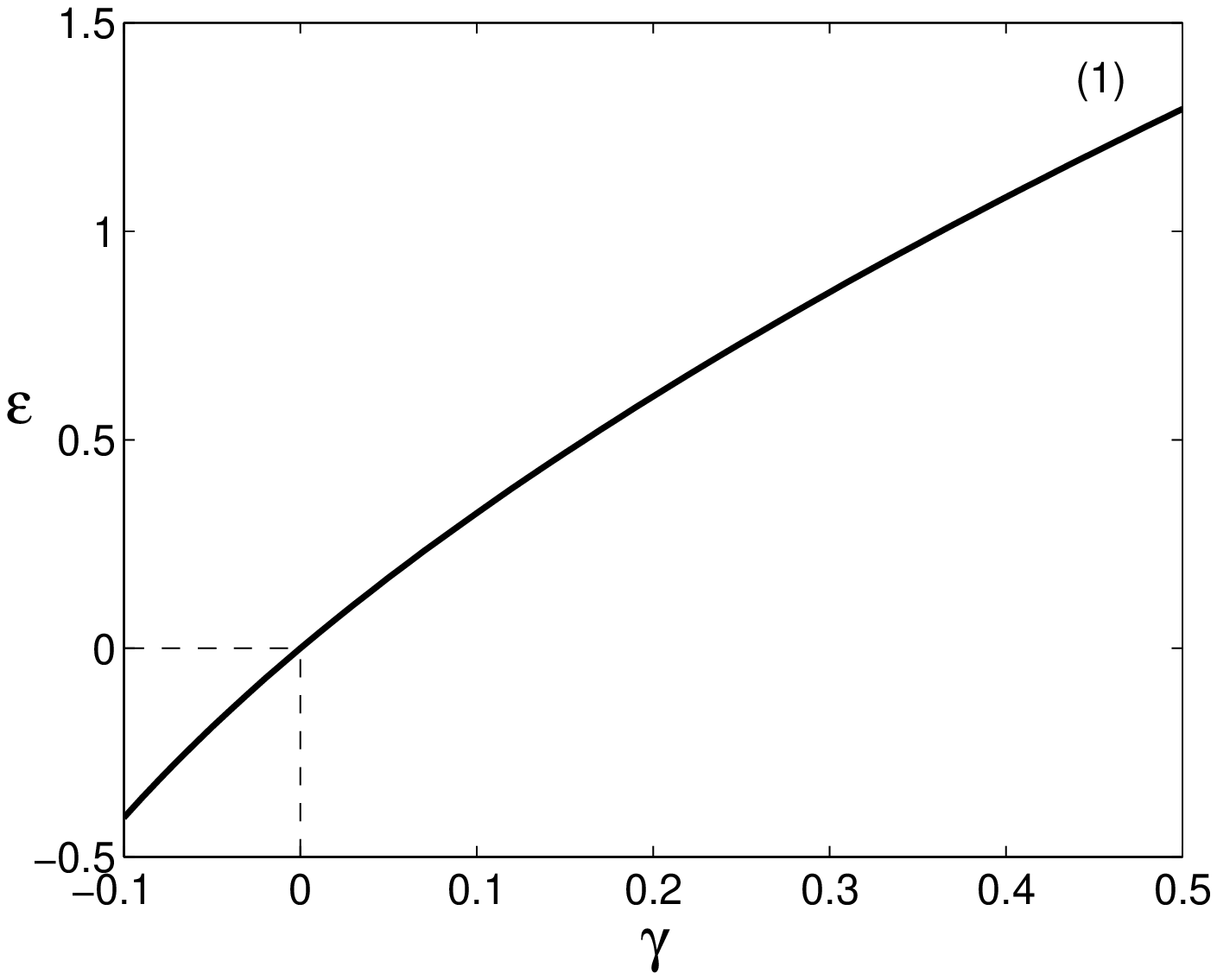}
\includegraphics[width=55mm,height=45mm]{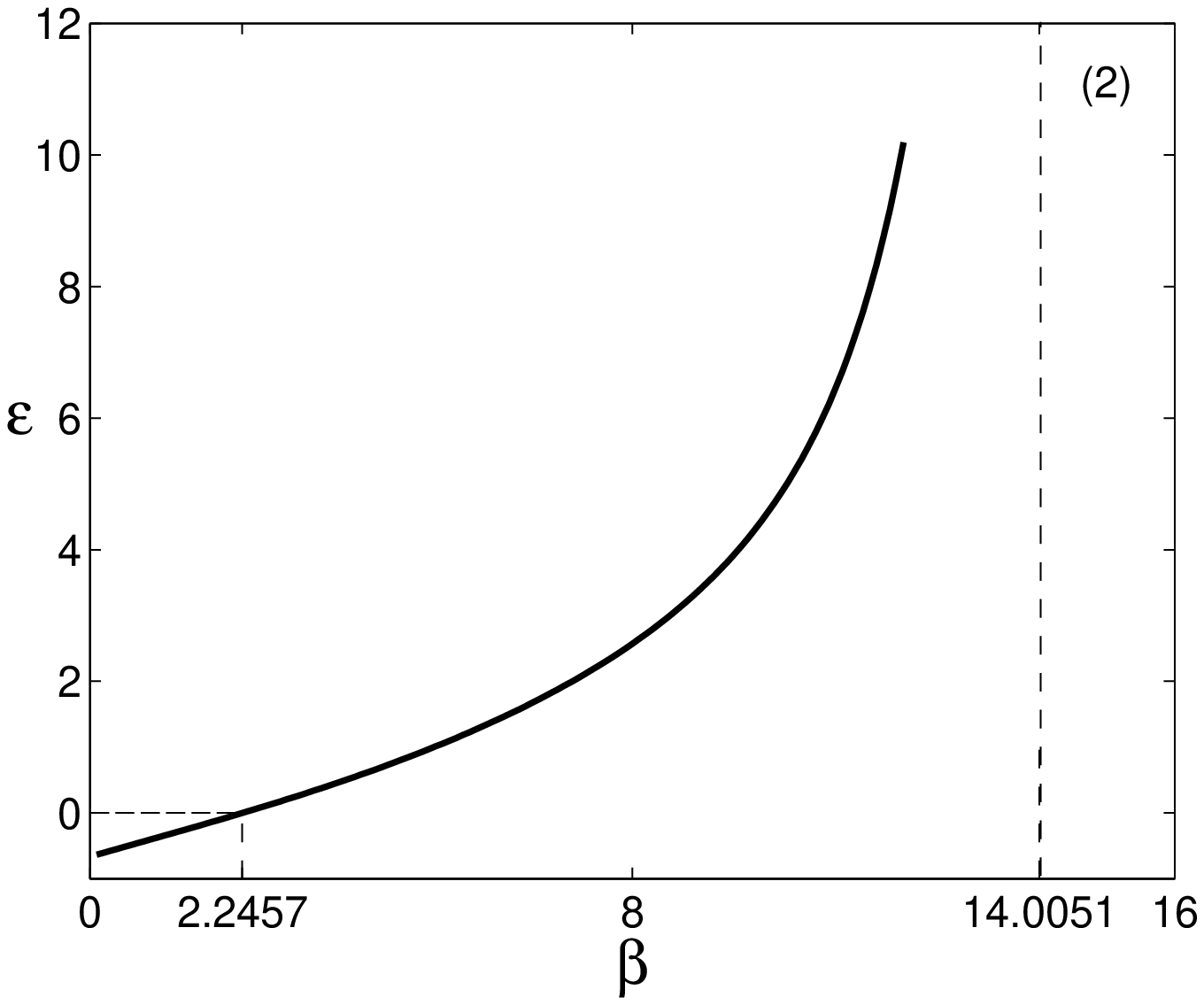}
\includegraphics[width=55mm,height=45mm]{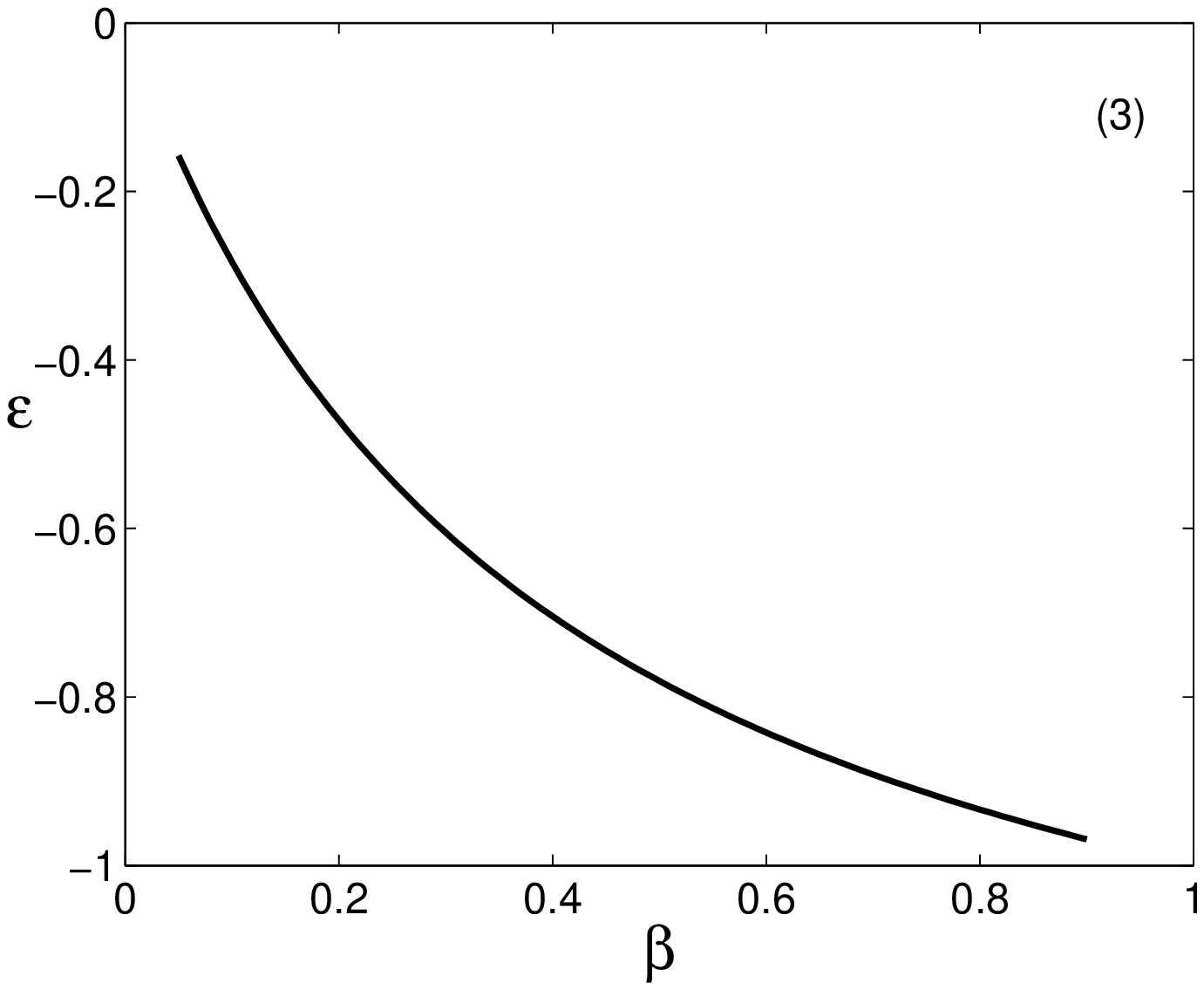}
\caption{\label{varepsilon}~ Graphs of $\varepsilon$ versus system
parameters for three different nonlinearities: (1) $\varepsilon$
verse $\gamma$ for the cubic-quintic nonlinearity (\ref{cqNL}); (2)
$\varepsilon$ verse $\beta$ for the exponential nonlinearity
(\ref{exp_non}); (3) $\varepsilon$ verse $\beta$ for the saturable
nonlinearity (\ref{sat_non}).}
\end{figure*}

First we consider the cubic-quintic nonlinearity (\ref{cqNL}). When
$\alpha<0$, we found that $P_\beta$ is always negative, i.e. the
solitary wave is always linearly unstable. Thus we only consider the
$\alpha>0$ case below. In this case, it is easy to see from Eqs.
(\ref{eqn:2}) and \ref{cqNL}) that by a rescaling of variables, we
can make $\alpha=\beta_0=1$. Thus the only remaining parameter for
this nonlinearity is $\gamma$. Using the analytical formula
(\ref{Pformula}) for $P$, we can obtain the dependence of
$\varepsilon$ on $\gamma$, which is plotted in Fig.
\ref{varepsilon}(1). From this graph, we see that $\varepsilon>0$
when $\gamma>0$, and $\varepsilon<0$ when $\gamma < 0$. Thus
\emph{fractal structures in this cubic-quintic model can appear only
when $\gamma>0$, not when $\gamma<0$}. If $\gamma=0$, this
cubic-quintic model reduces to the integrable NLS equation, and the
dynamical equations (\ref{Dyreduce}) reduce to the integrable case
(with $\varepsilon=0$) studied in Sec. 5. In this integrable case,
there is of course no fractal dependence in solitary wave
interactions.

Next, we consider the exponential nonlinearity (\ref{exp_non}). In
this case, the solitary wave depends only on the propagation
constant $\beta$, thus $\varepsilon$ depends only on $\beta$ as
well. The analytical expression for function $\varepsilon(\beta)$ is
not available, but this function can be easily determined by
numerical methods, and its graph is plotted in
Fig.\ref{varepsilon}(b). It is seen that this graph has two critical
propagation constants, $\bar{\beta}_a=2.2457$ where $\varepsilon=0$,
and $\bar{\beta}_b=14.0051$ where $\varepsilon=+\infty$. When
$\beta< \bar{\beta}_a$, $\varepsilon < 0$, thus fractal structures
do not exist; when $\bar{\beta}_a < \beta <\bar{\beta}_b$, thus
fractal structures can appear (see previous sections); when
$\beta>\bar{\beta}_b$, $\varepsilon < -1$, thus the solitary wave is
linearly unstable.

Next, we consider the saturable nonlinearity,
\begin{equation} \label{sat_non}
F(|U|^2)=1-\frac{1}{1+|U|^2},
\end{equation}
which is common in optics (for instance, in photorefractive crystals
\cite{Segev_PRE}). Here one is added in the above formula to make
$F(0)=0$ (this does not affect the solitary waves and their
interaction dynamics). In this case, $\varepsilon$ also depends only
on the propagation constant $\beta$. This dependence is computed
numerically and plotted in Fig.\ref{varepsilon}(c). We find that
$\varepsilon$ is negative for all values of $\beta$, thus fractal
structures can not exist in weak interactions of solitary waves for
this saturable nonlinearity. This conclusion is consistent with our
earlier results for the cubic-quintic nonlinearity, as the saturable
nonlinearity (\ref{sat_non}) resembles the cubic-quintic
nonlinearity (\ref{cqNL}) with $\alpha >0$ and $\gamma<0$. It is
noted, however, that for the saturable nonlinearity, weak
interactions of solitary waves can still exhibit some interesting
structures as shown in Figs. \ref{ga2zero}(5, 6), but these
structures are not fractal structures.

From the above three examples (as well as the previous section), we
see that the reduced ODE model (\ref{Dyreduce}) enables us to
accurately predict when and where fractal structures and hill
sequences appear in the space of initial parameters of solitary
waves. Based on this reduced model, a global and universal
understanding on weak interactions of solitary waves has been
achieved for the generalized NLS equations (\ref{eqn:1}) with
arbitrary forms of nonlinearity.

\section{conclusion and discussion}

In this paper, we have analyzed weak interactions of solitary waves
in the generalized nonlinear Schr\"{o}dinger equations with general
forms of nonlinearity. We have shown that these interactions exhibit
similar fractal dependence on initial conditions for different
nonlinearities. To analytically explain these universal fractal
structures, we derived a set of fourth-order dynamical equations for
the solitary-wave parameters using asymptotic methods. A remarkable
feature of these dynamical equations is that they contain only one
parameter, which is dependent on the specific form of nonlinearity.
When this parameter is zero, these dynamical equations are
integrable, and the exact analytical solutions are derived. When
this parameter is non-zero, the dynamical equations exhibit fractal
structures which match those in the original PDEs both qualitatively
and quantitatively. We have also investigated the origin of these
fractal structures, and found that they bifurcate from the
singularity points (i.e. initial conditions for singularity
solutions) in the integrable system. Based on this observation, an
analytical criterion for the existence and locations of fractal
structures is obtained. Lastly, we applied these analytical results
to the generalized nonlinear Schr\"{o}dinger equations with various
nonlinearities such as the cubic-quintic, exponential and saturable
nonlinearities, and predictions on their weak interactions of
solitary waves are presented.

Regarding the bifurcation of fractal structures from the integrable
dynamical equations, even though we have established that this
bifurcation occurs at the singularity points of the integrable
system, more challenging questions are to comprehensively analyze
how this bifurcation takes place, and to quantitatively predict the
detailed geometric structures inside these fractals. This has not
been done yet. Recently, Goodman and Haberman analyzed the
approximate ODE models for the collisions of solitary waves in three
physical systems where window sequences and fractal structures have
been reported \cite{Goodman_Haberman1, Goodman_Haberman2,
Goodman_Haberman3}. They found that the origin of window sequences
and fractal structures in these systems lies in the crossing of the
separatrix (homoclinic orbit). Analytical predictions on the
locations of window sequences in the ODE models were derived as
well. It is not clear at the moment whether similar analysis can be
performed for our system (\ref{Dyreduce}). This question is beyond
the scope of the present article, and will be left for future
studies.

\begin{acknowledgments}
The authors thank Drs. Richard Haberman, Roy Goodman and Meirong
Zhang for valuable discussions. This work was supported in part by
the Air Force Office of Scientific Research under grant USAF
9550-05-1-0379.
\end{acknowledgments}

\end{document}